\documentclass[11pt]{article}
\usepackage{latexsym,amsmath,natbib,psfig,fullpage,numinsec,amssymb,xcolor}
\usepackage{graphicx}
\usepackage{multirow}
\usepackage{booktabs,caption}
\usepackage[flushleft]{threeparttable}

\usepackage{algorithm2e}

\makeatletter
\RestyleAlgo{boxruled}
\renewcommand{\@algocf@capt@plain}{above}% formerly {bottom}
\makeatother

\newcommand{\real}{\ensuremath{\mathbb{R}}}
\newcommand{\bm}[1]{\boldsymbol{#1}}

\newtheorem{remark}{Remark}[section]
\newcommand{\ra}[1]{\renewcommand{\arraystretch}{#1}}

\usepackage{lipsum}

\newcommand\blfootnote[1]{%
  \begingroup
  \renewcommand\thefootnote{}\footnote{#1}%
  \addtocounter{footnote}{-1}%
  \endgroup
}

\title{High-Dimensional Spatial Quantile Function-on-Scalar Regression}
\author{Zhengwu Zhang, Xiao Wang, Linglong Kong, Hongtu Zhu}

\date{}

\begin{document}
\maketitle
\blfootnote{Zhengwu Zhang is  Assistant Professor of Biostatistics, Department of Biostatistics and Computational Biology, University of Rochester, Rochester, New York 14627. Xiao Wang is Professor of Statistics, Department of Statistics, Purdue University, West Lafayette, IN 47907. Linglong Kong is Associate Professor of Statistics, Department of Mathematical and Statistical Sciences, University of Alberta, Edmonton, AB Canada T6G 2G1. Hongtu Zhu is Professor of Biostatistics, Department of Biostatistics, University of North Carolina, Chapel Hill, NC  27599.}

\begin{abstract}
This paper develops a novel spatial quantile function-on-scalar regression model, which studies the conditional spatial distribution of a high-dimensional functional response given scalar predictors. With the strength of both quantile regression and copula modeling, we are able to explicitly characterize the conditional distribution of the functional or image response on the whole spatial domain. Our method provides a comprehensive understanding of the effect of scalar covariates on functional responses across different quantile levels and also gives a practical way to generate new images for given covariate values. Theoretically, we establish the minimax rates of convergence for estimating coefficient functions under both fixed and random designs. We further develop an efficient primal-dual algorithm to handle high-dimensional image data. Simulations and real data analysis are conducted to examine the finite-sample performance.\\

\end{abstract}

\section{Introduction}

Functional data analysis (FDA) has been an active area of research in the past decade.  Well-known monographs in this area include \cite{ramsay2005functional}, \cite{ramsay2007applied}, \cite{ bowman2010functional} and \cite{ferraty2006nonparametric}. Functional regression, particularly the functional linear regression model (FLM), has been extensively studied.  Functional data can be treated as
either response variables or covariate predictors \citep{greven2017general}. In FLM, the mean dependence is modeled through a linear model generalizing the standard multiple linear regression model. The literature on FLM is vast. For example, scalar-on-function regression (a continuous
response variable regressed on functional covariates) has been studied by \cite{tony2006}, \cite{crambes2009smoothing}, \cite{yuan2010reproducing}, \cite{hall2017} and \cite{wang2017generalized}.   Function-on-function regression (a functional response regressed on functional predictors) has been investigated by \cite{yao2005functional}, \cite{ramsay2005functional}, \cite{ivanescu2015penalized}, and \cite{sun2017optimal}. Function-on-scalar regression (a functional response regressed on scalar predictors) has been studied by Ramsay and Silverman (\cite{ramsay2005functional}, Ch13, 2005), \cite{reiss2010fast} and \cite{goldsmith2016assessing}. The aim of this paper is 
to develop a spatial quantile function-on-scalar regression model and investigate 
its related computational and theoretical issues. 

Different from the ordinary regression that only models conditional mean dependence, quantile regression as a more comprehensive procedure estimates conditional quantiles of the response variable. Since  \cite{koenker1978regression}, quantile regression has been an emerging field of statistical research and has been widely applied in many disciplines including neuroimaging, finance, and economics. Quantile regression is attractive because it estimates conditional quantiles and allows statistical inference on the response's entire conditional distribution.  Most importantly, quantile regression does not require a specified error distribution, providing a flexible framework for modeling complex non-Gaussian data.  To date, while many studies have examined different quantile regression models, a limited number of these consider functional data. \cite{cai2008nonparametric} studied quantile regression methods for a set of smooth coefficient time series models. \cite{koenker2004quantile} and \cite{yi2009median} developed quantile regression methods for longitudinal data. \cite{wang2009quantile} studied semi-parametric quantile auto-regression models in partially linear varying coefficient models using splines. \cite{li2007quantile} proposed a kernel quantile regression to estimate conditional quantile functions given training scalar covariates and responses. \cite{kato2012estimation} studied functional quantile regression with functional covariates and a scalar response.

In this paper, motivated by   medical imaging analysis,   we propose a novel spatial quantile function-on-scalar regression model (denoted by SQR) that studies the conditional spatial distribution of a functional response given scalar predictors. We are particularly interested in the case where the responses are high-dimensional functions or images obtained from $n$ independent subjects. Let the functional response be $\{Y(s): s\in {\cal S}\}$ on a field ${\cal S}$ and the scalar predictors be ${\bf x} \in \real^p$. Our SQR model includes two major components. The first component is to model the marginal conditional distribution of $Y(s)$ given ${\bf x}$ at a fixed location $s$. This task can be achieved by  assuming that the $\tau$-level conditional quantile of $Y(s)$ is assumed to be a linear function of ${\bf x}$ such that $Q_{Y(s)}(\tau|x) = {\bf x}^T \boldsymbol{\beta}_\tau(s)$, where $Q_{Y(s)}(\cdot|{\bf x})$ is the conditional quantile function of $Y(s)$ given ${\bf x}$ at a location $s$ and $\boldsymbol{\beta}_\tau(s) = (\beta_{\tau 1}(s),...,\beta_{\tau p}(s))^T$
contains $p$ unknown coefficient functions.   The second component is to model the joint conditional distribution of $Y(s)$ among spatial locations via a copula model. We adopt a parametric copula to specify the joint distribution of $F_{(s,\bf x)}(Y(s))$, where $F_{(s,\bf x)}(\cdot)$ is the conditional cumulative distribution function (CDF) of $Y(s)$ given ${\bf x}$ at $s$ and $F_{(s,\bf x)}(Y(s))$ follows a uniform distribution marginally. With these two components, we are able to explicitly characterize the conditional joint distribution of $Y(s)$ on the whole spatial domain ${\cal S}$, forming a generative model that given $\bf x$, we can simulate the image data $Y$. The proposed method provides a framework to comprehensively understand the effects of scalar covariates (e.g., age, gender and disease status) on an image response,  and a practical way to generate new images given covariates. 

Although SQR has been studied in the literature \citep{reich2011, reich2012spatiotemporal,yang2015,su2007,kostov2009,hallin2009,lu2014}, our method differs from them in several important aspects.      First, 
our   SQR is  primarily developed for  handling  large-scale image data $Y(s)$ with dense grids across multiple subjects. 
In contrast, most existing works on SQR focus on the development of statistical methods for analyzing simple spatial or longitudinal data  \citep{reich2011, reich2012spatiotemporal,yang2015,hallin2009}. For instance, \cite{hallin2009} considered SQR for a strictly stationary   real  random field  and 
proposed a local linear estimator of  spatial quantile function.  
Second,  we employ a parametric  copula  model to directly delineate the spatial dependence
of image data within each subject. 
In contrast, most existing 
copula models  were applied to quantile regression for different purposes and data types \citep{chen2009, bouye2013, kraus2017,backer2017,wang2019}.  
For instance, in  \cite{wang2019},   the copula was used to model the temporal dependence of longitudinal data, while  it is assumed a linear quantile regression model with  $Q_{Y(t)}(\tau|{\bf x}) = {\bf x}^T \boldsymbol{\beta}_0(\tau)$ at a fixed time point $t$ and for given covariates ${\bf x}$.  However, our model assumes that the components of $\boldsymbol{\beta}_\tau(s)$  are functions of both $\tau$ and the spatial location. 
Third, 
the unknown coefficient functions $\boldsymbol{\beta}_\tau(s)$  are assumed to reside in a reproducing
kernel Hilbert space and estimate them through minimizing a quantile check function \citep{koenker1978regression} plus a roughness penalty.  
Furthermore, we  estimate the unknown parameters of the copula model by using generalized least squares \citep{cressie1985fitting,genton1998variogram}.  
Most of the aforementioned papers involving coefficient function estimation utilize a finite-dimensional approximation, such as Fourier basis or through a roughness penalty, to
regularize model complexity.  It often leads to reasonable functional estimate, but at a price of complicating theoretical investigation.

Our main contributions are summarized as follows. 
First, it allows us to establish a Representation Theorem stating that, although the optimization is defined on an infinite-dimensional function space, its solution
actually resides in a data-adaptive finite-dimensional subspace. This result guarantees
an exact solution when optimization is carried out on this finite dimensional subspace. 
Second, we develop and implement an efficient primal-dual algorithm to handle large image data efficiently, as computation is generally very challenging with complex, high-dimensional images in quantile regression. We use extensive numerical studies to demonstrate the computational advantages of our method over other popular quantile regression optimization solvers such as the alternating direction method of multipliers (ADMM) algorithm \citep{boyd2011distributed}.
Third, an important theoretical result of our work establishes the optimal convergence rate of the error 
 in estimating these coefficient functions under both fixed and random designs. This generalizes the results in \cite{cai2012minimax} and  \cite{du2014penalized} for functional linear regression with a scalar response to the functional
response scenario.

%In summary, our method has the following
%distinguishing features: (i) The SQR framework characterizes the whole spatial distribution of the image response given the scalar predictor; (ii) Benefited from the Representer Theorem, we develop a numerically reliable algorithm that has
%remarkable performance in simulations; (iii) We show theoretically the estimator achieves the
%optimal minimax convergence rate in quantile prediction.

The rest of this paper is organized as follows. In Section \ref{sec:model}, we introduce the mathematical details of the proposed SQR model, including the quantile regression and copula models. Section \ref{sec:algorithm} introduces an efficient primal-dual algorithm to solve the high-dimensional  quantile regression model and a generalized least square method to estimate parameters in the copula model. In Section \ref{sec:convergence}, we studied the optimal convergence rate
of the error in estimating the coefficient functions under both fixed and random designs. In Section \ref{sec:numerical}, extensive numerical studies are used to illustrate the advantages of the proposed method.

\section{Spatial Quantile Regression and Copula Models}
\label{sec:model}

The SQR model studies the conditional spatial distribution of an imaging response $\{Y(s): s\in {\cal S}\}$ given scalar predictors ${\bf x} \in \real^p$.
The conditional quantile function of $Y(s)$ given ${\bf x}$ for a fixed location $s$ at any quantile level $\tau\in (0, 1)$ is assumed to be
\begin{equation}\label{equ:model}
Q_{Y(s)}(\tau| {\bf x}) = {\bf x} ^T {\bm \beta}_\tau(s),
\end{equation}
where $ {\bm \beta}_\tau (s) = (\beta_{\tau 1} (s), \ldots, \beta_{\tau p}(s) )^T\in {\cal F}^p$ is the vector of the unknown coefficient functions with each component of ${\bm\beta}_{\tau}$ residing in the function space ${\cal F}$. Assume that ${\cal F}$ is a reproduce kernel Hilbert space (RKHS) ${\cal H}({ K})$ with a reproducing kernel $K$. Equation (\ref{equ:model}) completely specifies the conditional marginal distribution of $Y(s)$ given $\bf x$ and $s$. 

%On the other hand, we model the spatial dependence of $m$ intensities $Y(s_1), \cdots, Y(s_m)$ at $s_1, \ldots, s_m$ by a copula model.  Let $F_{Y(s_i)}(\cdot)$ be the cdf of $Y(s_i)$. We apply a parametric copula for 
%\begin{equation}\nonumber
%\Big(U(s_1), U(s_2), \ldots, U(s_m)\Big) = \Big(F_{Y(s_1)}(Y(s_1), F_{Y(s_2)}(Y(s_2), \ldots, F_{Y(s_m)}(Y(s_m)\Big),
%\end{equation}
%which has uniformly distributed marginals. The copula contains all information on the dependence structure between the components of $Y(s_1), \cdots, Y(s_m)$, whereas the marginal quantiles contain all information on the marginal distributions. 

%The proposed spatial quantile regression (SQR) model studies the conditional spatial distribution of imaging response $\{Y(s): s\in {\cal S}\}$ given scalar predictors ${\bf x} \in \real^p$:
%$Q_{Y(s)}(\tau| {\bf x}) = {\bf x} ^T {\bm \beta}_\tau(s)$,
%where $\tau\in (0, 1)$ is the quantile level, and $ {\bm \beta}_\tau (s) = (\beta_{\tau 1} (s), \ldots, \beta_{\tau p}(s) )^T\in {\cal F}^p$ is the vector of the unknown coefficient functions with each component $\beta_{\tau i}$ resides in the function space ${\cal F}$. We assume ${\cal F}$ is a reproduce kernel Hilbert space (RKHS) ${\cal H}({ K})$ with an reproducing kernel $K$. Equation (\ref{equ:model}) completely specifies the conditional marginal distribution of $Y(s)$ given $\bf x$ and $s$. 

In practice, the image response is often observed at discrete locations $s_1, \ldots, s_m$ in ${\cal S}$, and the image values across different points in ${\cal S}$ are highly correlated. The conditional marginal distribution of $Y(s)$ is fully captured by (\ref{equ:model}), however, the spatial dependence among $Y(s)$ at different points  in ${\cal S}$ is not modeled.  To capture the joint distribution of $(Y(s_1),\ldots, Y(s_m))$ given scalar covariates $\bf x$, we propose using copula models.  Let the conditional  CDF of $Y(s_j)$ given $\bf x$ be $F_{(s_j,{\bf x})}(y) = \mathbb P(Y(s_j)\le y| {\bf x})$, $j=1, \ldots, m$. The random variable $U_{\bf x}(s_j) =F_{(s_j,{\bf x})}(y)$  has a uniform marginal distribution.  
%
%We apply a parametric copula for 
%\begin{equation}\nonumber
%\Big(U(s_1), U(s_2), \ldots, U(s_m)\Big) = \Big(F_{Y(s_1)}(Y(s_1), F_{Y(s_2)}(Y(s_2), \ldots, F_{Y(s_m)}(Y(s_m)\Big).
%\end{equation}
%
The conditional copula of $(Y(s_1), \ldots, Y(s_m))$ given ${\bf x}$ is defined as the conditional joint CDF of $(U_{\bf x}(s_1),  \ldots, U_{\bf x}(s_m))$ given by
\begin{equation}\label{equ:copula}
C_{{\bm \theta}}(u_1, \ldots, u_m) = \mathbb P\Big( U_{\bf x}(s_1)\le u_1, \ldots, U_{\bf x}(s_m)\le u_m \Big| {\bf x} \Big).
\end{equation}
%For simplicity, we assume that the copula $C$ in (\ref{equ:copula}) does not depend on $X$.
The copula $C$ parameterized by ${\bm \theta}$ contains all information on the spatial dependence structure of $Y(\cdot)$. In this paper, we adopt a parametric family of copulas $C_{{\bm \theta}}$ with ${\bm \theta} = {\bm \theta}({\bf x})$ to characterize the joint conditional distribution of $(Y(s_1), \ldots, Y(s_m))$.
%
%, meaning that the joint distribution image response depends on the observed covariates. 
%
Therefore, the conditional joint distribution of $(Y(s_1), \ldots, Y(s_m))$ given ${\bf x}$ can be written as
\[
\mathbb P\Big(Y(s_1)\le y_1, \ldots, Y(s_m)\le y_m \Big| {\bf x}\Big) = C_{{\bm \theta}({\bf x})}\Big(F_{(s_1,{\bf x})}(y_1), \ldots, F_{(s_m,{\bf x})}(y_m))\Big).
\]

{There are many choices of copula models such as the Student-t copula \citep{demarta2005t}, the non-central $\chi^2$-copula \citep{bardossy2006copula}, and the Gaussian copula \citep{Kazianka2010Copula}. It is a challenging task to determine the optimal choice of the copula. In our numerical analysis, we have examined 21 different copula models for the real diffusion tensor imaging (DTI) data set from the ADNI and found that the Student-t copula fits the data well in most of the times. Therefore, we adopt the Student-t copula in this paper, and denote it by 
\[
C_{\bm\theta}\Big(u_1, \ldots, u_m\Big) = {\bm t}_{\varrho, \boldsymbol{\Sigma}_{\bm \theta}}\Big(t_{\varrho}^{-1}(u_1), \ldots, t_{\varrho}^{-1}(u_m)    \Big),
\]
where  ${\bm t}_{\varrho, \bm \Sigma_{\bm\theta}}$ represents the multivariate t-distribution function with parameter $(\bm \Sigma_{\bm\theta}, \varrho, {\bf 0})$, and $t_{\varrho}^{-1}$ is the inverse CDF of t-distribution with $\varrho$ degrees of freedom. Note that $\varrho/(\varrho-2)\bm \Sigma_{\bm\theta}$ is the covariance matrix of the t-distribution for $\varrho>2$ in t-distribution. Furthermore, we use the Mate\'rn family \citep{matern2013spatial,guttorp2006studies} to specify a class of isotropic correlation functions given by
\[
\mathrm{Corr}\Big(t_{\varrho}^{-1}(U(s+h)), t_{\varrho}^{-1}(U(s)) ~\Big|~{\bf x}\Big) =  M(h)
\]
that depends on the spatial separation parameter $h$, where 
\begin{equation} \label{eqn:mh}
M(h) = {2^{1-\nu}\over \Gamma(\nu)} (\alpha  \|h\|)^\nu K_\nu(\alpha  \|h\|).
\end{equation}
Here $K_\nu$ is the modified Bessel function of the second kind, $\nu>0$ is a smoothness parameter defining the smoothness of the sample path, and $\alpha>0$ is a spatial scale parameter. For our numerical studies, the parameters $\varrho$ and $\nu$ are selected and kept fixed, and the parameter $\alpha$ is treated as unknown. We assume that $\alpha$ depends on ${\bf x}$. With an abuse of notation, we write the scale parameter as $\exp({\bm \alpha}^T {\bf x})$, where ${\bm \alpha} \in \mathbb R^p$. }

%To better illustrate our copula model, consider a special case where $(Y(s_1), \ldots, Y(s_m))$ given ${\bf x}$ is from a Gaussian process with mean ${ \mu}_{\bf x}$ and covariance ${\bm \Sigma}_{\bf x}$. The marginal distribution of $Y(s_j)$ given $\bf x$ is normal with mean ${ \mu}_{\bf x}(s_j)$ and variance $\sigma^2_{\bf x}(s_j)$. The $\tau$th conditional quantile is
%\[
%Q_{Y(s_j)}(\tau) = \mu_{\bf x}(s_j) + \sigma_{\bf x}(s_j) \Phi^{-1}(\tau), ~~~j=1, \ldots, m.
%\]  
%The model (\ref{equ:model}) corresponds to the case when both $\mu_{\bf x}(s_j)$ and $\sigma_{\bf x}(s_j)$ are linear functions of ${\bf x}$. The conditional CDF of $Y(s_j)$ is $F_{(s_j,{\bf x})}(y) = \Phi((y-\mu_{\bf x}(s_j))/\sigma_{\bf x}(s_j)  )$. The spatial dependence among $Y(s_1), \ldots, Y(s_m)$ is captured by the copula model (\ref{equ:copula}), indicating that $(\Phi^{-1}(U_{\bf x}(s_1)), \ldots, \Phi^{-1}(U_{\bf x}(s_m)))$ given ${\bf x}$ follows a multivariate normal distribution with the correlation matrix specified by the Mate\'rn model.

Suppose that we observe $({\bf x}_i, Y_i(s_{ij}))$ for subject $i$ and location $s_{ij}$, $i=1, \ldots, n$ and $j=1, \ldots, m$. In this paper, we consider two different designs for the location points $\{s_{ij}\}$.  The first one is called a {\it fixed design}, where the functional response  are observed at the same locations across curves or images. That is, $s_{1j}=s_{2j}=\cdots=s_{nj}:=s_j$ for $j=1,\ldots, m$. A second design, called {\it random design}, occurs when $\{s_{i1}, \ldots, s_{im}\}$ are independently sampled from a distribution $\{\pi(s): s\in {\cal S}\}$. Our goal is to estimate the coefficient functions $\beta_{\tau k}(\cdot)$, $k=1,\ldots, p$ and the parameters in the copula model. There are two main challenges related to this problem. The first one is to construct statistically efficient estimates of unknown parameters, in particularly the infinite dimensional coefficient functions. The second one is to develop a numerically efficient algorithm to estimate unknown parameters to handle the high dimensionality of functional or image data.

\section{Algorithms}
\label{sec:algorithm}

\subsection{Efficient Primal-Dual Algorithm for Estimating Coefficient Functions}

For given observations  $({\bf x}_i, Y_i(s_{ij}))$, {we estimate the unknown parameters in the spatial quantile regression model (\ref{equ:model}) by solving  the following optimization problem for a given $\tau\in (0, 1)$},
\begin{equation} \label{equ:obj1}
\min_{{\bm \beta} \in {\cal F}^p}  ~~\sum_{i=1}^n\sum_{j=1}^m \rho_\tau( Y_i(s_{ij}) - {\bf x}_i^T {\bm \beta}_\tau(s_{ij})  ) + {\lambda\over 2}\sum_{k=1}^p \mathcal R(\beta_{\tau k}),
\end{equation}
where $\rho_\tau(r) = \tau r I(r>0) -(1-\tau) r I(r\le 0)$ is the check function \citep{koenker1978regression} and $I(\cdot)$ is the indicator function,  ${\bm\beta}_\tau(s) = (\beta_{\tau 1}(s), \ldots, \beta_{\tau p}(s))^T$ are the coefficient functions,  $\mathcal R$ is a roughness penalty on ${\bm \beta}_\tau$, and $\lambda>0$ is a parameter controlling the smoothness penalty. We let $\mathcal R(\beta_{\tau k}) = \|\beta_{\tau k}\|_K^2$, where $\| \cdot \|_K$ is a semi-norm in the RKHS ${\cal H}(K)$ \citep{wahba1990spline}. For simplicity, we also assume that the null space of ${\cal H}(K)$ is $\{0\}$. In the following derivations, we illustrate the algorithm in the case of fixed design, so that $Y(s_{ij})$ can be denoted as $Y(s_j)$. An extension to the case of a random design is straightforward.  

{Throughout the paper,  Gaussian kernels are used for $K(\cdot, \cdot)$ just for simplicity. Other kernels can be easily incorporated. For example, some common options include the Laplace kernel $K(x, y) = \exp(-\sigma\|x-y\|_1)$, the polynomial kernel $K(x, y) = (\langle x, y\rangle+\sigma^2)^d$, and the inverse-quadratic kernel  
$K(x, y) = \sigma^2/(\sigma^2 + \|x- y\|_2^2)$. Learning kernels is definitely a non-trivial question. A common way for learning kernels is to combine  different kernels to improve them. For example, let $K(x, y) = c_1 K_1(x, y) + c_2 K_2(x, y)$ with $c_1, c_2 \ge 0$, where $K_1$ and $K_2$ are potential kernels. The hyperparameters $c_1$ and $c_2$ can be set by cross-validation. Moreover, based on   both simulations and real data analysis reported below, the use of Gaussian kernels leads to accurate estimation results. }

Let $\widehat {\bm\beta}_\tau = (\widehat \beta_{\tau 1}, \ldots, \widehat\beta_{\tau p})$ be the optimal solution of (\ref{equ:obj1}). For notational simplicity, we drop the subscript $\tau$ from the coefficient functions $\bm \beta$ when the context is clear.  It is straightforward to establish a  Representation Theorem 
\citep{wahba1990spline} stating that $\widehat {\bm \beta}$ actually resides in a finite-dimensional subspace of ${\cal H}$, which facilitates computation by reducing an infinite-dimensional optimization problem to a finite-dimensional one. Specifically,
\begin{equation} \label{eqn:repbeta}
\widehat\beta_k(s) = \mu_k +{\bm b}_k^T {\bf k}_s, 
\end{equation}
for $k = 1, \ldots, p$, where $\mu_k$ is a scalar, ${\bm b}_k=(b_{k1}, \ldots, b_{km})^T\in \mathbb R^m$, and ${\bf k}_s=(K(s, s_1), \ldots, K(s, s_m))^T \in \mathbb R^m$.  
The spatial smoothness of $\widehat\beta_k(s)$ comes  from the  RKHS assumption  and the nature of image  data.   The use of  regularization term $\lambda {\cal R}(\beta_k)$ is to regularize the coefficients $\{b_{kj}\}_{j=1}^m$ to avoid overfitting and the use of GACV allows us to implicitly incorporate spatial correlations to select the tuning parameter $\lambda$. 
   Specifically, if there is high correlation among observations on the same subject, 
   then GACV would lead to the selection of a large $\lambda$.

The challenges of solving (\ref{equ:obj1}) under the parameterized form in (\ref{eqn:repbeta}) come from several aspects. First, the number of parameters to be estimated is generally large in medical imaging applications. In particular, each of the $p$ coefficient functions is represented by $m+1$ unknown parameters ($m$ depends on the dimensionality of $Y$). For example, if $Y$ is a 2D image with a resolution of $50 \times 50$, we have $m = 2500$, and so we will need to estimate $p(2500+1)$ parameters. Second, the non-differentiability of the check function significantly increases the difficulty of the optimization problem. A straightforward optimization strategy for (\ref{equ:obj1}) is to use the popular ADMM algorithm \citep{boyd2011distributed} that divides the optimization (\ref{equ:obj1}) into a few simpler optimization subproblems that can be solved iteratively. Details of the ADMM algorithm applied for solving (\ref{equ:obj1}) are included in the Supplement III. However, the convergence of ADMM can be very slow, and when the dimension of $Y$ is large, the computational cost of some subproblems of ADMM can be huge due to the large matrix inversion.  

{
To overcome this computational challenge, we propose a primal-dual algorithm that converges much faster than ADMM and avoids large matrix inversion in each optimization iteration. The primal-dual kind of optimization approaches has been applied in the field of quantile regression. For instance,  \cite{koenker1996} described a primal-dual approach for computation of nonlinear
quantile regression estimators based on the interior point method.  \cite{portnoy1997a} and \cite{portnoy1997b} compared the interior point
method with existing simplex-based methods for quantile regression, and showed that the interior point approach
was competitive and exhibited a rapidly increasing advantage for large problems. In our problem, to estimate coefficient function $\widehat\beta_k$ corresponding to the covariate ${ x}_k$ for $k=1, \ldots, p$,  we rely on the formula (\ref{eqn:repbeta}). However, the traditional interior point algorithm for quantile regression \citep{koenker1996,portnoy1997a,portnoy1997b} does not involve $\{{\bf b}_k\}$ that are difficult to solve.

We make a critical observation that the primal variable ${\bf\mu}=(\mu_1, \ldots, \mu_p)^T$ is also a dual variable for the dual
problem, which leads to the following efficient optimization strategy.
  Similar to \cite{li2007quantile}, we introduce positive residuals $ \{\xi_{ij} \in \real_+ \}$ and negative residuals $\{ \zeta_{ij} \in \real_+\}$ such that $-\zeta_{ij} \leq Y_i(s_j) - \widehat{Y}_i({s_j})\leq \xi_{ij}$, where $\widehat{Y}_{i}(s_j) = {\bf x}_i^T \widehat {{\bm \beta}}(s_{j})$. We obtain an optimization problem that is equivalent to (\ref{equ:obj1}) as follows: }
\begin{align*}
\min_{{\bm \xi}, {\bm \zeta}, {\bm \mu}, {\bm b}} & ~~ \tau \sum_{i=1}^n\sum_{j=1}^m \xi_{ij} + (1-\tau)\sum_{i=1}^n\sum_{j=1}^m \zeta_{ij} + {\lambda\over 2}\sum_{k=1}^p {\bm b}_k^T {\bm \Sigma} {\bm b}_k\\
s.t. & ~~ Y_{i}(s_j) = {\bf x}_i^T {\bm \mu} + \sum_{k=1}^p x_{ik} {\bf k}_{s_j}^T {\bm b}_k + \xi_{ij} - \zeta_{ij},  ~i=1, \ldots, n, ~j=1, \ldots, m, ~({\bm \xi}, {\bm \zeta})\in \mathbb R_+^{nm},
\end{align*}
where ${\bm  \Sigma} =(K(s_i, s_j)) \in \mathbb R^{m\times m}$, ${\bm \mu}=(\mu_1, \ldots, \mu_p)^T\in \mathbb R^p$, $x_{ik}$ is the $k$-th element of ${\bf x}_i$, and $\xi_{ij}$ and $\zeta_{ij}$ are elements of ${\bm \xi}$ and ${\bm \zeta}$, respectively. The primal Lagrangian function is defined as
\begin{align*}
\Lambda &= \tau \sum_{i=1}^n\sum_{j=1}^m \xi_{ij} + (1-\tau)\sum_{i=1}^n\sum_{j=1}^m \zeta_{ij} + {\lambda\over 2}\sum_{k=1}^p {\bm b}_k^T {\bm \Sigma} {\bm b}_k\\
& - \sum_{i=1}^n \sum_{j=1}^m d_{ij} (  {\bf x}_i^T {\bm \mu} + \sum_{k=1}^p x_{ik} {\bf k}_{s_j}^T {\bm b}_k  + \xi_{ij} - \zeta_{ij} - Y_{i}(s_j) )-\sum_{i=1}^n\sum_{j=1}^m \kappa_{ij}\xi_{ij}-\sum_{i=1}^n\sum_{j=1}^m \tilde\kappa_{ij}\zeta_{ij},
\end{align*}
where $d_{ij}, \kappa_{ij} \geq 0,$ and  $\tilde\kappa_{ij} \geq 0$ are Lagrange multipliers. Setting the derivatives of $\Lambda$ to zero leads to
\begin{align}
{\partial\over \partial {\bm \mu}}: & ~~~\sum_{i=1}^n\sum_{j=1}^m d_{ij}  {\bf  x}_i = {\bf 0}, \label{equ:dij1}\\
{\partial\over\partial {\bm b}_k}: & ~~~{\bm b}_k = {1\over \lambda}\sum_{i=1}^n\sum_{j=1}^m d_{ij} x_{ik} {\bm \Sigma}^{-1} {\bf k}_{s_j},  \label{equ:bk}\\
{\partial\over\partial \xi_{ij}}: & ~~~d_{ij} = \tau - \kappa_{ij}  \label{equ:d1}\\
{\partial\over\partial \zeta_{ij}}: & ~~~d_{ij} = -(1-\tau) + \tilde\kappa_{ij}, \label{equ:d2}
\end{align}
and the Karush-Kuhn-Tucker conditions are
\begin{align}
d_{ij} (  {\bf x}_i^T {\bm \mu} + \sum_{k=1}^p x_{ik} {\bf k}_{s_j}^T {\bm b}_k  + \xi_{ij} - \zeta_{ij} - Y_{i}(s_j) ) =& 0,\\
\kappa_{ij}\xi_{ij} =&0,\\
\tilde\kappa_{ij}\zeta_{ij} =& 0.
\end{align}
Since  $\kappa_{ij}$ and  $\tilde\kappa_{ij}$ must be non-negative, following (\ref{equ:d1}) and (\ref{equ:d2}), we have $d_{ij}\in [-(1-\tau), \tau]$. Furthermore,  we have 
\[
d_{ij}\in (-(1-\tau), \tau) \Rightarrow \kappa_{ij}, \tilde\kappa_{ij}>0 \Rightarrow \xi_{ij}=\zeta_{ij} =0 \Rightarrow Y_{i}(s_j) = {\bf x}_i^T {\bm \mu} + \sum_{k=1}^p x_{ik} {\bf k}_{s_j}^T {\bm b}_k.
\]
Define $Se$ as the index set
\[
Se = \{(i, j): d_{ij}\in (-(1-\tau), \tau), i=1, \ldots, n,~j=1, \ldots, m\}.
\]
If $(i, j)\in Se$, we have 
\begin{align}\label{equ:mu}
Y_{i}(s_j) &= {\bf x}_{i}^T {\bm \mu} + \sum_{k=1}^p x_{ik} {\bf k}_{s_{j}}^T {\bm b}_k = {\bf x}_{i}^T {\bm \mu} + {1\over \lambda} \sum_{i'=1}^n \sum_{j'=1}^m d_{i'j'} {\bf x}_{i'}^T {\bf x}_{i} K(s_{j'}, s_{j}).  
\end{align}
If the $\{d_{ij}\}$ are known, then we may solve (\ref{equ:obj1}) by obtaining $\bm \mu$ and ${\bm b}_k$ through (\ref{equ:mu}) and (\ref{equ:bk}). However, it is difficult to obtain $d_{ij}$ in the presence of linear constraints (\ref{equ:dij1}). 

Next, we  consider the Lagrange dual problem where the dual function is defined as $\inf_{ {\bm \xi},{\bm  \zeta}, {\bm \kappa}, \tilde {\bm \kappa}} \Lambda$. To simplify the notation, we vectorize our data by letting ${\bf y} = (Y_i(s_j)) \in \mathbb R^{nm}$, ${\bm d}=(d_{ij}) \in \mathbb R^{nm}$, and ${\bf X} = [{\bf x}_1; ...;{\bf x}_n] \in \real^{p \times n}$. With these notations, from (\ref{equ:dij1})-(\ref{equ:d2}), the dual problem  can be written as 
\begin{align}\label{equ:dij2}
\max_{\bm d}&~~-{1\over 2\lambda} {\bm d}^T {\bf Q} {\bm d} + {\bm d}^T {\bf y},\\
s.t.&~~ {\bm d} \in [-(1-\tau),  \tau]^{nm}, ~~\widetilde {\bf X} {\bm d} = {\bf 0}, \nonumber
\end{align}
where ${\bf Q} = {\bm \Sigma} \otimes {\bf X}^T {\bf X} \in \mathbb R^{nm \times nm}$ and $\widetilde {\bf X} = (e\otimes {\bf X}^T)^T \in \mathbb R^{p \times nm}$ with $e = [1,..,1] ^T \in  \mathbb R^{m}$ and $\otimes$ indicating the Kronecker product . The optimization problem (\ref{equ:dij2}) is a high-dimensional quadratic programming problem with both box and linear constraints. Because of the linear constraints, we may not be able to solve this problem efficiently. However, these linear constraints can be removed using the primal information because the primal variable $\bm \mu$ is also a dual variable for the dual problem. Hence, if $ \bm \mu$ is known, the dual variable $\bm d$ can be obtained by solving
\begin{align}\label{equ:dij3}
\max_{\bm d}&~~-{1\over 2\lambda} {\bm d}^T {\bf Q} {\bm d} + {\bm d}^T ({\bf y}- \widetilde {\bf X}^T {\bm \mu}),\\
%\max_d&~~-{1\over 2\lambda} d^T Q d + d^T ( Y -\widetilde X^T\mu) \\
s.t &~~{\bm d} \in [-(1-\tau),  \tau]^{nm} \nonumber
\end{align}
The quadratic program (\ref{equ:dij3}) with the box constraints can be solved efficiently  with the algorithm introduced in \cite{de1997quadratic}. Therefore, our optimization strategy for (\ref{equ:obj1}) is to solve the primal and dual problems alternatively until convergence.

%In Algorithm 1, we summarize the complete primal-dual algorithm. 

%\begin{algorithm} 
%\label{alg:pri}
%\caption{The primal-dual algorithm to estimate the coefficient functions.}
%\SetKwInOut{Input}{Input} \SetKwInOut{Output}{Output} \SetAlgoLined
%%\begin{algorithmic}[1]
%\Input{Training samples $(Y_i(s_j), {\bf x}_i)$, $i=1, \ldots, n$; smoothing
%parameter $\lambda$; and quantile level $\tau$}%

%Initialize ${\bm \mu}^{0}\in\mathbb{R}^{p}$;
%\While{not converged}{
%\begin{enumerate}
%\item Solve (\ref{equ:dij3}) for ${\bm d}^i$; 
%\item Solve (\ref{equ:bk}) for ${\bm b}^i_1, \ldots, {\bm b}_p^i$;
%\item Update ${\bm \mu}^{i}$ using (\ref{equ:mu});
%\item Update $i=i+1$ and go to Step 1;
%\end{enumerate}
%}
%\Output{${\bm \mu}$, ${\bm d}$, and ${\bm b}_k$ for $k=1, \ldots, p$} 
%\end{algorithm}

%\subsubsection{Effective Degrees of Freedom and Smoothing Parameter Selection}

Next, we discuss the effective degrees of freedom of our model and  smoothing parameter selection. The divergence 
\[
\mbox{div}(\widehat Y) = \sum_{i=1}^n\sum_{j=1}^m {\partial \widehat Y_{i}(s_j)\over \partial Y_{i}(s_j)}
\]
has been used by many authors \citep{efron1986biased,meyer2000degrees,Koenker2015book,li2007quantile} to estimate the effective dimension for a general modeling procedure. This idea arises from the framework of Stein's unbiased risk estimation theory \citep{stein1981estimation}. Under the setting of nonparametric regression with homoscedastic normal errors, divergence is an unbiased estimate of the sum of the covariance between individual fitted values and the corresponding observed values. \cite{koenker1994quantile} heuristically argued that, under the one-dimensional nonparametric quantile smoothing spline model, the number of interpolated observations is a plausible estimate for the effective dimension of the fitted model. \cite{li2007quantile} proved that, under the one-dimensional nonparametric kernel quantile regression model, the divergence is exactly equal to the number of interpolated observations. In this paper, we formally prove that the divergence is exactly the same as the the number of interpolated $Y_{i}(s_j)$'s, thus justifying its use for the selection of $\lambda$. 
 
\begin{thm}\label{thm:df}
Let $Se = \{(i,j): Y_{i}(s_j) = \widehat Y_{i}(s_j)\}$. For any fixed $\lambda>0$ and any $Y_{i}(s_j)$, we have
\[
\mbox{div}(\widehat Y) = |{Se}|.
\]
\end{thm}

The proof is presented in the Appendix. The choice of the smoothing parameter $\lambda$ is a critical but difficult question. Commonly used criteria for quantile regression include the Schwarz information criterion (SIC) \citep{schwarz1978estimating,koenker1994quantile} and the generalized approximate cross validation (GACV) \citep{yuan2006gacv}. We adopt the GACV criteria in this paper to select $\lambda$:
\begin{equation}
\mbox{GACV}(\lambda) = {\sum_{i=1}^n\sum_{j=1}^m\rho_\tau(Y_i(s_{j}) - {\bf x}_i^T\widehat {\bm \beta}(s_{j}))\over nm- df},
\end{equation}
where $df$ is a measure of the effective dimensionality of the fitted model that can be unbiasedly estimated by the divergence as defined in Theorem \ref{thm:df}.

\subsection{Generalized Least Squares for the Copula}

The quantile regression model in (\ref{equ:model}) only gives the marginal distribution of $Y(s_j)$ given $\bf x$. To obtain the joint distribution of $(Y(s_1),\ldots,Y(s_m))$ given $\bf x$, we utilize the copula model.  { Let $U_{\bf x}(s_j)= F_{(s_j,{\bf x})} (Y(s_j)) $, $j=1,\ldots, m$. We characterize the joint distribution of $(U_{\bf x}(s_1),\ldots,U_{\bf x}(s_m))$ using the Student-t copula model
\[
C_{\bm \theta} \Big(u_1, \ldots, u_m\Big|{\bf x}\Big) = \mathbb P \Big( U_{\bf x} (s_1)<u_1,\ldots, U_{\bf x} (s_m)<u_m \Big | {\bf x}\Big) = {\bm t}_{\varrho, {\bm \Sigma}_{{\bm \theta}({\bf x} )}}\Big(  t_{\varrho}^{-1}(u_1), \ldots, t_{\varrho}^{-1}(u_m)   \Big).
\]
The covariance function follows the Mate\'rn model \citep{matern2013spatial, guttorp2006studies}, which specifies a class of isotropic correlation functions 
\[
\mathrm{Corr}(t_{\varrho}^{-1}(U(s_j)), t_{\varrho}^{-1}(U(s_k))|{\bf x}) = M(s_j - s_k) , ~~~s_j\neq s_k.
\]
For the Mate\'rn model, we can easily verify that   
\begin{align*}
 \mathrm{Var}\Big(t_{\varrho}^{-1}(U(s_j)) - t_{\varrho}^{-1}(U(s_k))   \Big) = 2\varrho/(\varrho-2)(1 - M(h)),
\end{align*}
where $h = s_j - s_k$. We denote $N(h)=\{(s_j, s_k): s_j-s_k=h; j,k=1,\ldots, m\}$ to be the set of all pairs of locations having lag difference $h$. Denote $\gamma(h) = 1- M(h)$. \cite{matheron1963principles} defined the earliest unbiased nonparametric variogram estimator of $\gamma(h)$ for a fixed lag $h$ as
\[
2 \widehat{\gamma} (h) = {\varrho -2 \over \varrho|N(h)|} \sum_{(s_j, s_k)\in N(h)} \Big|t_{\varrho}^{-1}(U(s_j)) - t_{\varrho}^{-1}(U(s_k))\Big|^2.
\] 
}
%But this estimator does not ensure non-negative definiteness.

The method of generalized least squares determines an estimator $\widehat {\bm \theta}$ by minimizing
\[
G({\bm \theta}) = (2\widehat\gamma - 2\gamma({\bm \theta}))^T \Omega^{-1} (2\widehat\gamma - 2\gamma({\bm \theta})).
\]
\cite{cressie1985fitting} studied the method of generalized least squares for variogram fitting in the case where the variance-covariance matrix $\Omega$ is diagonal, leading to the method of weighted least squares (WLS). \cite{genton1998variogram} discussed a more general framework to approximate the variance-covariance matrix and significantly improved the fit.  
We first discuss how to construct pseudo copula observations. Pseudo copula observations $\tilde U_{{\bf x}_i}(s_j) $ can be constructed as $\widehat F_{(s_j,\bf x)}( Y_i(s_j))$, for $i=1, \ldots, n$, $j=1, \ldots, m$, where $\widehat F_j$ is the estimated conditional CDF of $Y_i(s_j)$ given ${\bf x}_i$. The information contained in the dual problem is the key to obtain these pseudo copula observations.
Let $\tilde {d}_{ij} = {d}_{ij} + (1-\tau)$ such that $\tilde {d}_{ij} (\tau; {\bf x}_j, s_j): [0, 1]\rightarrow [0, 1]$.  {This} ${\tilde d}_{ij}$ plays a crucial role in connecting the statistical theory of quantile regression to the classical theory of rank tests \citep{gutenbrunner1992regression,gutenbrunner1993tests}. In particular, let $\widehat Y_i(s_j) = {\bf x}_i^T \widehat{\bm \beta}_\tau(s_j)$ be the fitted value. We have
\begin{equation}
\tilde d(\tau; {\bf x}_i, s_j) =\left\{\begin{array}{ll}
1 & \mbox{if } ~ Y_i(s_j) > \widehat Y_i(s_j),\\
(0, 1) & \mbox{if } ~ Y_i(s_j) = \widehat Y_i(s_j),\\
0 & \mbox{if } ~ Y_i(s_j) < \widehat Y_i(s_j).
\end{array}
\right.
\end{equation}  
The integral $\int_0^1 \tilde d(\tau; {\bf x}_i, s_j)d\tau$ provides a natural estimate of conditional quantile level of the observed response $Y_i(s_j)$ given ${\bf x}_i$, that is
\begin{equation}\label{equ:pseudo}
\tilde U_{{\bf x}_i}(s_j) = \int_0^1 \tilde d(\tau; {\bf x}_i, s_j)d\tau, ~~~i=1, \ldots, n~\mbox{and} ~j=1, \ldots, m.
\end{equation}
We denote different lags $h$ by $h_k$ for $k = 1,\ldots,K$, and let $g_{\bf x}(\bm \theta)=(2\widehat\gamma_{\bf x}(h_1) - 2\gamma_{\bf x}(h_1, \bm \theta), \ldots, 2\widehat\gamma_{\bf x}(h_K) - 2\gamma_{\bf x}(h_K, \bm \theta))^T$. We utilize Genton's method \citep{genton1998variogram} to estimate $\Omega$  in the generalized least squares. Let $V_{\bf x}( \bm \theta)$ be the inverse of the dispersion matrix of the sample variogram $\mathbb E(g_{\bf x}({\bm \theta})g_{\bf x}({\bm \theta})^T|{\bf x})$.  We can obtain $\bm \theta$ via:
\begin{equation}
\widehat {\bm \theta} = \arg\min_{\bm \theta} \sum_{i=1}^n g_{{\bf x}_i}^T(\bm \theta) V_{{\bf x}_i}(\bm \theta) g_{{\bf x}_i}(\bm \theta).
\end{equation}
The asymptotic properties of this estimator have been established in \cite{cressie1985fitting,Cressie2015} and \cite{lahiri2002asymptotic}. We omit the details here.

\section{Optimal Rate of Convergence}
\label{sec:convergence}

In this section, we establish the minimax rate of convergence of estimating the coefficient functions. We consider the two different designs introduced previously.  For the fixed design, the spatial functional response is observed at the same locations across curves, that is, $m_1=m_2=\cdots =m_n:=m$ and $s_{1j}=s_{2j}=\cdots=s_{nj}:=s_j$ for $j=1,\ldots, m$. Assume that, as $m\rightarrow \infty$, the empirical distribution of $s_j$'s converges to a fixed distribution $\pi(s)$. For the random design, the $s_{ij}$ are independently sampled from a distribution. With an abuse of notation, we also denote it by $\pi(s)$. For any two $p$-dimensional vector functions ${\bm f}_1, {\bm f}_2\in {\cal F}^p$, we define the $\mathbb{L}_2$-distance as
\[
\Big\| {\bm f}_1- {\bm f}_2\Big\|_{s,2}^2 = %\left\{\begin{array}{ll}
%{\displaystyle{1\over m}\sum\limits_{j=1}^m \sum_{k=1}^p (f_{1k}(s_j) - f_{2k}(s_j))^2} & \mbox{for the fixed design,}\\
{\displaystyle\int_{\cal S} \sum_{k=1}^p(f_{1k}(s) - f_{2k}(s))^2 \pi(s) ds}. %& \mbox{for the random design.}
%\end{array}
%\right.
\]
For any fixed quantile level $\tau\in (0, 1)$, let $\beta_{\tau k}$ be the true coefficient function. We measure the accuracy of the estimation of $\widehat\beta_{\tau k}$ by 
\[
{\cal E}_{\tau}(\widehat\beta_{\tau k}, \beta_{\tau k}) =   \big\|\widehat\beta_{\tau k} - \beta_{\tau k} \big\|_{s,2}^2.
\]
The rate of convergence of ${\cal E}_{\tau}(\widehat\beta_{\tau k}, \beta_ {\tau k})$ as the sample size $n$ and the location sampling frequency $m$
increase reflects the difficulty of the estimation problem.

\subsection{Minimax lower bound}

The following result establishes the minimax lower bound for estimating ${\bm  \beta}_\tau$ over ${\cal F}^p$ under both fixed and random designs.
The minimax lower bound is given in the following theorem.
\begin{thm}\label{thm:lower} Fix $\tau\in (0, 1)$, and suppose the eigenvalues $\{\rho_k:k\ge 1\}$ of the reproducing kernel $K$ satisfies $\rho_k\asymp k^{-2r}$ for some constant $0<r<\infty$. Then
\begin{enumerate}
\item[(a).] for the fixed design,
\begin{equation}\label{equ:low1}
\lim_{a_\tau \rightarrow 0}\lim_{n,m\rightarrow\infty}\inf_{\widetilde {\bm \beta}_\tau \in {\cal F}^p }\sup_{{\bm \beta}_{\tau}\in {\cal F}^p}\mathbb P\Big(  {\cal E}_{\tau}(\tilde {\bm \beta}_\tau, {\bm \beta}_{\tau}) \geq a_\tau(n^{-1}+m^{-2r})\Big) = 1;
\end{equation}
\item[(b).] and for the random design,
\begin{equation}\label{equ:low2}
\lim_{a_\tau\rightarrow 0}\lim_{n,m\rightarrow\infty}\inf_{\widetilde {\bm \beta}_\tau \in {\cal F}^p}\sup_{{\bm \beta}_{\tau}\in {\cal F}^p}\mathbb P\Big(  {\cal E}_{\tau}(\tilde {\bm \beta}_\tau, {\bm \beta}_{\tau}) \geq a_\tau((nm)^{-{2r\over 2r+1}} + n^{-1})\Big) = 1.
\end{equation}
\end{enumerate}
The above infimums are taken over all possible estimators $\widetilde {\bm \beta}_\tau$ based on the training data.
\end{thm}

\begin{remark}
The lower bounds established in Theorem \ref{thm:lower} depend on the decay rate of the eigenvalues of the reproducing kernel $K$.
The lower bounds are different between the fixed design and the random design. For both designs, when the number of locations $m$ is large, it has no effect on the rate of convergence and the optimal rate is of order $n^{-1}$. On the other hand, a phrase transition phenomenon happens when $m$ is of order $n^{1/2r}$. If $m$ is below this order, the optimal rate for the fixed design is of order $m^{-2r}$ and for the random design is of order $(mn)^{-2r/(2r+1)}$. We may conclude that the random design leads to a better result in terms of the rate of convergence. Similar phenomenon has been studied when estimating the mean of functional data \citep{cai2012minimax}. 
\end{remark}

\begin{remark} The constants $a_\tau$ in (\ref{equ:low1}) and (\ref{equ:low2}) depend on the quantile level $\tau$. If we assume $\tau\in {\cal T}\subset (0, 1)$ with ${\cal T}$ being a compact subset of $(0, 1)$, it is possible to extend Theorem \ref{thm:lower} to hold uniformly for all $\tau$, that is, we may choose a constant $a$ in both  (\ref{equ:low1}) and (\ref{equ:low2}) that does not depend on $\tau$. This result can be established by a slight modification of the proof of Theorem \ref{thm:lower}.
\end{remark}

\subsection{Minimax upper bound}

In this section, we consider the upper bound for the minimax risk and construct specific rate optimal estimators under both designs. The upper bound shows that the rates given in Theorem \ref{thm:lower} are sharp. 
Specifically, 
we adopt the roughness regularization method to estimate the coefficient function vector ${\bm \beta}_\tau \in {\cal F}^p$ by minimizing
\begin{equation}\label{equ:objmin}
\sum_{i=1}^n \sum_{j=1}^{m} \rho_\tau\big( Y_i(s_{ij}) - {\bf x}_i^T {\bm \beta}_\tau(s_{ij}) \big) + \lambda \sum_{k=1}^p \|{\beta}_{\tau k}\|_K^2,
\end{equation}
where $\lambda>0$ is a tuning parameter balancing fidelity to the data and smoothness of the estimate. Let $\widehat {\bm \beta}_
\tau$ be the estimate from (\ref{equ:objmin}).

We now introduce the following main assumptions, which put some constraints on the conditional CDF of $Y(s)$ given $\bf x$ and $s$ and the design matrix from the covariates:
\begin{enumerate}
\item[A1.]  Let the CDF of $Y(s)$ given $\bf x$ and $s$ be $F_{(s,\bf x)}(y) = \mathbb P( Y(s) \le y | {\bf x} )$. Assume that there exist constants $c_0>0$ and $c_1>0$ such that for any $u$ satisfying $|u|\le c_0$,
$\big|F_{(s,\bf x)}(u) - F_{(s,\bf x)}(0)\big| \ge c_1 |u|^2$. 
\item[A2.] Assume ${\bf x}$ belongs to a compact subset of $\mathbb R^p$ and that the eigenvalues of $\mathbb E ({\bf x} {\bf x}^T)$ are bounded below and above by some positive constants $c_2$ and $1/c_2$, respectively.
\end{enumerate}

\begin{thm}\label{thm:upper}
Fix $\tau\in (0, 1)$. Suppose the eigenvalues $\{\rho_k:k\ge 1\}$ of the reproducing kernel $K$ satisfies $\rho_k\asymp k^{-2r}$ for some constant $0<r<\infty$. If A1 and A2 hold, then
\begin{enumerate}
\item[(a).] for the fixed design,
\begin{equation}
\lim_{A_\tau\rightarrow\infty}\lim_{n,m\rightarrow \infty} \sup_{ {\bm \beta}_{\tau}\in {\cal F}^p} \mathbb P\Big(  {\cal E}_n(\widehat {\bm \beta}_\tau, {\bm \beta}_{\tau}) \ge A_\tau(n^{-1} + m^{-2r})  \Big) = 0
\end{equation}
for any $\lambda = O(nm(n^{-1} + m^{-2r}))$;
\item[(b).]  and for the random design,
\begin{equation}
\lim_{A_\tau\rightarrow\infty}\lim_{n,m\rightarrow \infty} \sup_{{\bm \beta}_{\tau}\in {\cal F}^p} \mathbb P\Big(  {\cal E}_n(\widehat {\bm \beta}_\tau, {\bm \beta}_{\tau}) \ge A_\tau((nm)^{-{2r\over 2r+1}} + n^{-1})  \Big) = 0
\end{equation}
for any $\lambda = O(nm(n^{-1} + m^{-2r/(2r+1)}))$.
\end{enumerate}
\end{thm}

\begin{remark}
Combining Theorems \ref{thm:lower} and  \ref{thm:upper} demonstrates that $\widehat {\bm \beta}_\lambda$ is rate-optimal. Similar to Theorem \ref{thm:lower}, the rate of convergence shows both similarities and significant differences between fixed and random designs. In particular, for the random design, the optimal rate is $(mn)^{-2r/(2r+1)}$ when $m=O(n^{1/2r})$ and it is interesting to note that the dependency among $Y_i(s_{ij})$ does not affect the rate of convergence.
\end{remark}

It is possible to extend Theorem 2 to hold uniformly for all $\tau$ in some domain of interest. In the following, we assume that $\tau \in {\cal T}$, with ${\cal T}$ being a compact subset of $(0, 1)$.

\begin{coro}
Assume $\tau\in {\cal T}$, and assume the same conditions of Theorem \ref{thm:upper}.
\begin{enumerate}
\item[(a).] For the fixed design, we have 
\begin{equation}
\sup_{\tau\in {\cal T}} \big\|\widehat {\bm \beta}_\tau - {\bm \beta}_{\tau}\big\|_{s, 2}^2 = O_p\Big( n^{-1} + m^{-2r}  \Big)
\end{equation}
for any $\lambda = O(nm(n^{-1} + m^{-2r}))$;
\item[(b).] For the random design, we have  
\begin{equation}
\sup_{\tau\in {\cal T}} \big\|\widehat {\bm \beta}_\tau - {\bm \beta}_{\tau}\big\|_{s, 2}^2 = O_p\Big((nm)^{-{2r\over 2r+1}} + n^{-1})  \Big)
\end{equation}
for any $\lambda = O(nm(n^{-1} + m^{-2r/(2r+1)}))$.
\end{enumerate}
\end{coro}

\section{Numerical Analysis} 
\label{sec:numerical}
{In the following section, we evaluate the finite sample performance of our method by using both simulated and real data sets, and compare with other spatial quantile regression methods.}

\subsection{Algorithm Complexity} \label{subsec:algcomplexity}
We first study the computational complexity of the proposed primal-dual algorithm for SQR. As a comparison, we developed another optimization algorithm based on the popular ADMM \citep{boyd2011distributed}. Please refer to Supplement III for more details on the ADMM algorithm.  The iterative ADMM algorithm divides the optimization problem in (\ref{equ:obj1}) into multiple sub-problems, each of which has an explicit solution. In the high-dimensional scenario, however, explicitly solving each step is not computationally easy since it requires the multiplication and inversion of large matrices. Another disadvantage of using ADMM is  its convergence rate, which can be poor when high accuracy is desired  \citep{boyd2011distributed}.  In contrast, the proposed primal-dual algorithm  converges very quick  and does not require matrix inversion. We implemented and compared both algorithms in MATLAB on a MacBook Pro with a 2.5 GHz Intel Core i7 CPU and 16 GB of RAM. 
{In addition, an Bayesian spatial quantile regression (BSQR) method proposed in \cite{reich2011} was implemented in R and compared with our method. In our implementation, $5000$ MCMC samples were drawn and the final results were summarized based on a burn-in of first $1000$ samples.   }

Simulation data sets were generated from the following model:
$$Y_i (s_j) = x_{i1}\beta_{\tau1}(s_j) + x_{i2}\beta_{\tau2}(s_j) + x_{i3}\beta_{\tau3}(s_j) + \eta_i(s_j,\tau), $$
for $ i=1,\ldots,n;$ $j=1,\ldots,m,$ where $x_{i1}=1$, $x_{i2} \sim Bernoulli(0.5)$, $x_{i3} \sim Uniform(0,1)$, and $\{s_j\}$ are evenly sampled on $[0,1]$.  We also set $\eta_i(s_j) = v_i(s_j) + \epsilon_i(s_j)$, where $\epsilon_i(s_j)\sim N(0,0.1)$, and $(v_i(s_1),\ldots,v_i(s_m))$ follows a multivariate normal distribution with zero mean and a covariance matrix taking the form $\mbox{Cov}(v_i(s_j),v_i(s_l)) = a*\exp(-((s_j - s_l)/h)^2)$ with $a = 0.6$ and $h=0.8$. We first construct $\eta_i(s_1),\ldots,\eta_i(s_m)$ for each $i$, and then let $\eta_i(s_j,\tau) = \eta_i(s_j) - F^{-1}(\tau)$ with $F$ being the marginal density function of $\eta_i(s_j)$ to make the $\tau$-th quantile of $\eta_i(s_j,\tau)$ zero for identifiability. We set  $\beta_{\tau1}(s) = 5s^2$, $\beta_{\tau2}(s) = 2(1-s)^4$, and $\beta_{\tau3} = 2+20\sin(6s) + 2s^3$ for $s\in[0,1]$. A Gaussian kernel with $\sigma = 0.2$ is used for ${\cal {H}}(K)$. In all simulations,  we fix the smoothing parameter $\lambda =1 $. 

To compare the computational complexity, we generated different data sets with a combination of $n=\{ 50, 100, 200, 500\}$ and $m = \{50, 100, 1000\}$. {For each simulated data set, we ran each algorithm for $100$ times and computed the averages and standard errors of the elapsed CPU times. Table \ref{tab:rt_individual} summarizes  the results at $\tau = 0.1, 0.5$ and $0.8$. Note that we did not directly compare with the BSQR in Table \ref{tab:rt_individual}, which as a Bayesian method estimates $\boldsymbol{\beta_\tau}$ at $
L$ quantile levels simultaneously. Using the same computer, it took BSQR about $1860$ seconds to get 5000 MCMC samples when $m=50$ and $n=50$. The computational time significantly increases as $m$ increases (i.e., $>3000$ seconds when $m=100$). }

\begin{table*} \centering
\caption{Elapsed CPU times of ADMM and primal-dual algorithms. Mean and standard deviation (in bracket) are displayed in each entry.}
\label{tab:rt_individual}
\begin{threeparttable}
%\ra{1.0}
\begin{tabular}{@ {}lcrrrcrrr@{}} \toprule
& & \multicolumn{3}{c}{ Primal-dual } &  \phantom{abc} & \multicolumn{3}{c}{ADMM} \\
\cmidrule{3-5}  \cmidrule{7-9}
  (n,m)&& $\tau = 0.1$  & $\tau = 0.5$  & $\tau = 0.8$ && $\tau = 0.1$  & $\tau = 0.5$  & $\tau = 0.8$ \\ \midrule
(50,50) &&  $.042_{(.009)}$   &   $.045_{(.008)}$  & $.059_{(.065)}$ &&    $18.5_{(.221)}$ &  $18.1_{(.595)}$ &   $19.10_{(.468)}$\\
(50,100) && $.130_{(.008)}$    &  $.152_{(.007)}$   & $.131_{(.012)}$ && $41.1_{(.409)}$   &  $41.41_{(.433)}$   &  $41.97_{(.646)}$ \\
(100,50) &&  $.133_{(.007)}$  &   $.148_{(.007)}$  & $.126_{(.005)}$ &&    $39.01_{(.920)}$  &  $38.98_{(.909)}$ &   $38.98_{(.908)}$ \\
(100,100) && $.293_{(.037)}$    &  $.400_{(.089)}$   & $.358_{(.068)}$ && $89.57_{(.273)}$   &  $89.61_{(.292)} $  &  $89.61_{(.292)}$ \\
(200,50) &&  $.295_{(.036)}$  &   $.450_{(.088)}$  & $.330_{(.043)}$ &&   $82.49_{(.356)}$   &  $82.51_{(.340)}$ &   $82.45_{(.339)}$ \\
(200,100) && $.703_{(.046)}$    &  $.849_{(.095)}$   & $.737_{(.096)}$ && $178.80_{(323)}$   & $ 179.45_{(.400)}$  &  $179.20_{(.399)}$ \\
(500,50) &&  $1.10_{(0.076)}$   &   $1.24_{(.137)}$  & $1.07_{(.058)}$ &&    $201.62_{(.211)}$  &  $209.07_{(.349)}$ &   $200.07_{(.349)}$ \\
(500,100) && $2.71_{(0.230)}$    &  $3.11_{(.544)}$   & $2.92_{(.562)}$ && $440.60_{(.490)}$   &  $444.49_{(.552)} $  &  $443.40_{(.553)}$ \\
(500,1000) && $120_{(1.36)}$    &  $118_{(.798)}$   & $119_{(.570)}$ && *   &  *  &  *  \\
\bottomrule
\end{tabular}
\begin{tablenotes}
\small
\item * indicates missing value due to significant computing time. The unit is second.
\end{tablenotes}
\end{threeparttable}
\end{table*}

To compare the estimation precision, we use the root mean integrated squared error (RMISE), which is defined as 
$$\text{RMISE}_\tau = \left(  {m}^{-1}\sum_{j=1}^m (\widehat{\beta}_{\tau k}(s_j) - \beta_{\tau k}(s_j))^2 \right)^{1/2} \text{for } k =1,2,3. $$
Table \ref{tab:rmise_individual} reports RMISE, averaged over 100 runs, and its standard deviation (SD). { Table 1.1 in Supplment II reports RMISEs in a more noisy scenario.} 
\begin{table}[h]
\footnotesize
\setlength\tabcolsep{1.5pt} % let LaTeX determine whitespace between columns

\begin{threeparttable}
\caption{Comparison of the RMISE of the estimators obtained from different methods. Mean and standard deviation (in bracket) are displayed in each entry.}
\label{tab:rmise_individual}
\ra{1.0}
\begin{tabular}{@ {}lcrrrcrrrcrrr@{}} 
\toprule
& & \multicolumn{3}{c}{ Primal-dual ($\tau = 0.5$)} &  \phantom{abc} & \multicolumn{3}{c}{ADMM ($\tau = 0.5$)} &  \phantom{abc} & \multicolumn{3}{c}{BSQR ($\tau=0.5$))}\\
\cmidrule{3-5}  \cmidrule{7-9}  \cmidrule{11-13}
  (n,m)&& $\beta_1$  & $\beta_2$  & $\beta_3$ && $\beta_1$  & $\beta_2$  & $\beta_3$ && $\beta_1$  & $\beta_2$  & $\beta_3$ \\ 
  \midrule
(50,50) &&  $.065_{(.034)}$   &   $.092_{(.041)}$  & $.272_{(.069)}$ &&    $.282_{(0.071)}$  &  $.210_{(.067)}$  &   $.560_{(.112)}$ && $.482_{(.097)}$ & $.340_{(.081)}$& $.584_{(.191)}$ \\

(50,100) && $.042_{(.027)}$    &  $.054_{(.030)}$   & $.150_{(.056)}$ && $.078 _{(.034)}$  & $.148_{(.046)}$   &  $.212_{(.069)}$ && $.206_{(.141)}$ & $.342_{(.163)}$& $.542_{(.123)}$ \\

(100,50) &&  $.053_{(.035)}$   &  $.063_{(.034)}$  & $.189_{(.056)}$ &&    $.105_{(.035)}$  &  $.205_{(.065)}$ &   $.311_{(.064)}$ && $.235_{(.055)}$ & $.200_{(.144)}$& $.337_{(.123)}$ \\

(100,100) && $.028_{(.019)}$    &  $.025_{(.017)}$   & $.087_{(.032)}$ && $.033_{(.019)}$   &  $.136_{(.045)}$   &  $.151_{(.065)}$ && $.231_{(.024)}$ & $.157_{(.072)}$& $.354_{(.143)}$ \\

(200,50) &&  $.039_{(.021)}$   &  $.029_{(.018)}$  & $.110_{(.031)}$ &&    $.067_{(.033)}$  &  $.164_{(.047)}$ &   $.227_{(.053)}$ && $.231_{(.097)}$ & $.128_{(.056)}$& $.310_{(.091)}$ \\

(200,100) && $.029_{(.013)}$    &  $.018_{(.010)}$   & $.079_{(.020)}$ && $.025_{(.020)}$   &  $.122_{(.053)} $  &  $.082_{(.042)}$ && $.137_{(.025)}$ & $.069_{(.015)}$& $.251_{(.025)}$ \\

(500,50) &&  $.036_{(.012)}$   &   $.019_{(.010)}$  & $.090_{(.019)}$ &&    $.028_{(.019)}$  &  $.166_{(.057)}$ &   $.083_{(.043)}$ && $.077_{(.039)}$ & $.083_{(.009)}$& $.157_{(.041)}$ \\

(500,100) && $.029_{(.007)}$    &  $.018_{(.007)}$   & $.078_{(.011)}$ && $.020_{(.018)}$   &  $.097_{(.029)}$   &  $.062_{(.035)}$ && $.073_{(.017)}$ & $.098_{(.007)}$& $.143_{(.073)}$ \\

\bottomrule
\end{tabular}
%\begin{tablenotes}
%\small
%\end{tablenotes}
\end{threeparttable}
\end{table}

{The results in Tables \ref{tab:rt_individual} and \ref{tab:rmise_individual} clearly indicate that the primal-dual algorithm outperforms the ADMM algorithm and BSQR in both computational efficiency and accuracy}. Also the proposed algorithm is robust to high noise based on the results in Table 1.1 in Supplement II.   {When the sample size $n$ and dimension of data $m$ are large, it takes an extremely long time for  ADMM to obtain a solution, whereas the primal-dual algorithm is much more  efficient. Such computational efficiency is particularly important for our high-dimensional neuroimaging applications. } Moreover, given a fixed $m$ and an increasing number of the observations $n$, the estimated $\bm \beta$ gets better, following our theoretical results. Experiments with the random design also produce similar results. The computational bottleneck of our current implementation is the quadratic programming with box constraints, which can be improved by using C++ or parallel computing. 

\subsection{Simulation Study on Copula Model}
\begin{figure}
\begin{center}
\begin{tabular}{ccc}
\includegraphics[height=1.2in]{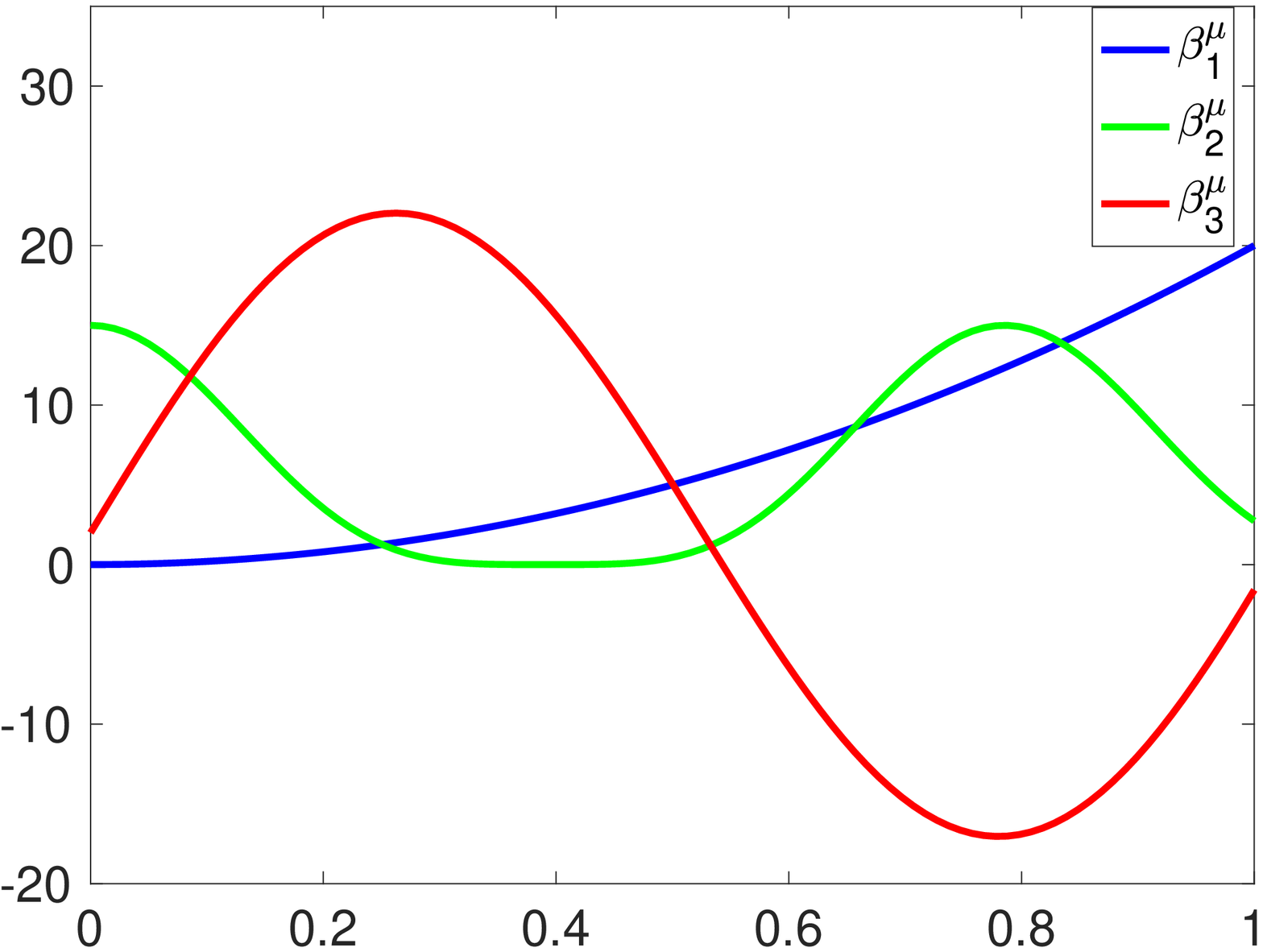} &
\includegraphics[height=1.2in]{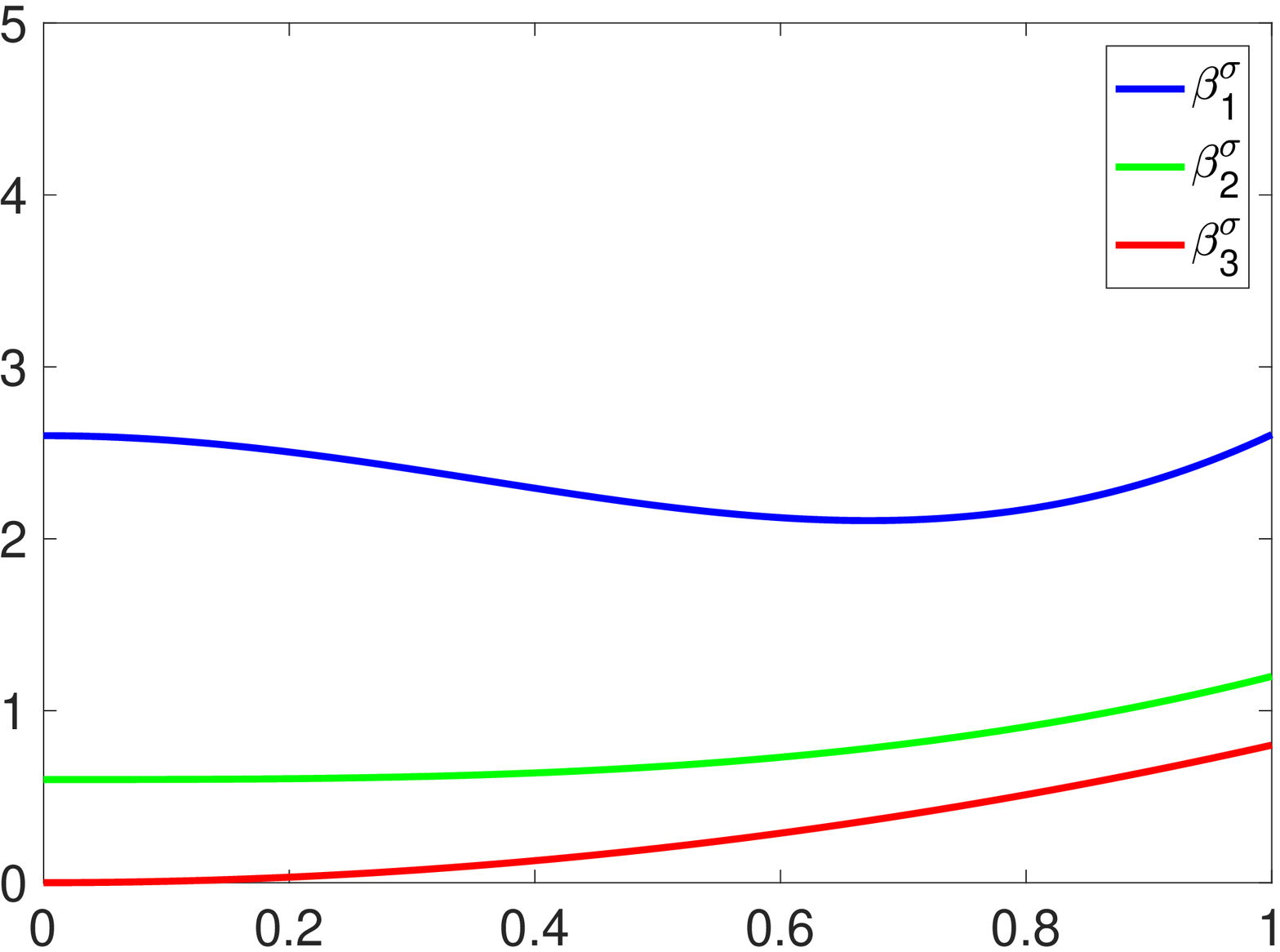}&
\includegraphics[height=1.2in]{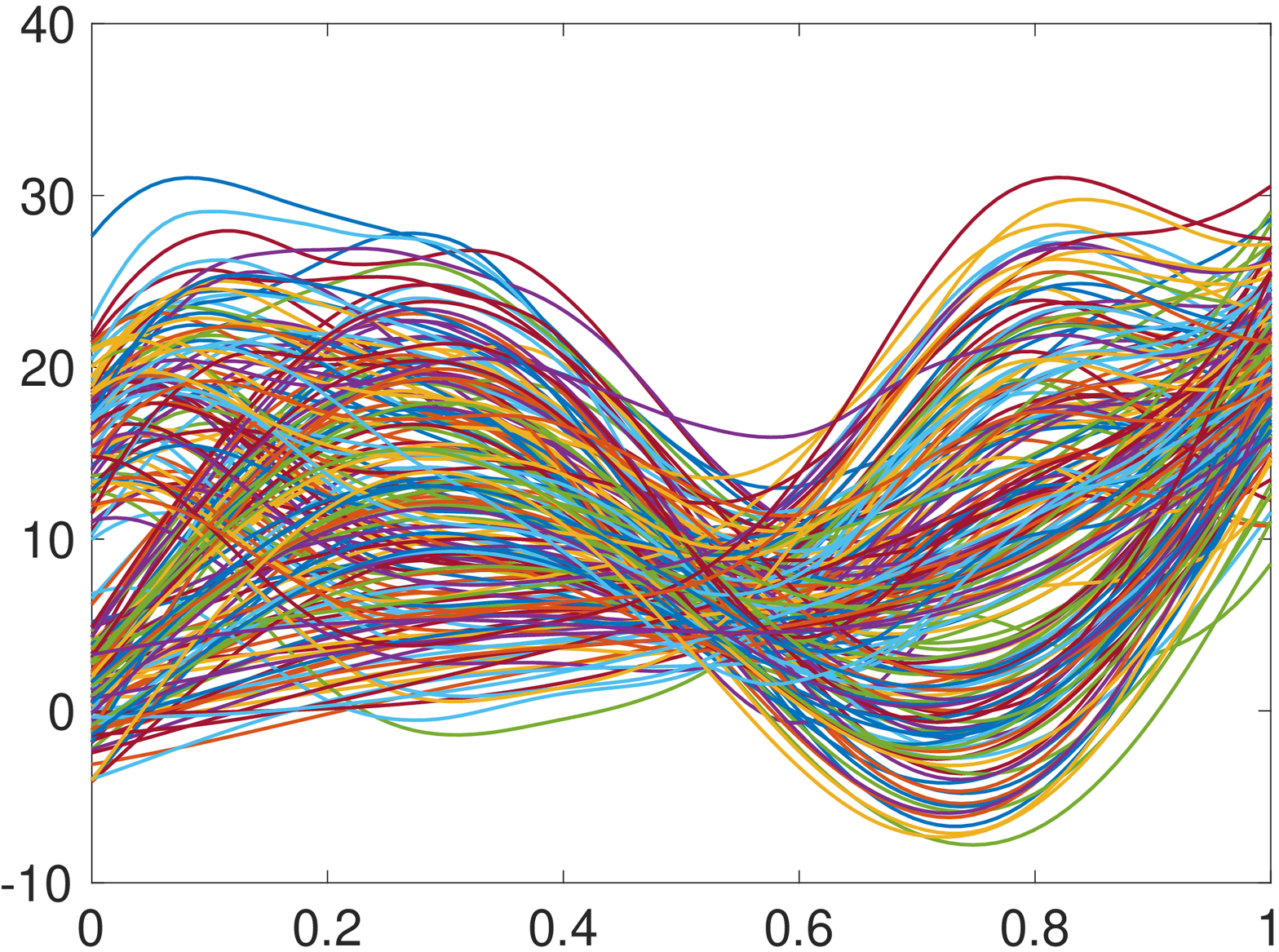}\\

(a) Simulated $\beta^\mu$'s & (b) Simulated $\beta^\sigma$'s & (c) Simulated $Y$ under the copula model \\ 
\end{tabular}
\caption{ ${\bm \beta}^\mu$, ${\bm \beta}^\delta$ and simulated $Y$ in the copula model.} \label{fig:simulation_beta}
\end{center}
\end{figure}

 In this section,  we evaluate the finite sample performance of the proposed copula model in capturing the joint distribution of $\{Y(s_j)\}$ given $\bf x$. Similar to Section \ref{subsec:algcomplexity}, we simulate ${\bf x} = [x_1, x_2,x_3]^T \in \real^3 $ according to $x_{1}=1$, $x_{2} \sim Bernoulli(0.5)$, and $x_{3} \sim Uniform(0,1)$. To adapt the copula model, we simulate the response image $Y$ using a different procedure. We let $Y(s)$ conditional on $\bf x$ be a normal distribution with mean ${\mu}_{\bf x}(s)$  and variance $\sigma_{\bf x}^2(s)$ and the quantile of $Y(s)$ at different $s$ follow a Gaussian process. More specifically, the $\tau$-th quantile of $Y(s_j)$  is given as
$$ F^{-1}_{(s_j,{\bf x})}(\tau) = \mu_{\bf x} (s_j) + \sigma_{\bf x}(s_j)\Phi^{-1}(\tau). $$ 
Since in our mode we assume linear relations, both $\mu_{\bf x}(s_j)$ and $\sigma_{\bf x}{(s_j)}$ should be linear functions of $\bf x$ in the simulation. We let $\mu_{\bf x}(s_j) = x_{1}\beta_1^{\mu}(s_j) + x_{2}\beta_2^{\mu}(s_j) + x_{3}\beta_3^{\mu}(s_j)$ and $\sigma_{\bf x}(s_j) = x_{1}\beta_1^{\sigma}(s_j) + x_{2}\beta_2^{\sigma}(s_j) + x_{3}\beta_3^{\sigma}(s_j)$. {That is, we write the $\tau$-th quantile of $Y(s_j)$ as $F^{-1}_{(s_j,{\bf x})}(\tau) = {\bf x}^T{\bm \beta}_\tau(s_j)$, where ${\bm \beta}_\tau(s) = (\beta_1^{\mu}+\beta_1^{\sigma}\Phi^{-1}(\tau),\ldots,\beta_3^{\mu}+\beta_3^{\sigma}\Phi^{-1}(\tau))^T$.}  Figure \ref{fig:simulation_beta} plots the simulated $\bm \beta$'s. 
To simulate $Y(\cdot)$ with spatial dependence, we use the introduced copula model. We let $\Phi^{-1}(U(s))$ with $U(s)  = F_{(s,\bf x)}(Y(s))$ be a Gaussian process generated from the Mat\'ern family
$$(\Phi^{-1}(U(s_1)),\ldots,\Phi^{-1} (U(s_m) )|{\bf x} \sim N({\bf 0}, {\bm \Sigma}_{({\bf x},{\bm \alpha})}),$$
with $ {\bm \Sigma}_{({\bf x},{\bm \alpha})}(s,s) = \sigma^2_{({\bf x},{\bm \alpha})}(s)= (x_{1}\beta_1^{\sigma}(s) + x_{2}\beta_2^{\sigma}(s) + x_{3}\beta_3^{\sigma}(s))^2$, ${\bm \Sigma}_{({\bf x},{\bm \alpha})}(s_i,s_j) = \sigma_{({\bf x},{\bm \alpha})}(s_i)\sigma_{({\bf x},{\bm \alpha})}(s_j)$
$M(|s_i-s_j|)$, and $M(h) = {2^{1-\nu}\ \Gamma(\nu)} (\exp({\bm \alpha}^T{\bf x})  \|h\|)^\nu K_\nu(\exp({\bm \alpha}^T{\bf x})  \|h\|)$. We set $\nu = 5/2$, ${\bm \alpha} = (0.8,0.8,0.8)^T$. 
We now can simulate $(Y(s_1),\ldots,Y(s_m))$ from this model.  Figure \ref{fig:simulation_beta} (c) shows $200$ simulated $Y$'s. 

\begin{figure}
\begin{center}
\begin{tabular}{c}
\includegraphics[height=2in]{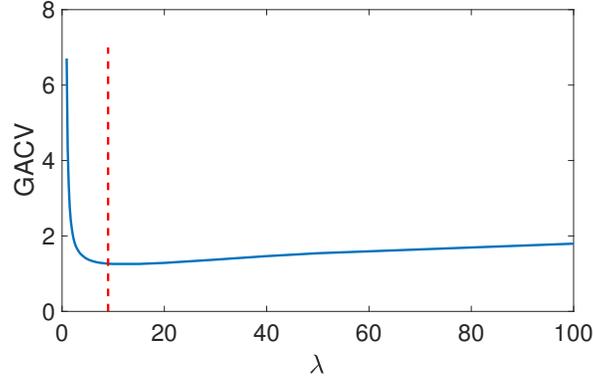} 
\end{tabular}
\caption{ GACV values versus $\lambda$.} \label{fig:simulation_gacv}
\end{center}
\end{figure}

We generated 200 training observations from the above model, along with 500 observations for validation and 500 for testing. We used a fixed Gaussian kernel with $\sigma = 0.2$ for ${\cal H}(K)$. First, we evaluate the $\lambda$ selection criterion. { Specifically,  we compare GACV with a gold standard method that minimizes the sum of RMISEs (sRMISE) defined as:
$$ \text{sRMISE}(\tau) = \sum_{k=1}^p \left(  {m}^{-1}\sum_{j=1}^m( \widehat{\beta}_{\tau k}(s_j) - \beta_{\tau k}(s_j)  )^2 \right)^{1/2} .$$
Figure \ref{fig:simulation_gacv} shows the selected $\lambda$ versus the corresponding GACV value. We then compare parameter estimation accuracy under GACV and the gold standard. 
Table \ref{tab:simulated_beta_gcv} shows the $\mathbb{L}_2$ distance between $\beta_{\tau i}$ and $\widehat{\beta}_{\tau i}$ across different $\tau$ values. It is not surprising   that the gold standard is better than   GACV since we used the ground truth information in the gold standard method. However, we also can observe that   GACV has very similar performances as the gold standard method, especially at $\tau=0.5$, where we have more data available for parameter estimation in this simulation.  The smoothness level goes up from $\beta_1$ to $\beta_3$ in our simulation. We observe that GACV tends to select a simpler model than does the gold standard - that is the smoothest $\beta_1$ is best estimated with GACV.}

%However, $\beta_1$ under GACV is better estimated, where GACV selected a larger $\lambda$ compared with golden standard. In our simulation study, the three functional coefficient $\beta$'s have very different smoothness, and therefore a signal $\lambda$ may not give good estimation for all the $\beta$'s. 

\begin{table*} \centering
\caption{Comparison of different methods on selecting $\lambda$, evaluated through $\|\beta_i - \widehat{\beta}_i\|$. The mean and standard deviation (in bracket) are displayed in each entry.}
\ra{1.0}
\begin{tabular}{@ {}lcrrrcrrr@{}} \toprule
&& \multicolumn{3}{c}{ GACV} &  \phantom{abc} & \multicolumn{3}{c}{ Gold Standard } \\
\cmidrule{3-5}  \cmidrule{7-9}
  (n=200,\\m=100)&& $\tau = 0.1$  & $\tau = 0.5$  & $\tau = 0.8$ && $\tau = 0.1$  & $\tau = 0.5$  & $\tau = 0.8$ \\ \midrule
$\beta_1$ &&  $.139_{(0.050)}$   &   $.045_{(.035)}$  & $.061_{(.039)}$ &&    $.095_{(0.061)}$  &  $.054_{(.042)}$  &   $.054_{(.051)}$ \\
$\beta_2$ && $.233_{(0.058)}$    &  $.143_{(.043)}$   & $.199_{(.047)}$ &&  $.101_{(0.059)}$   &  $.073_{(.052)}$  &  $.041_{(.041)}$ \\
$\beta_3$ &&  $.191_{(0.075)}$   &  $.242_{(.069)}$  & $.302_{(.065)}$ &&    $.159_{(0.093)}$ & $.155_{(.092)}$ &   $.162_{(.091)}$ \\
%\hline
%sRMISE  &&  $0.343_{(0.112)$  &  $0.276_{(0.118)}$  & $0.354_{(0.136)}$ &&  $0.355_{(0.099)}$ & $0.281_{(0.113)}$ &  $0.257_{(0.129)}$\\
\bottomrule
\end{tabular}
\label{tab:simulated_beta_gcv}
\end{table*}

\begin{figure}
\begin{center}
\begin{tabular}{cccc}
\includegraphics[height=1.0in]{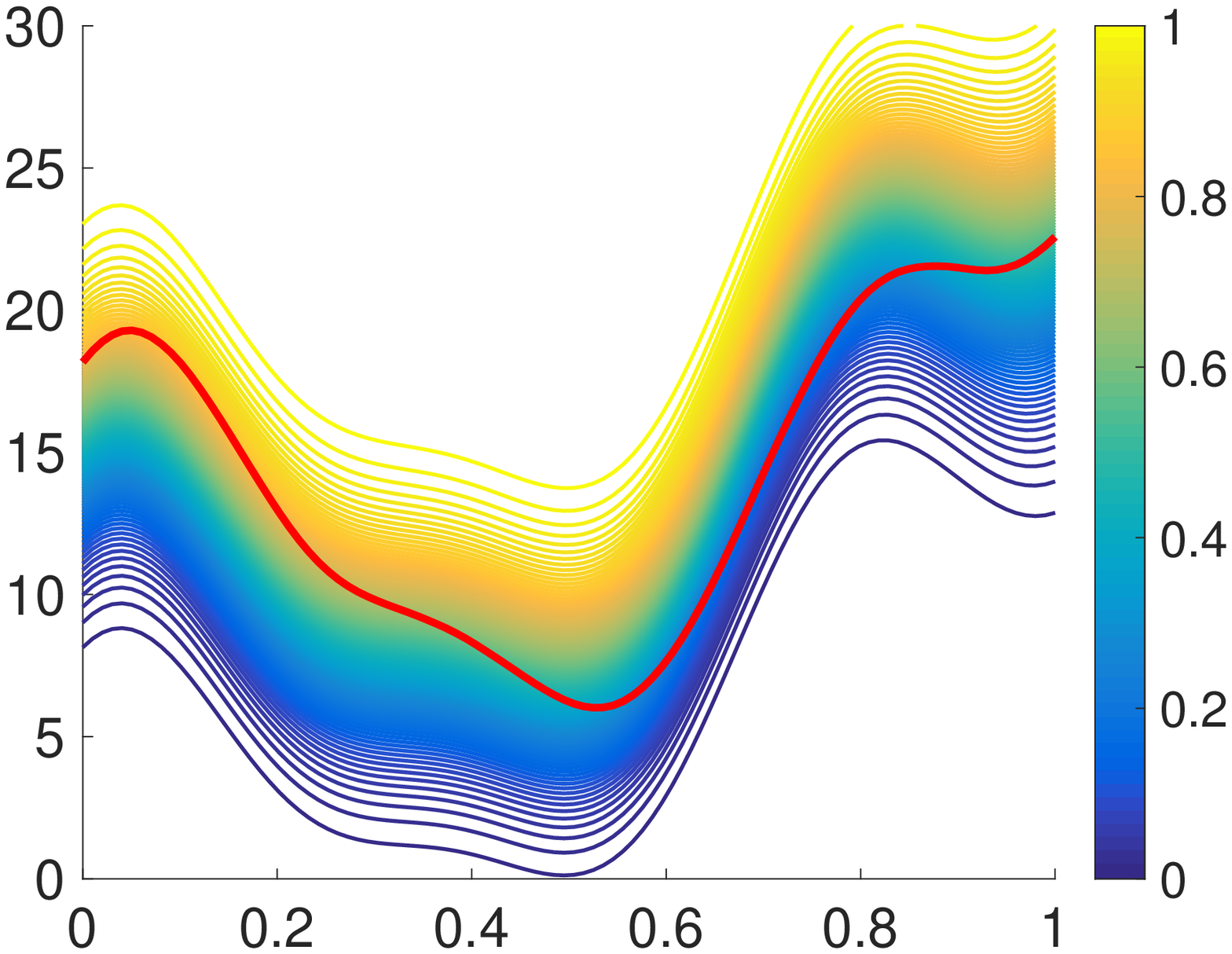} &
\includegraphics[height=1.0in]{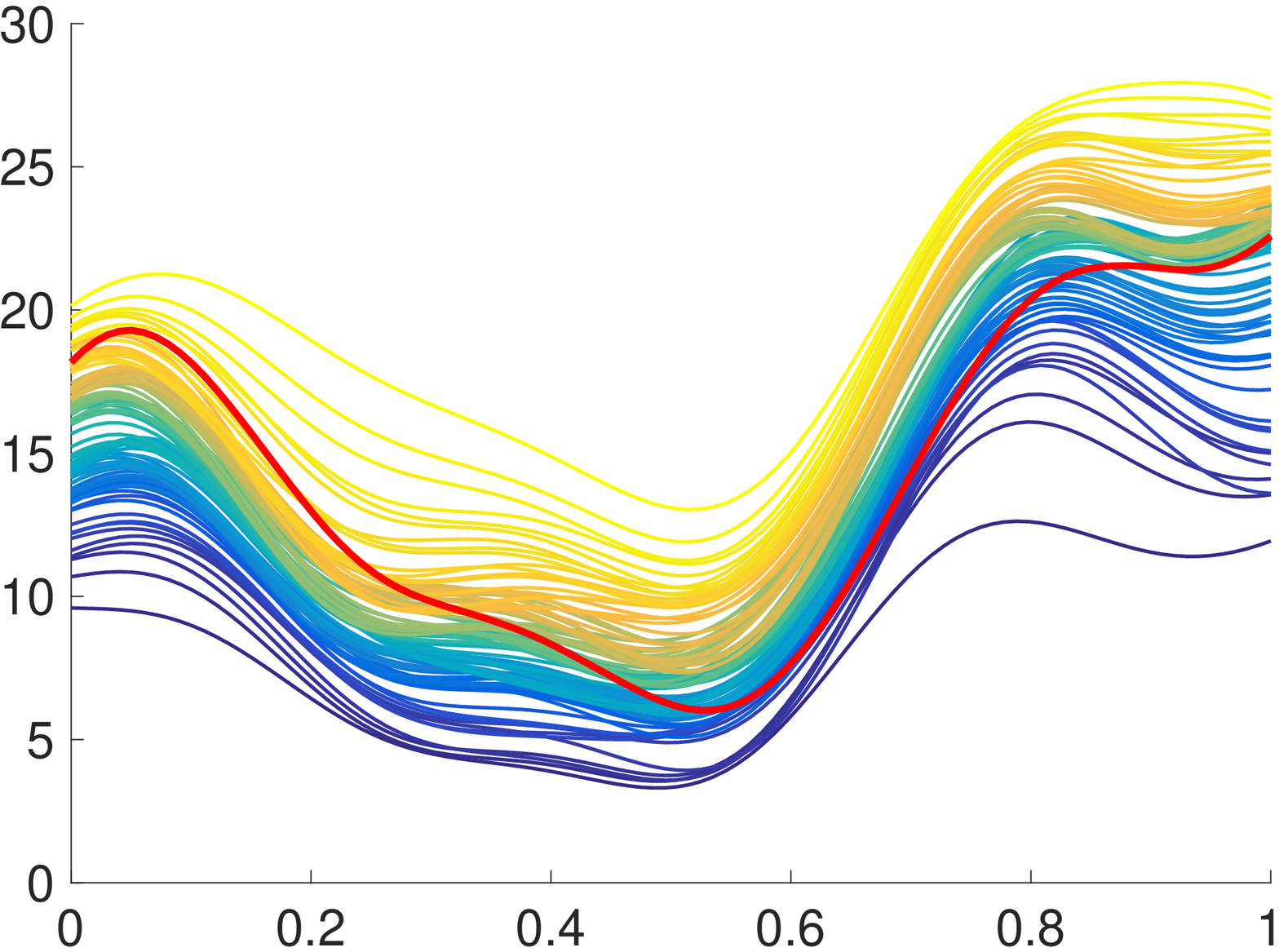}&
\includegraphics[height=1.0in]{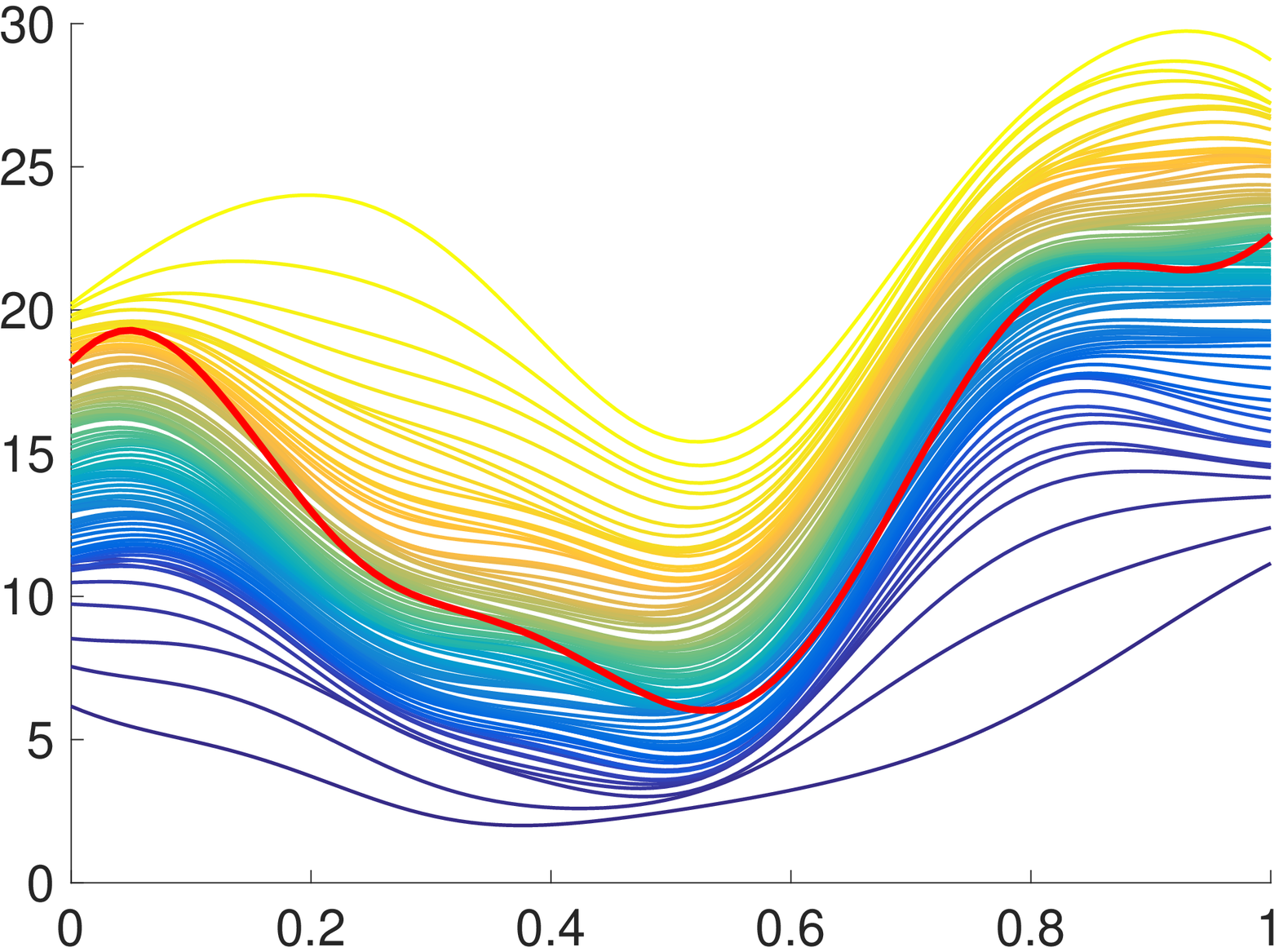} &
\includegraphics[height=1.0in]{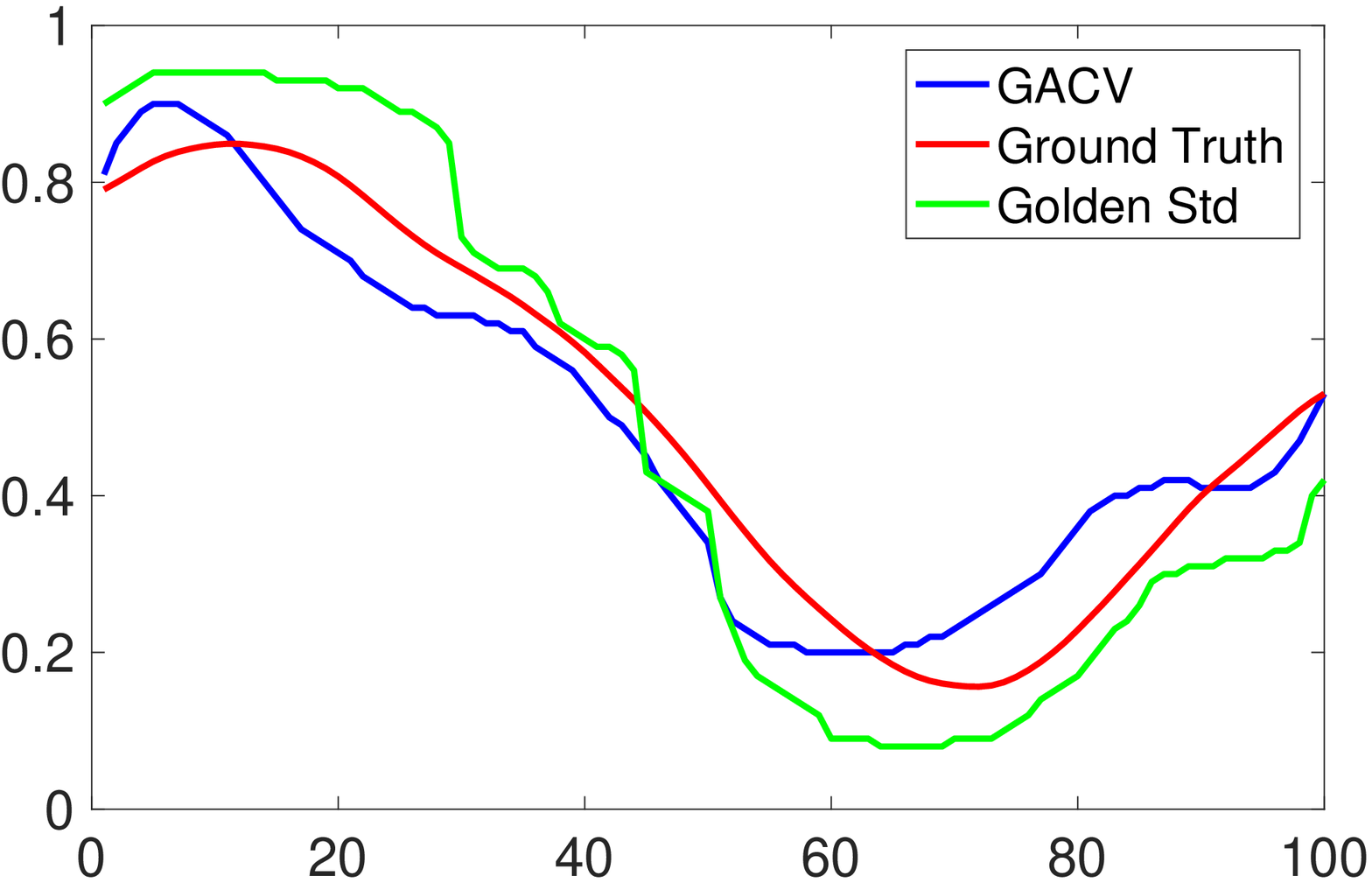}\\
(a) Ground truth & (b) Gold standard & (c) GACV & (d) Quantile functions
\end{tabular}
\caption{Comparison of marginal density estimation under different $\lambda$ selection criteria. Given an observation $(Y_i, {\bf x}_i = (1,1,0.2926)^T)$ ($Y_i$ is plotted in the red curve in panels (a), (b) and (c)), panels (a), (b) and (c) show the marginal distribution of $Y | {\bf x}_i$ with the ground truth parameters, parameters estimated based the gold standard method, and parameters  based on GACV, respectively. Panel (d) shows the quantile function $U(s)$ of the $Y_i(s)$ in the distributions in panels (a), (b) and (c). } \label{fig:simulation_margianl}
\end{center}
\end{figure}

Next, we consider the copula model. Based on a randomly selected observation pair $(Y_i,
{\bf x}_i = (1,1,0.2926)^T$), we obtained the following results in Figure \ref{fig:simulation_margianl}: (a) the ground truth marginal distribution given the ${\bf x}_i$  (different quantiles are plotted with different colors), (b) the estimated quantiles based on the $\lambda$ selected by the gold standard method, and (c) the estimated quantiles based on the GACV. The red curve in (a) (b) and (c) shows $Y_i$. In Figure \ref{fig:simulation_margianl} (d), we show the quantile functions $U(s)$ of the observation $Y_i$ in different scenarios (distributions) of (a),(b) and (c). 

To quantitatively evaluate the estimated quantile functions $U(s)$ in different $\lambda$ selection scenarios, we calculate its $\mathbb{L}^2$ distance to the ground truth quantile function. In average, we got $\|U(s)-\widehat{U}(s)\| =0.081 (\pm 0.0351)$ and $0.101 (\pm 0.0398)$ for GACV and gold standard methods, respectively. This result indicates that in terms of recovering the quantile function  $U(s)$, GACV is not worse than the gold standard. GACV is used in all of the following experiments. 

{We then compare different couple models on fitting the bivariate empirical observation pair $(\hat{F}_{(s_i,\bf{x})}(y_i),\hat{F}_{(s_j,\bf{x})}(y_j))$ with different $\Delta s$ $(\Delta s = \|s_j - s_i\|)$. Using the R package ``VineCopula" \citep{nikoloulopoulos2012vine}, we compared $21$ different copula models,  including the Gaussian copula and Student-t copula. The results are shown in Figure 2.2 in the Supplement II. Although the pair $(F_{(s_i,\bf{x})}(y_i),F_{(s_j,\bf{x})}(y_j))$ is simulated from Gaussian, t-distribution fits the empirical data better in most of the cases, justifying our choice of  t-copula. Note that we obtained similar conclusion with the real data (refer to Figure 2.1 in the Supplement II).} 

\begin{figure}
\begin{center}
\begin{tabular}{cccc}
\includegraphics[height=1.0in]{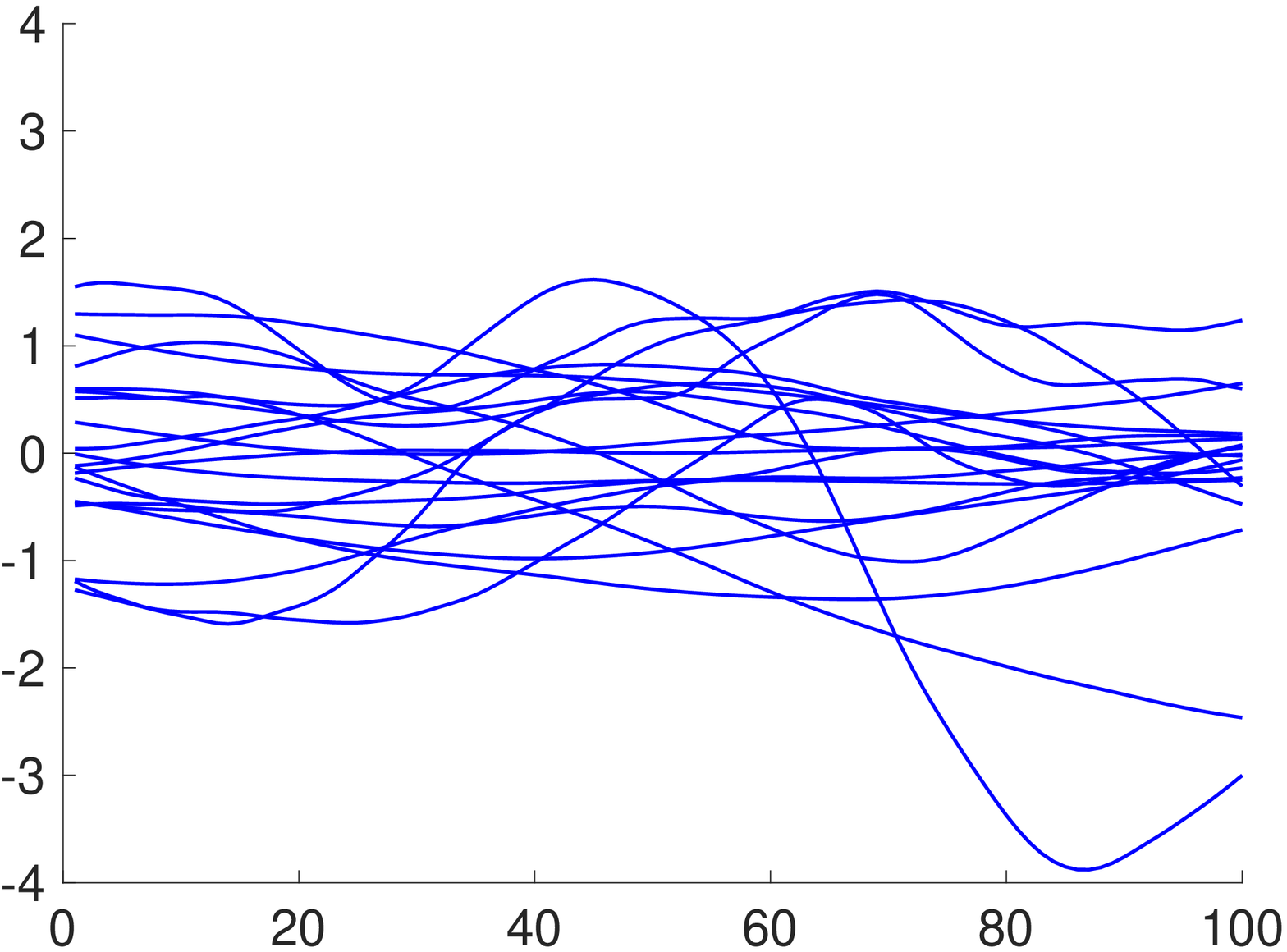} &
\includegraphics[height=1.0in]{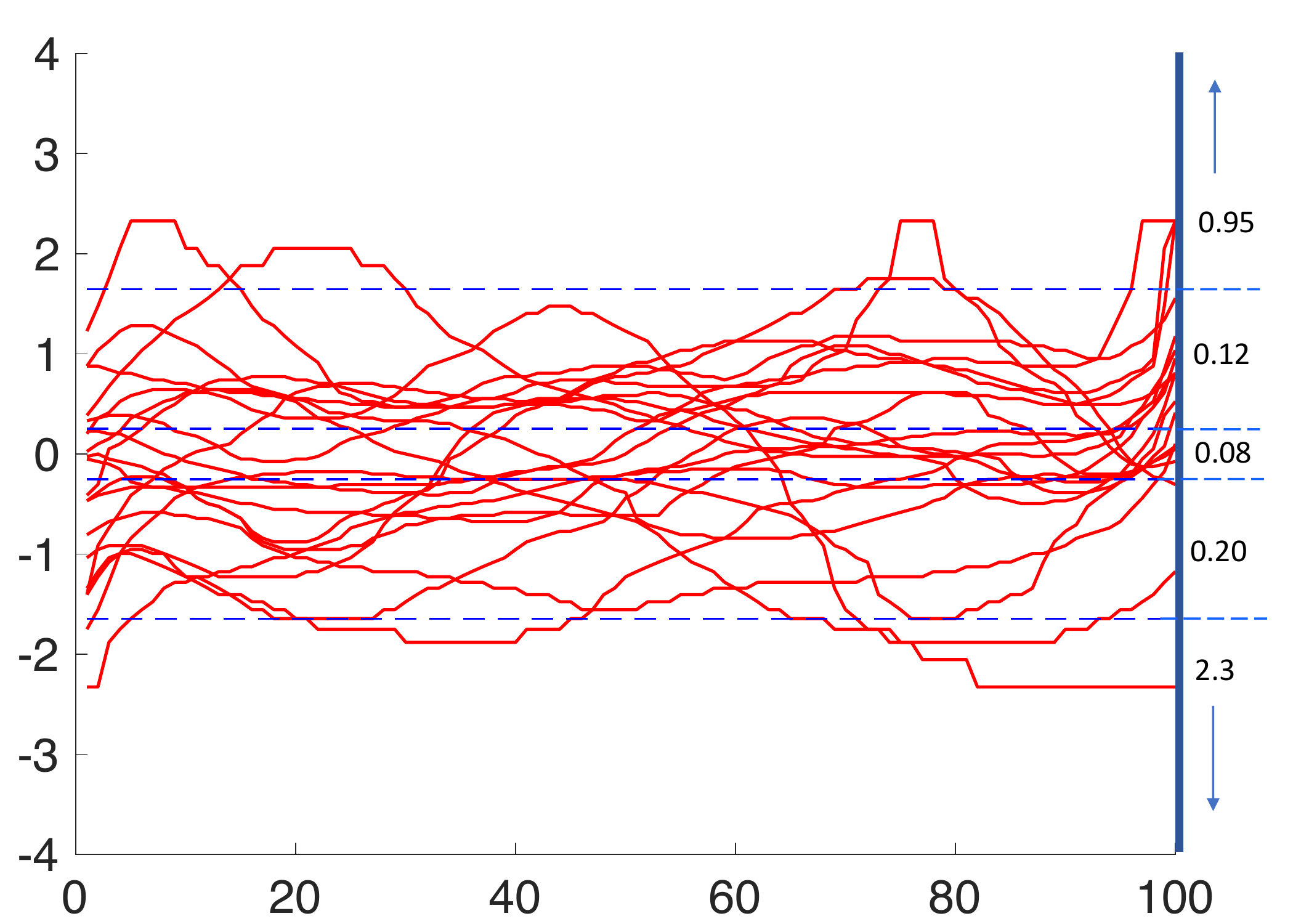}&
\includegraphics[height=1.0in]{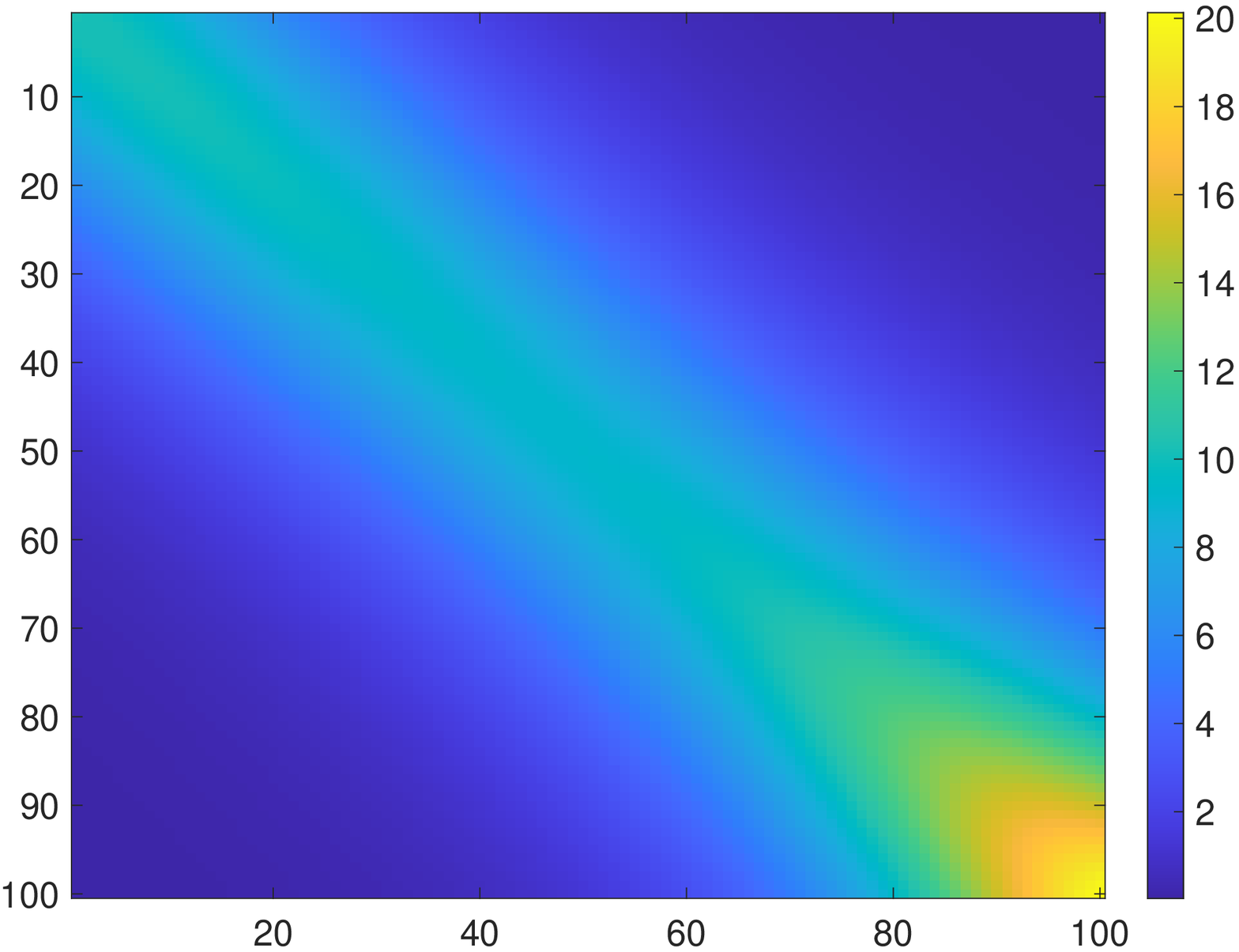} &
\includegraphics[height=1.0in]{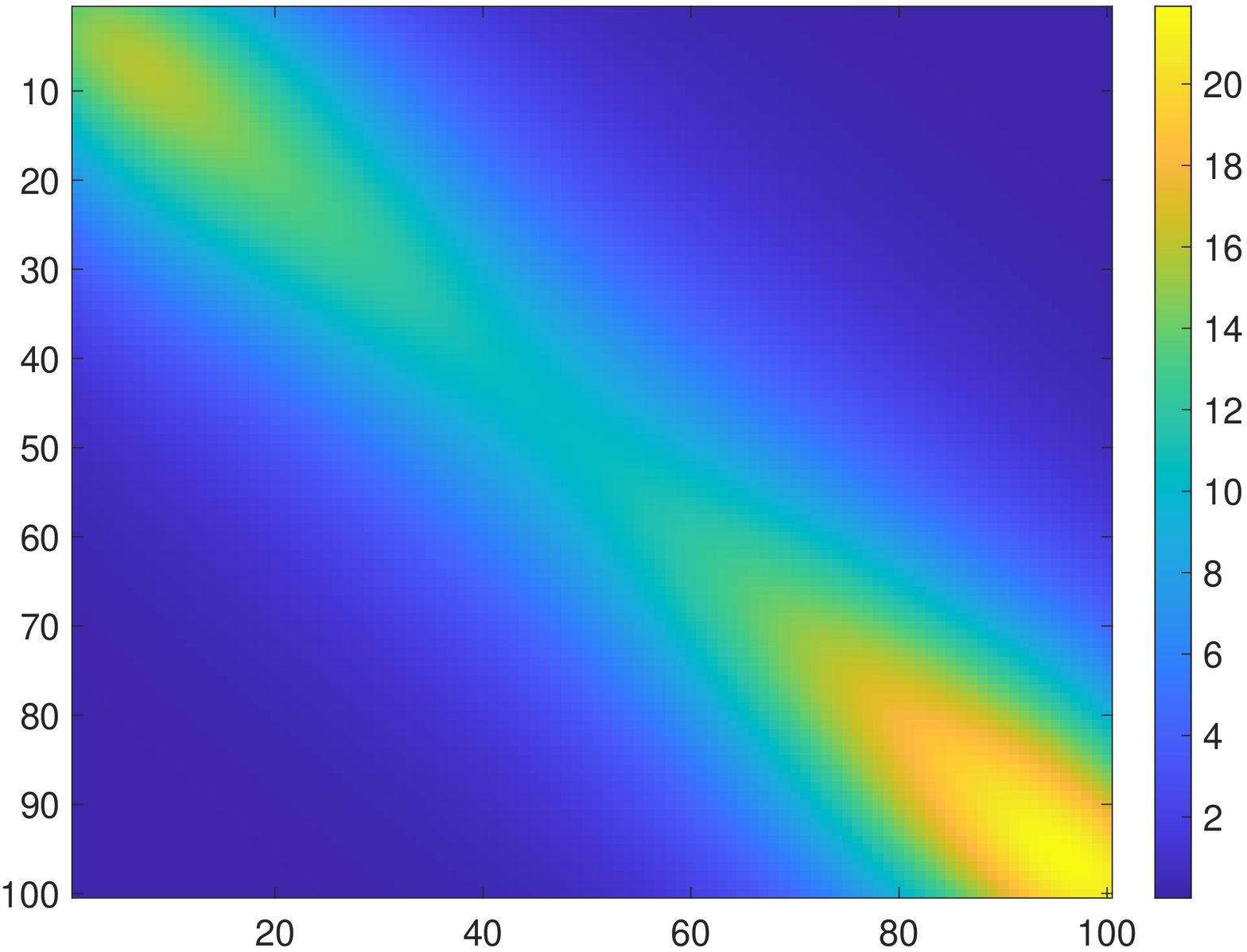}\\
(a) $\Phi^{-1}(U(s))$ & (b) $\widehat{\Phi}^{-1}(U(s))$  & (c)  truth cov  & (d)  estimated cov
\end{tabular}
\caption{Parameters estimation in the copula model.    Panels (a) and (b) show some ground truth quantile functions $\Phi^{-1}(U(s))$ and the corresponding estimated $\widehat{\Phi}^{-1}(U(s))$, respectively. The blue lines mark different quantile ranges for calculating difference between $\Phi^{-1}(U(s))$ and $\widehat{\Phi}^{-1}(U(s))$.   Panels (c) and (d) show the ground truth covariance and the estimated covariance, respectively. } \label{fig:simulation_marten}
\end{center}
\end{figure}

%In the copula model, $\Phi^{-1}(U(s))$ is a Gaussian process whose covariance is specified by the Mate\'rn function (where the parameter $\nu$ is fixed to be $5/2$ in all experiments). 
{We also evaluate the generalized least square algorithm for estimating parameters in the t-copula model, whose degree of free parameter $\varrho$ was estimated based on $(\hat{F}_{(s_i,\bf{x})}(y_i),\hat{F}_{(s_j,\bf{x})}(y_j))$ using the R package ``VineCopula". We fixed $\nu = 5/2$ in all experiments.    Figure \ref{fig:simulation_marten} (a) shows the $20$ ground truth  functions of $\Phi^{-1}(U(s))$, and (b) shows the corresponding estimated $\widehat{\Phi}^{-1}(U(s))$. Moreover, some flat regions in $\widehat{\Phi}^{-1}(U(s))$ are caused by the estimation precision because we only estimated quantiles from 1 to 99.  The average point-wise distances between $\Phi^{-1}(U(s))$ and $\widehat{\Phi}^{-1}(U(s))$ shown in Figure \ref{fig:simulation_marten} (b) show that the estimates are better in the middle  than in the tail.  We then calculated $t^{-1}_\varrho(U(s))$, and estimate  ${\bm \Sigma}_{\bm \theta(\bf x)}$ in our copula model. For ${\bf x} = (1,1,0.8507)^T$, Figure \ref{fig:simulation_marten} (c) and (d) show the ground truth covariance function and the estimated covariance function $\varrho/(\varrho-2)\widehat{{\bm \Sigma}}_{\bm \theta(\bf x)}$, respectively. } The estimation algorithm can work very well even when the domain is irregular (see additional simulation results   in the supplementary material).

\begin{figure}
\begin{center}
\begin{tabular}{cc}
\includegraphics[height=2in]{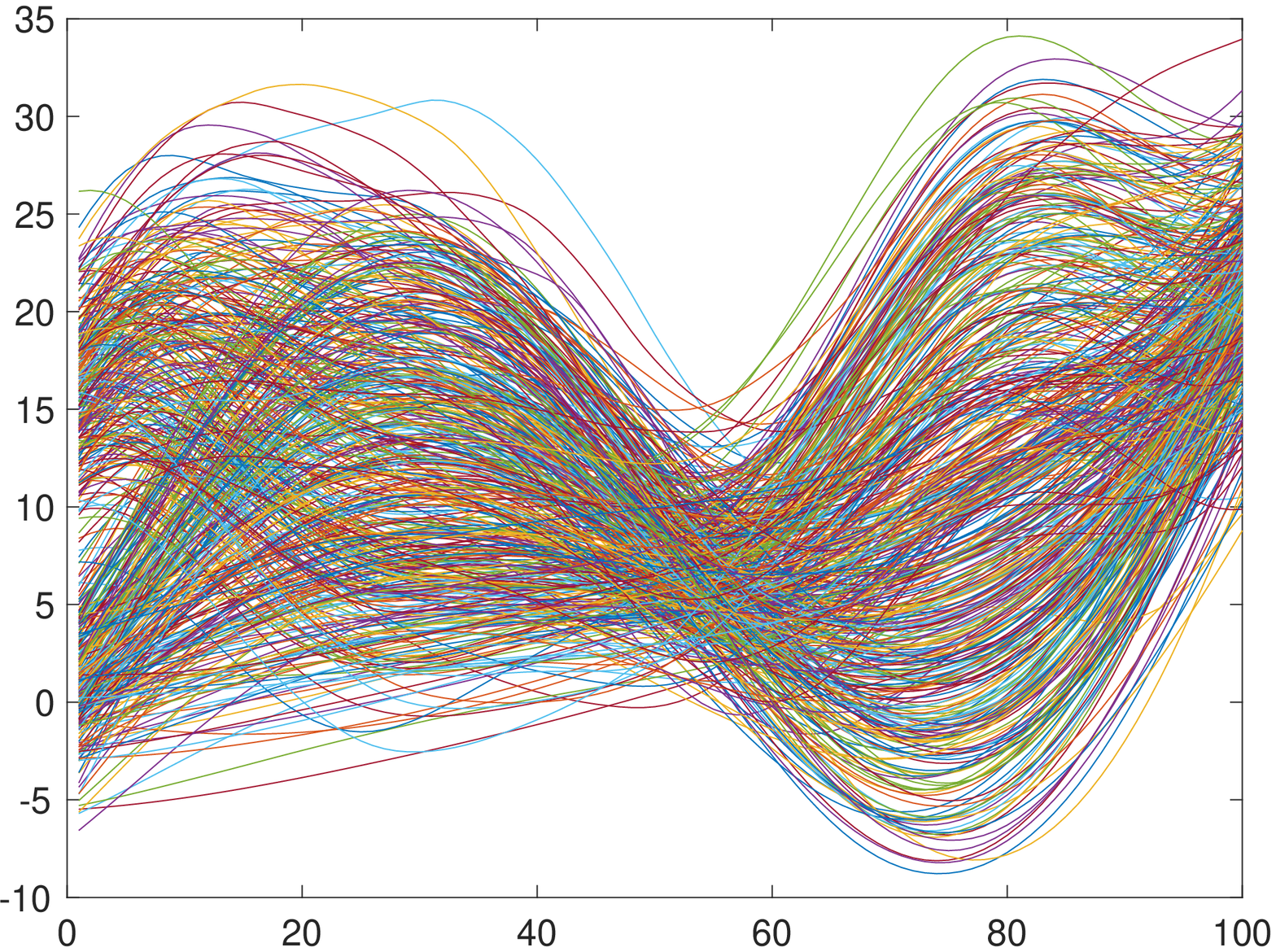} &
\includegraphics[height=2in]{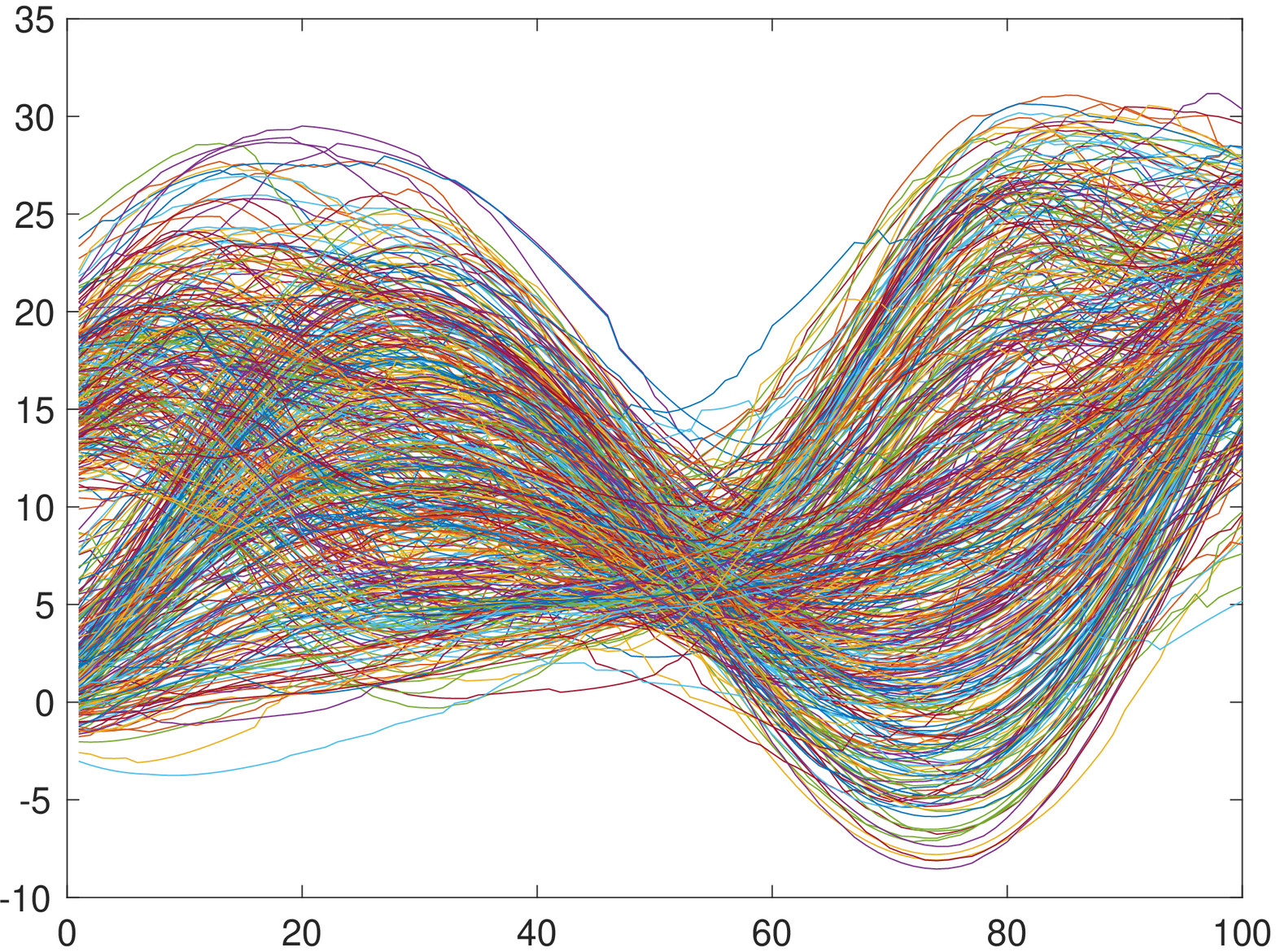}\\
(a) Samples using ground truth parameters & (b) Samples using estimated parameters  \\
\end{tabular}
\caption{Examples of simulating $\{Y_i\}$ with given $\{{\bf x}_i\}$ in the simulated test dataset.} \label{fig:sample_Y}
\end{center}
\end{figure}

{Next, we can obtain the joint distribution of $Y|{\bf x}$. Using $\{ {\bf x}_i \}$ ($i = 1,\ldots,500$) in the test dataset, we simulated ${Y_i}$. 
In Figure \ref{fig:sample_Y}, (a) shows the randomly sampled $\{Y_i\}$ using the ground truth parameters, and (b) shows the simulated $Y_i$ using estimated parameters based on GACV which we can see follow similar patterns of $\{Y_i\}$ sampled using the ground truth parameters.}

\subsection{Real Data Analysis}

\subsubsection{ADNI DTI Data}

We apply the proposed method to analyze the diffusion tensor imaging (DTI) data in the Alzheimer's Disease Neuroimaging Initiative (ADNI) study. In 2003, the ADNI was started by National Institute on Aging, the
National Institute of Biomedical Imaging and Bioengineering, the Food and Drug Administration, and some private pharmaceutical companies and non-profit organizations. This multisite study assesses clinical, imaging, genetic and biospecimen biomarkers through the process of normal aging to early mild cognitive impairment, to late mild cognitive impairment, to dementia or Alzheimer's disease (AD). Participants were recruited across North America to participant in the project and a variety of imaging and clinical assessments were conducted for each participant. Results were shared by ADNI through the Laboratory of Neuro Imaging's Image Data Archive (https://ida.loni.usc.edu/).

\begin{figure}
\begin{center}
\begin{center} $
\begin{array}{lll}
\includegraphics[height=1.4in]{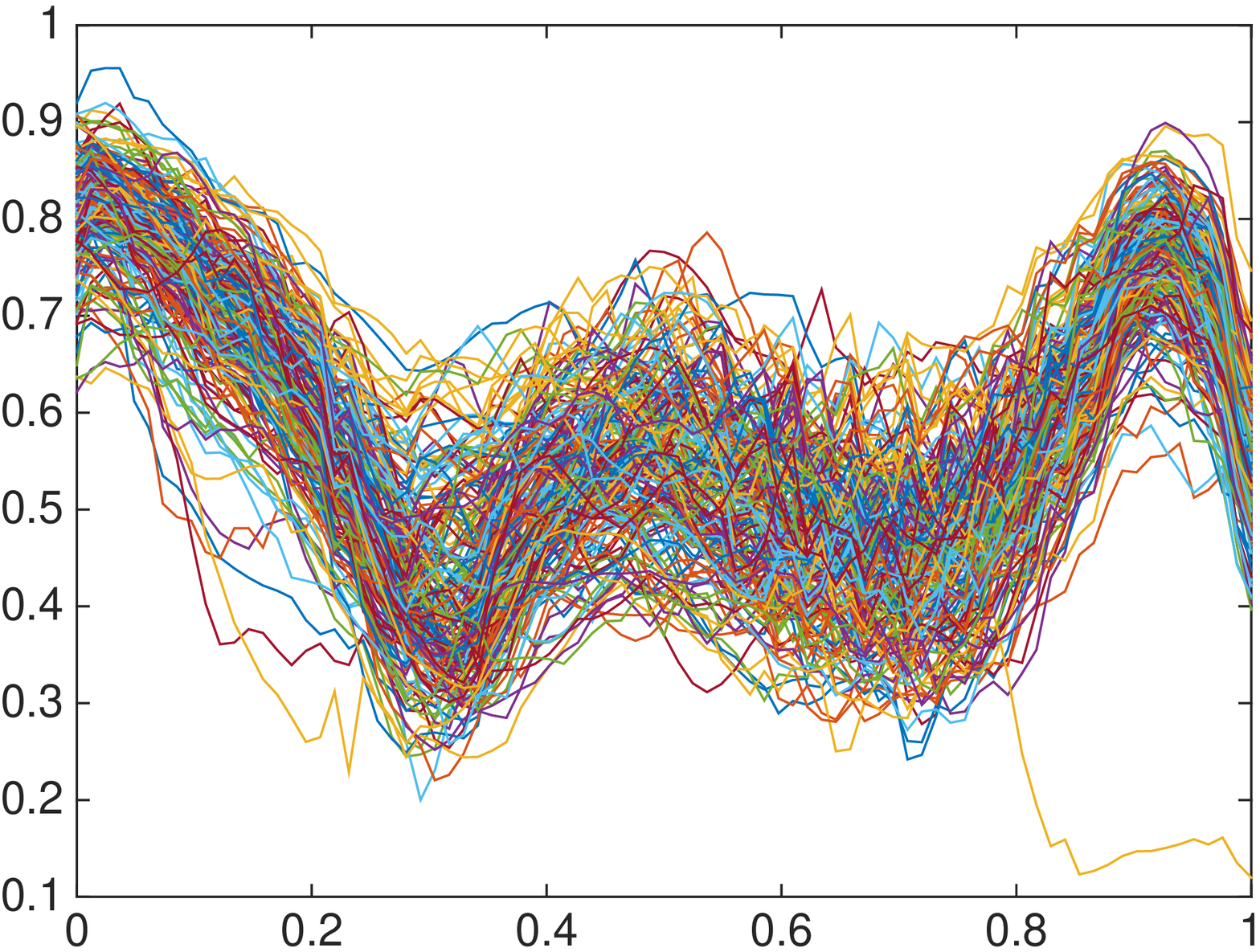} &
\includegraphics[height=1.4in]{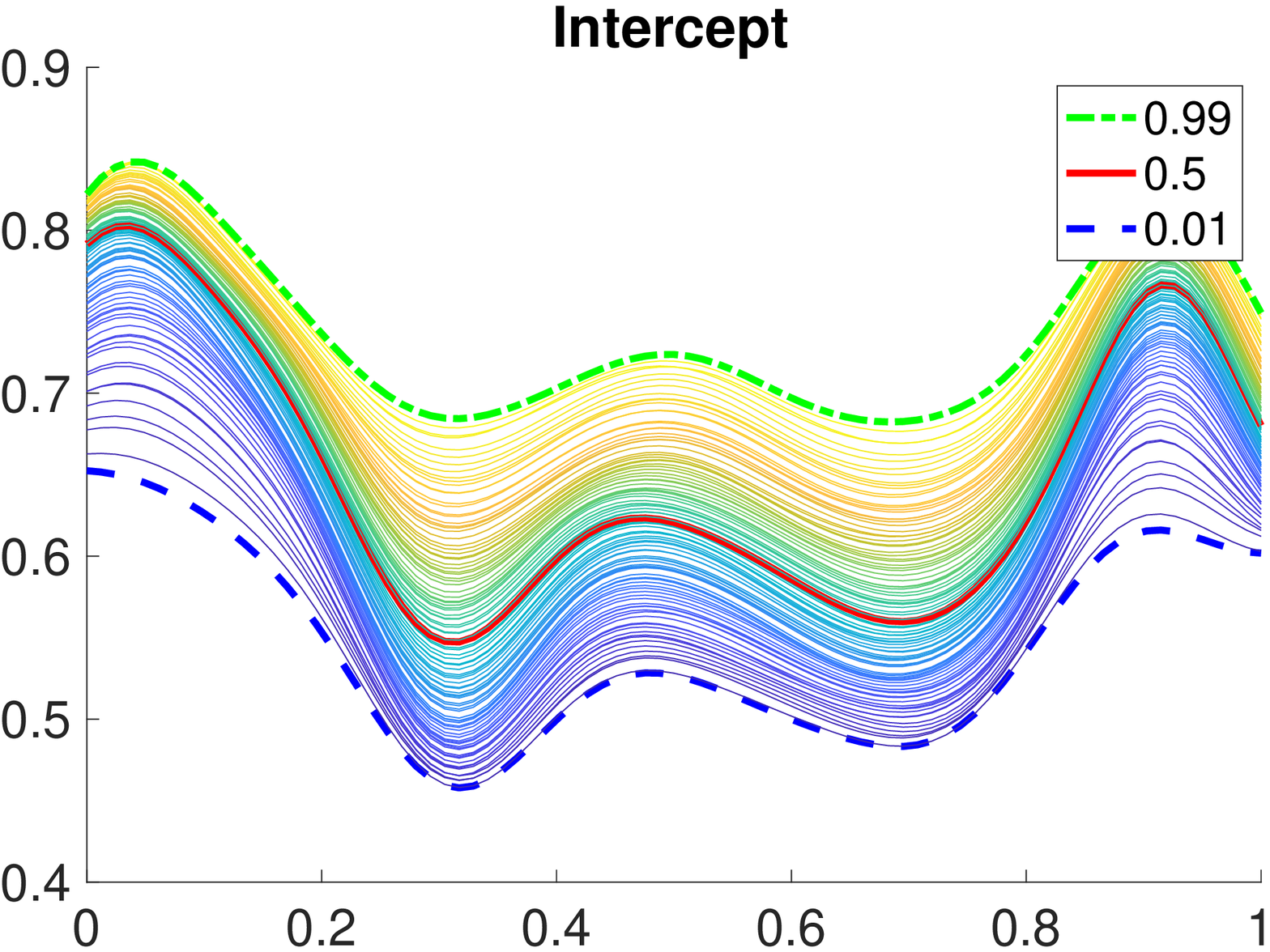} \\
~~~~~~~\text{ 203 subjects data           } & ~~~~~~~~~~~~~~\beta_{intercept}
\end{array} $
\end{center}

\begin{center}$
\begin{array}{lll}
\includegraphics[height=1.4in]{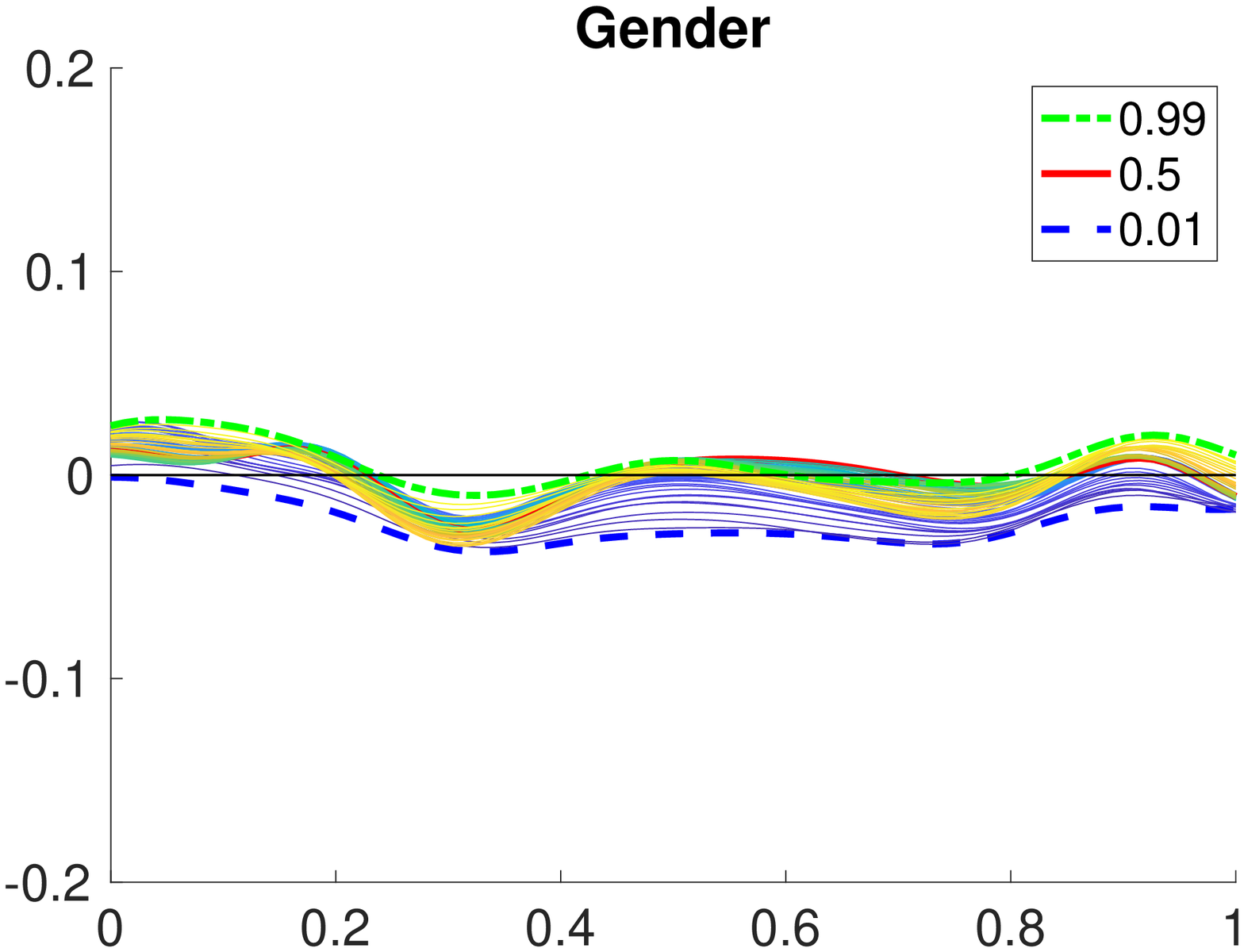} &
\includegraphics[height=1.4in]{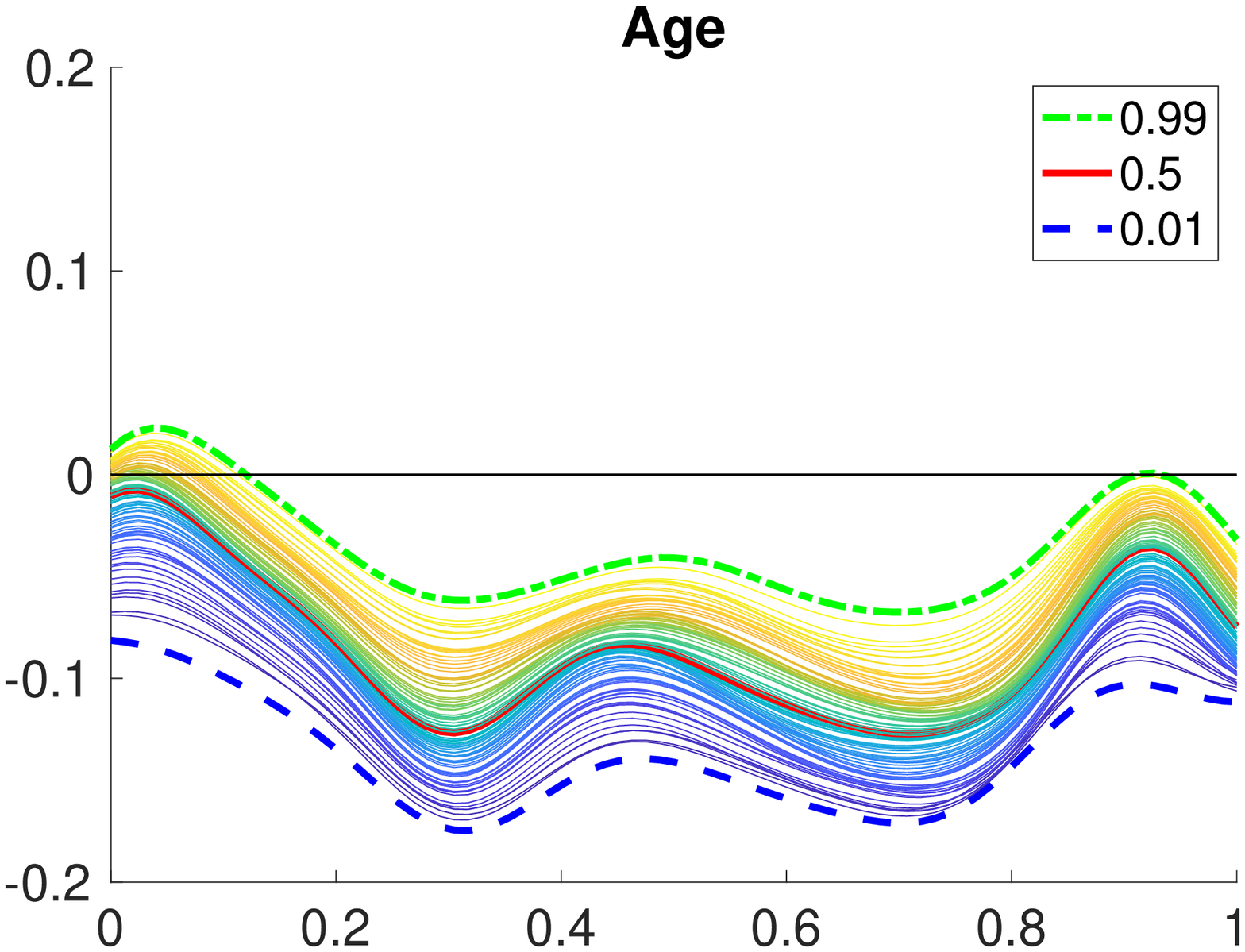} &
\includegraphics[height=1.4in]{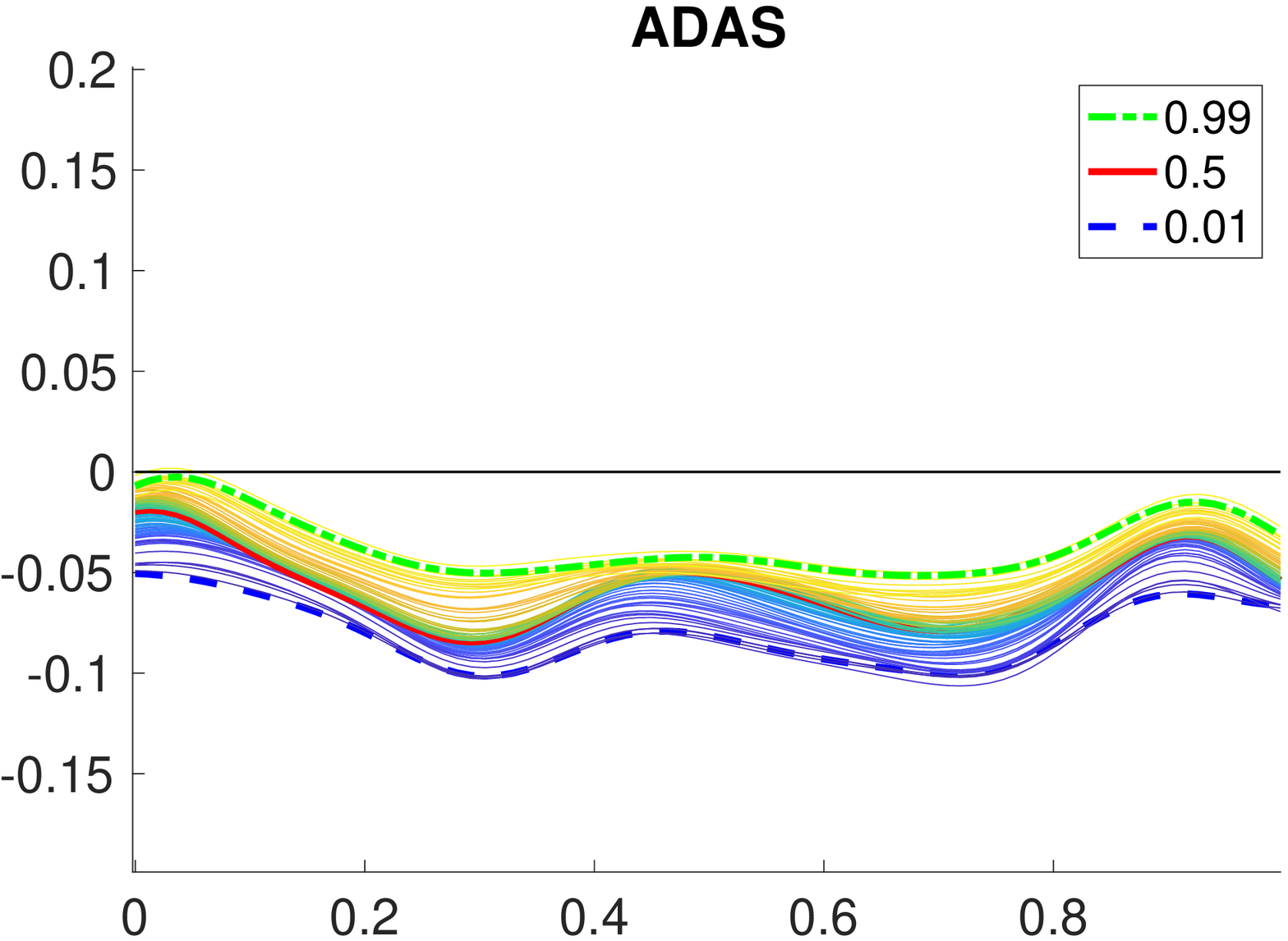} \\
~~~~~~~~~~~~~~\beta_{gender} &~~~~~~~~~~~~~~ \beta_{age} & ~~~~~~~~~~~~~~\beta_{ADAS}
\end{array}$
\end{center}
\caption{Our SQR results on corpus callosum FA curves from ADNI data. The top left panel shows the original FA curves of 203 subjects in ADNI. The remaining panels show our estimated ${\bm \beta}_\tau$ at different $\tau$ ($\tau$ increases from blue to yellow).} \label{fig:fittedfacurves}
\end{center}
\end{figure}

In our study, $203$ subjects' diffusion weighted MRI and demographic data were downloaded and processed. DTI data for each subject were extracted using two steps. First, estimate a diffusion tensor at each voxel using a weighted least square estimation \citep{koay2006unifying, zhu2007statistical}. Second, register DTI images from multiple subjects using the FSL TBSS pipeline \citep{smith2006tract} to create a mean image and a mean skeleton. To be more specific, after estimating the diffusion tensor, fractional anisotropy (FA), a scalar measure of the degree of anisotropy, was calculated for each voxel.  Next, FA maps of all subjects were fed into the TBSS tool in the FSL software and were aligned  non-linearly. The mean FA image was then calculated and thinned to obtain a mean FA skeleton representing the centers of all white matter tracts common to the group. Subsequently, each aligned FA data were projected onto this skeleton. We focus on the midsagittal corpus callosum skeleton and the associated FA curves from all subjects.  The corpus callosum is the largest fiber bundle in the human brain and is responsible for much of the communication between the two cerebral hemispheres. Figure \ref{fig:fittedfacurves} (a) shows the FA curves from all $203$ subjects.

\begin{figure}
\begin{center}
\begin{tabular}{ccc}
\includegraphics[height=1.4in]{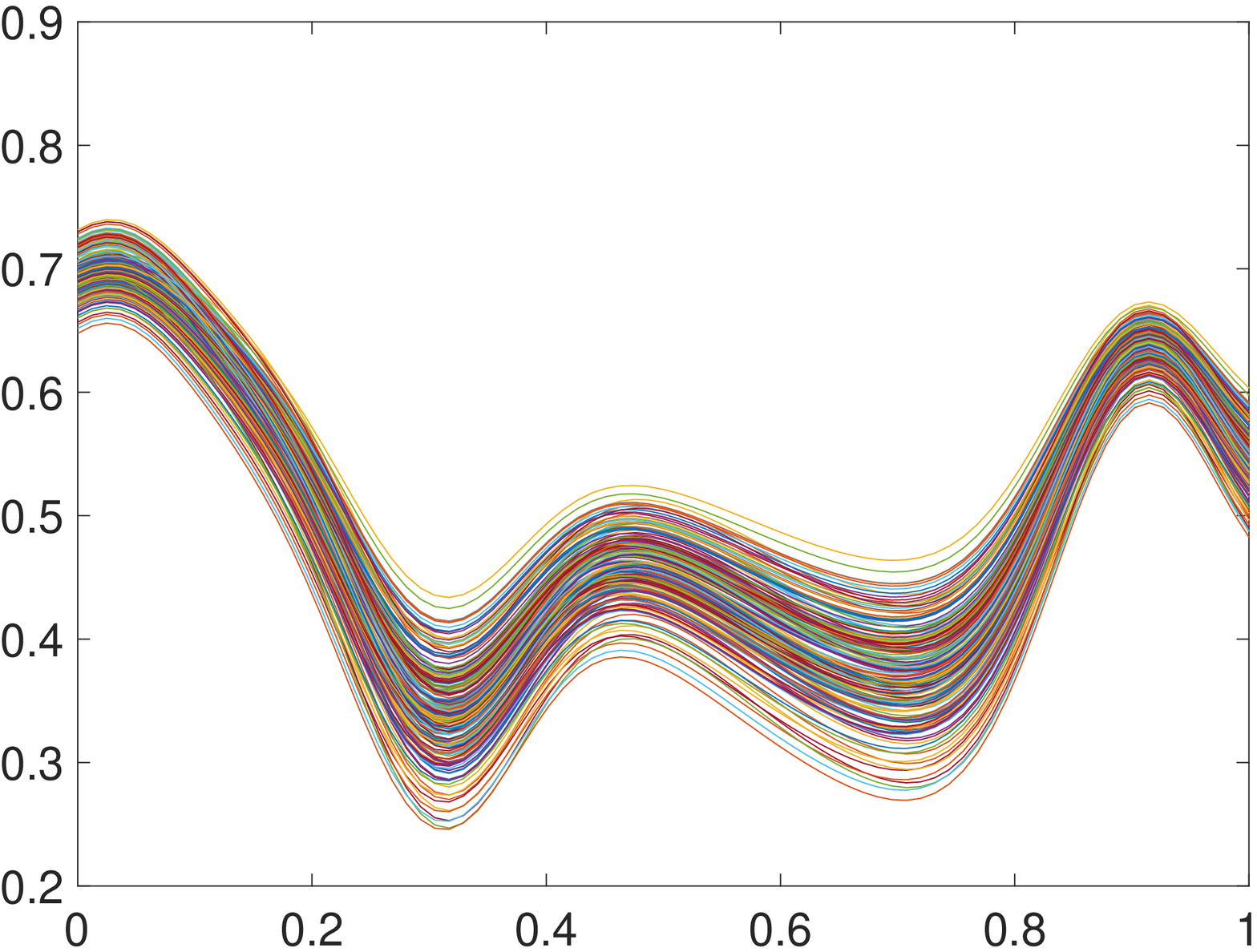} &
\includegraphics[height=1.4in]{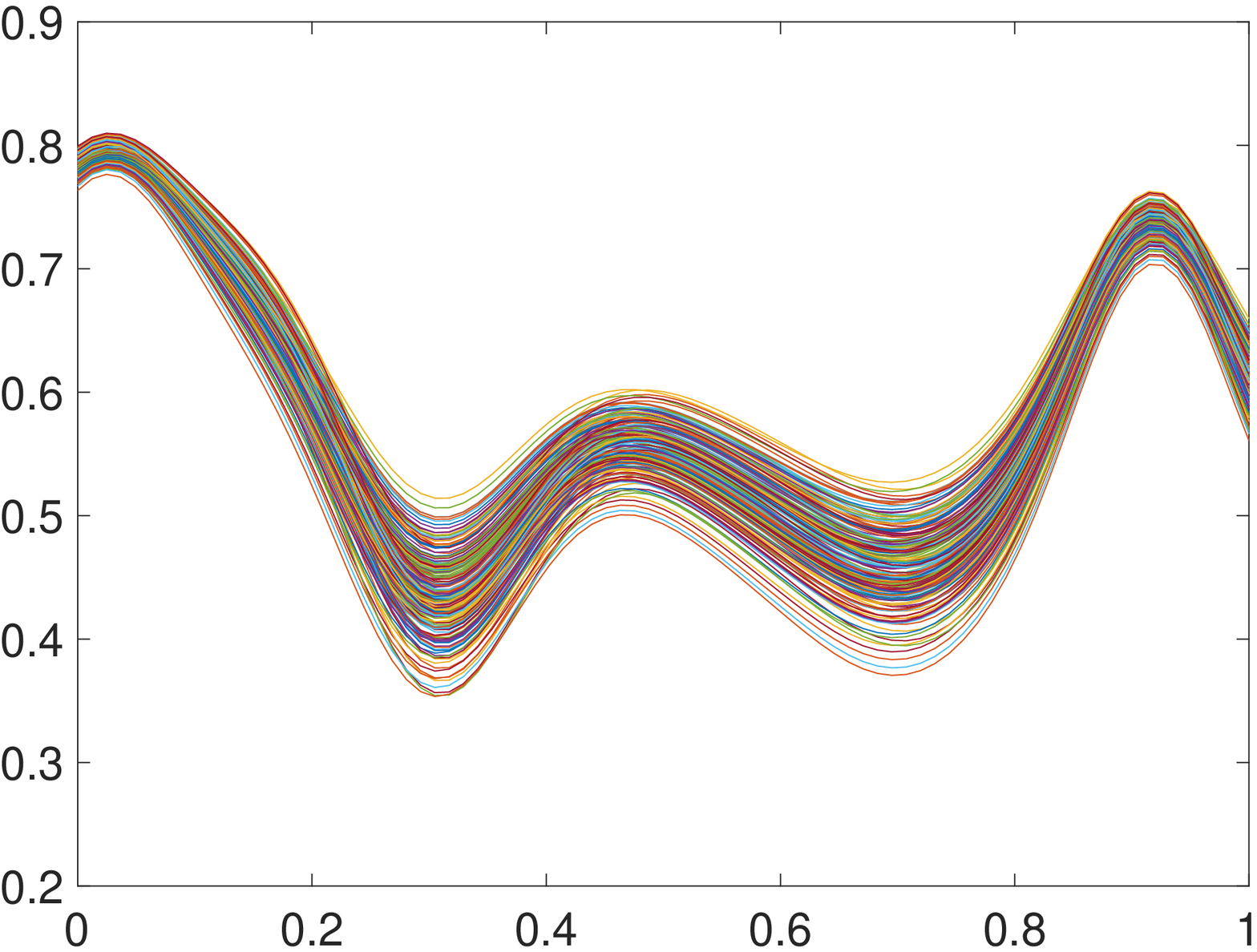}&
\includegraphics[height=1.4in]{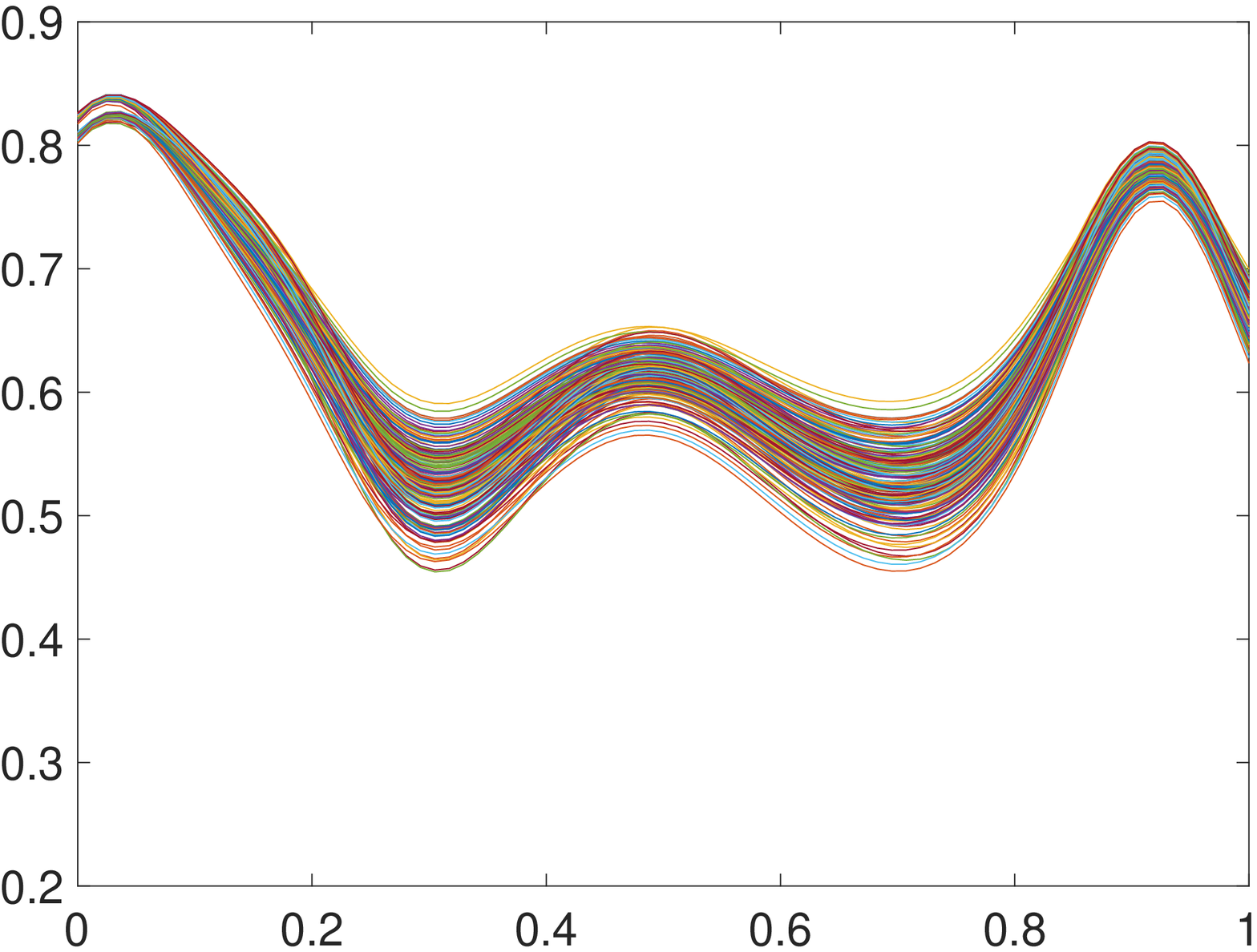}\\
(a) $\tau = 0.10$ & (b) $\tau = 0.50$ & (c) $\tau =0.80 $\\
\end{tabular}
\caption{Predicted FA curves ${\bf x}^T_i {\bm \beta}$ at different $\tau$ given ${\bf x}_i$ in ADNI DTI dataset.} \label{fig:predfacurves}
\end{center}
\end{figure}

{ We are interested in exploring the effects of gender, age, and Alzheimer's Disease Assessment Scale-Cognitive subscale (ADAS) \citep{rosen1984new} score on the diffusion properties along corpus callosum tract at different quantile levels.  Specifically,  we have ${\bf x}_i = (1, \text{ gender, age, ADAS} )^T$, and $Y_i$ representing an FA curve. For gender, males are set as $0$ and females are set as $1$.  To make comparisons with \cite{reich2011}, we standardized both age and ADAS to be in $(0,1)$. The fitted coefficient functions $\hat{\bm \beta}_\tau$  are shown in Figure \ref{fig:fittedfacurves} at different quantile levels ($\tau = 0.01,0.02,\ldots,0.99$). Figure \ref{fig:predfacurves} shows the predicted FA curves ${\bf x}^T_i \hat{\bm \beta}_\tau$ across $\tau = 0.1, 0.5$ and $0.8$.  It is well-known that aging  deteriorates brain structure \citep{wyss2016ageing}. The ADAS is widely used to detect cognitive deficits in people suffering from AD. The range of ADAS in our dataset is $(0, 51)$ with higher scores indicating greater degrees of cognitive deficit. From Figure \ref{fig:fittedfacurves}, it follows that both aging and ADAS have a negative effect on  the diffusivity on the midsagittal corpus callosum skeleton.}

{
Across different quantiles, we observe a nice layout of $\beta_{intercept}$ - the median is in the middle between quantile $1$ and quantile $99$. We also observe some interesting structures in the coefficient functions. From $\beta_{gender}$,  we see that there is not much difference between males and females at quantiles ranging from $10$ to $99$. However, if we consider lower quantiles (e.g., from $1$ to $10$), the FA for females along corpus callosum skeleton has smaller values compared with those for males. Biologically, this indicates in the population that the lower FA values along the midsagittal corpus callcosum in females are smaller than the lower FA values in males, but their means might not have much difference \citep{inano2011effects}. We utilized the package FADTTS \citep{zhu2011fadtts} to perform a mean regression and the results are presented in the Supplement II, where we can see that the coefficient function for gender fluctuates around $0$.  From $\beta_{age}$, we see that for people with worse FA values (at lower quantiles), the same amount of aging can contribute to more FA reduction (indicating worse white matter deterioration \citep{kochunov2007relationship}) than those at high quantiles. The mean regression results from FADTTS cannot give this information.
For $\beta_{ADAS}$, the deterioration of Alzheimer's Disease (measured by ADAS) is more similar across different quantiles, acting differently compared with age.}

\begin{figure}
\begin{center}
\begin{tabular}{cccc}
\includegraphics[height=1.2in]{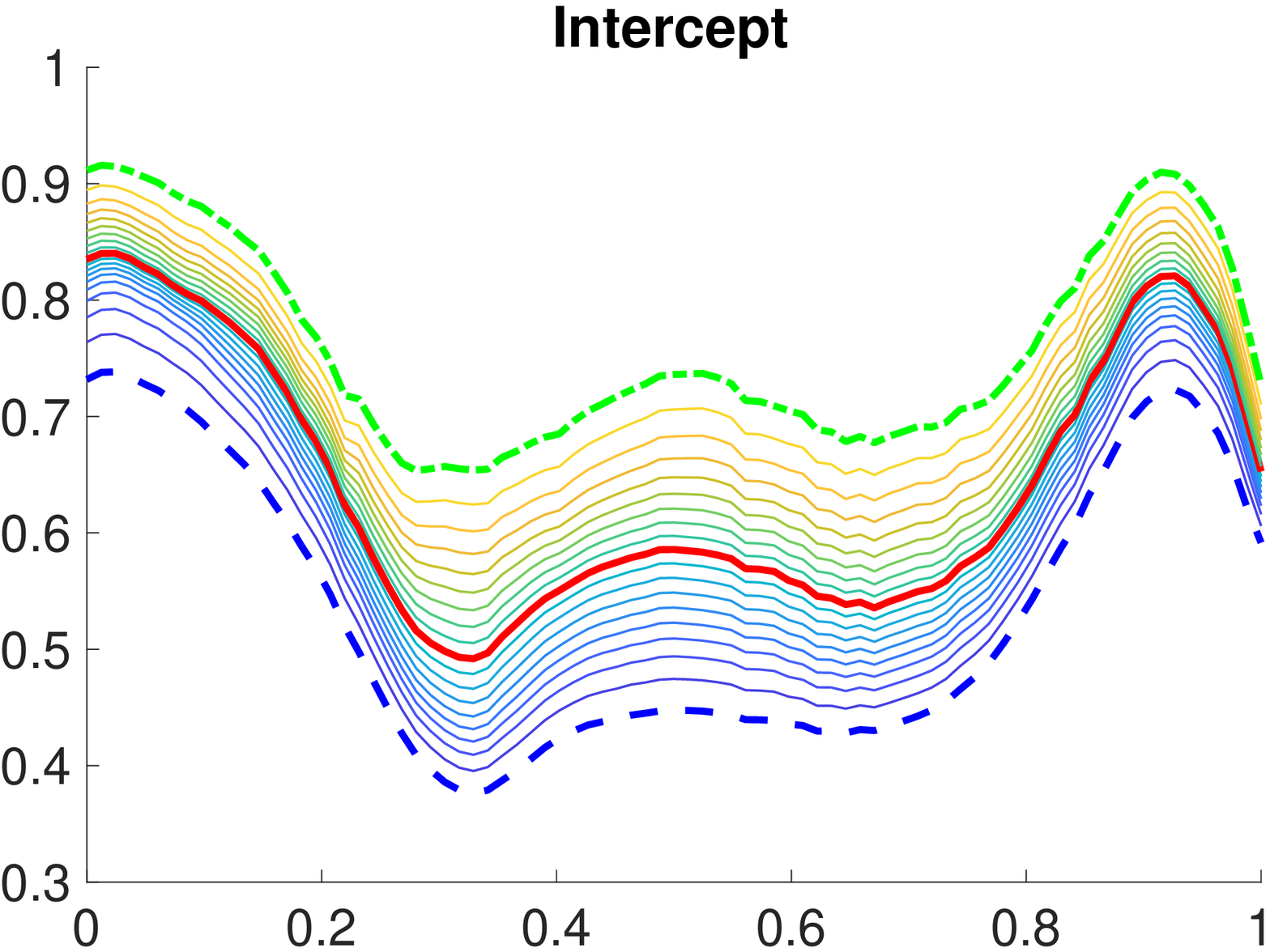} &
\includegraphics[height=1.2in]{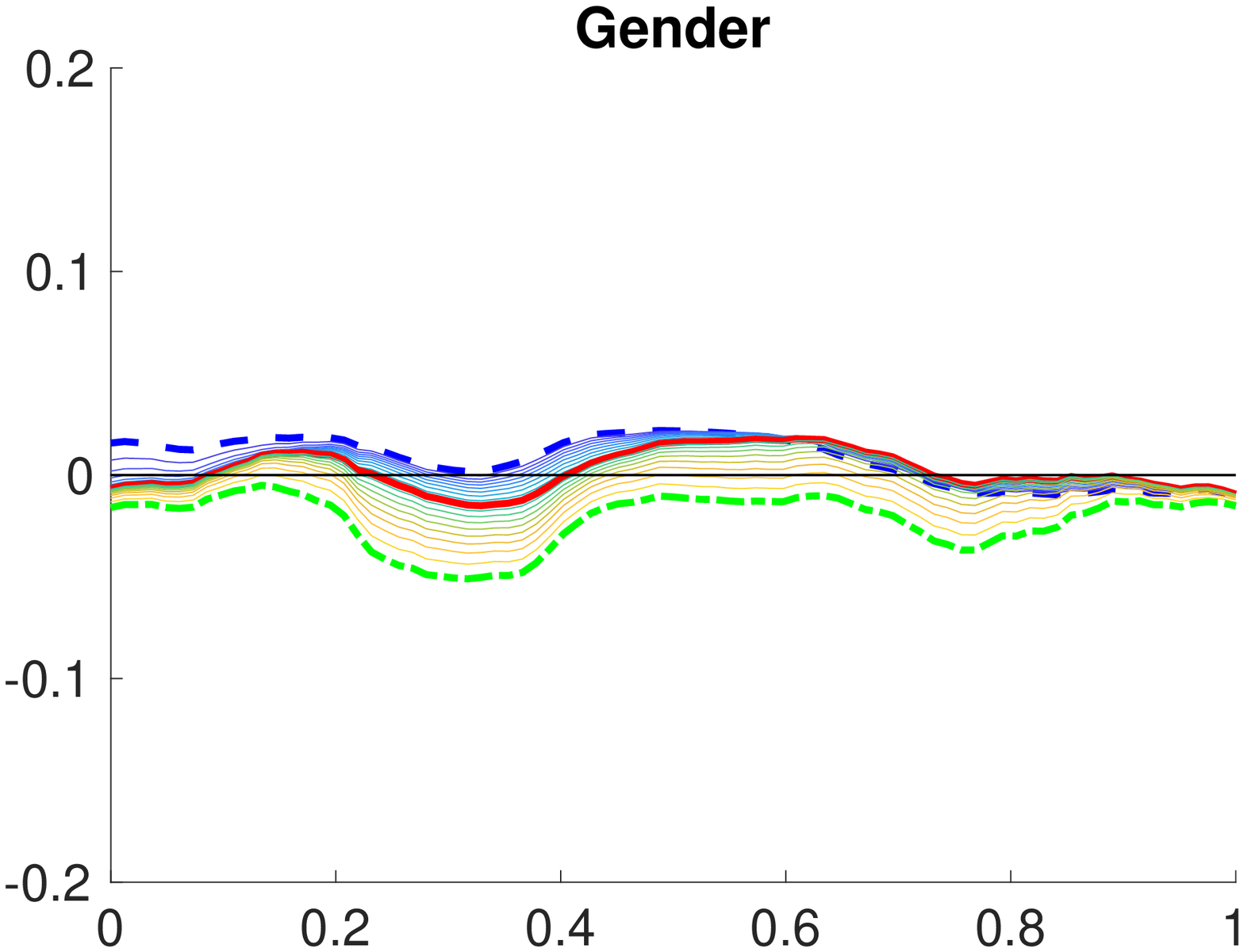}&
\includegraphics[height=1.2in]{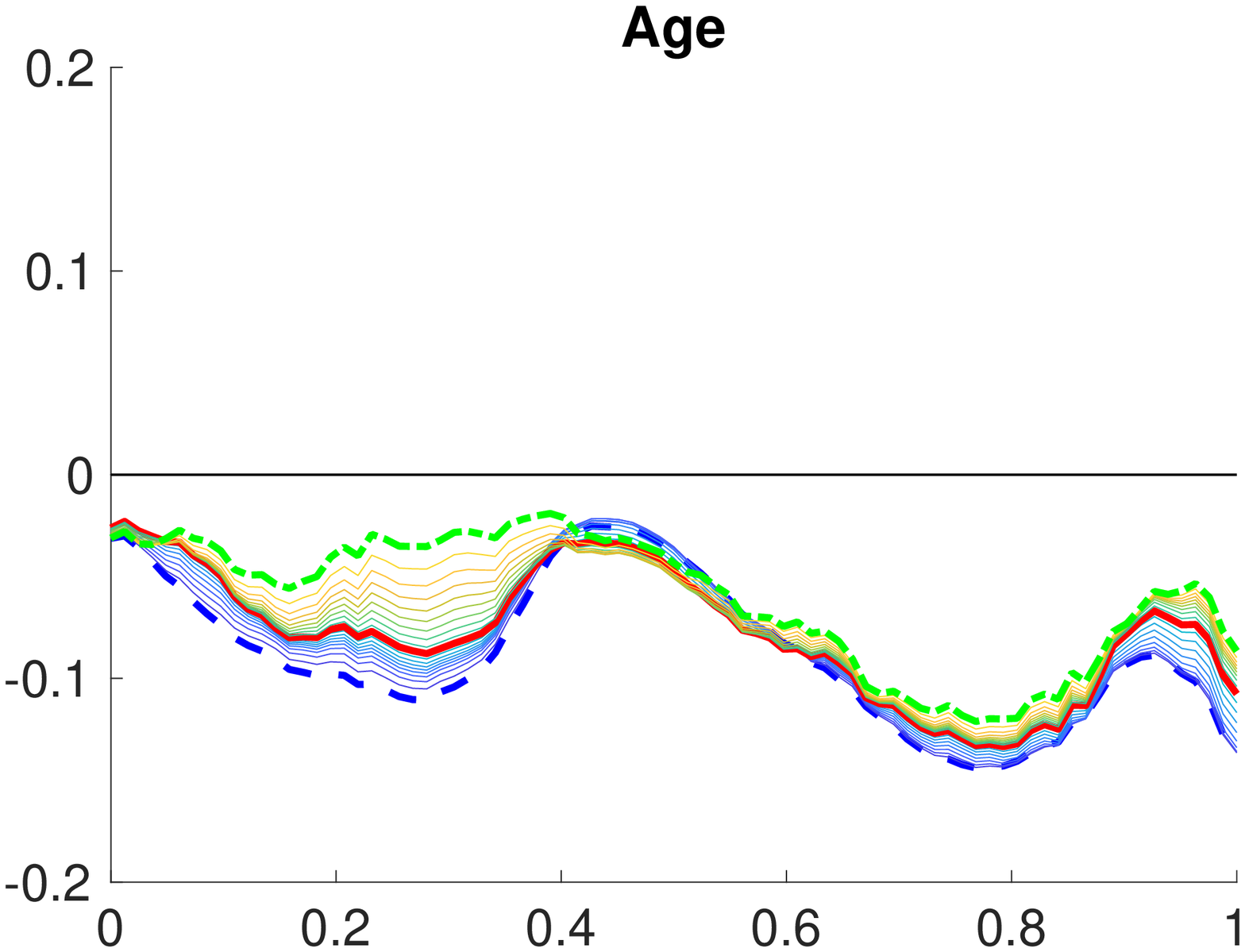}&
\includegraphics[height=1.2in]{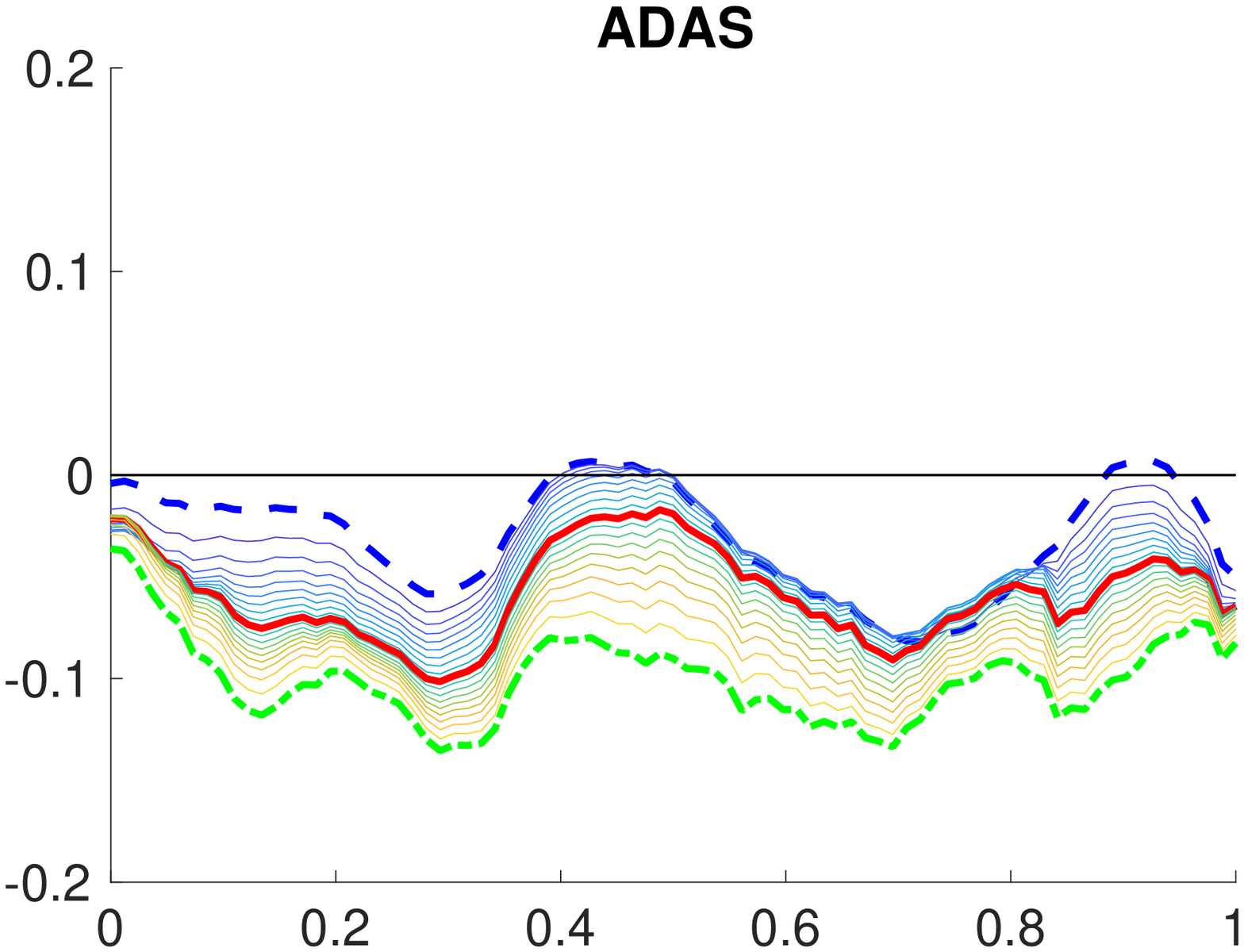}
\\
\end{tabular}
\caption{BSQR results on on corpus callosum FA curves from ADNI data.} \label{fig:bsqr}
\end{center}
\end{figure}

{As another comparison, we ran the BSQR model \citep{reich2011} on the DTI data. Figure \ref{fig:bsqr} shows the results at $20$ different quantile levels.  These results are based on $5000$ MCMC runs after burn-in of the first $1000$ samples. In a laptop with a 2.3 GHz Intel i9 CPU and 32 GB memory, it took about $1400$ seconds to get $\boldsymbol{\beta}_\tau$ at $20$ quantile levels with the BSQR, and 13 seconds to get $\boldsymbol{\beta}_\tau$ at $99$ quantile levels with our method. BSQR contains a two-stage approach to approximate $\boldsymbol{\beta}_\tau$. First, independent  quantile regressions at different locations are done to obtain  estimates of the quantile process and their asymptotic covariance. Next, a Gaussian process model is fitted based on the initial estimates of $\boldsymbol{\beta}_\tau(s)$ to introduce spatial dependence structure. As a consequence of this two-stage solution, we see from Figure \ref{fig:bsqr} that the estimated $\boldsymbol{\beta}_\tau$ can be rough and sub-optimal. }

\begin{figure}
\begin{center}
\begin{tabular}{cc}
\includegraphics[height=1.6in]{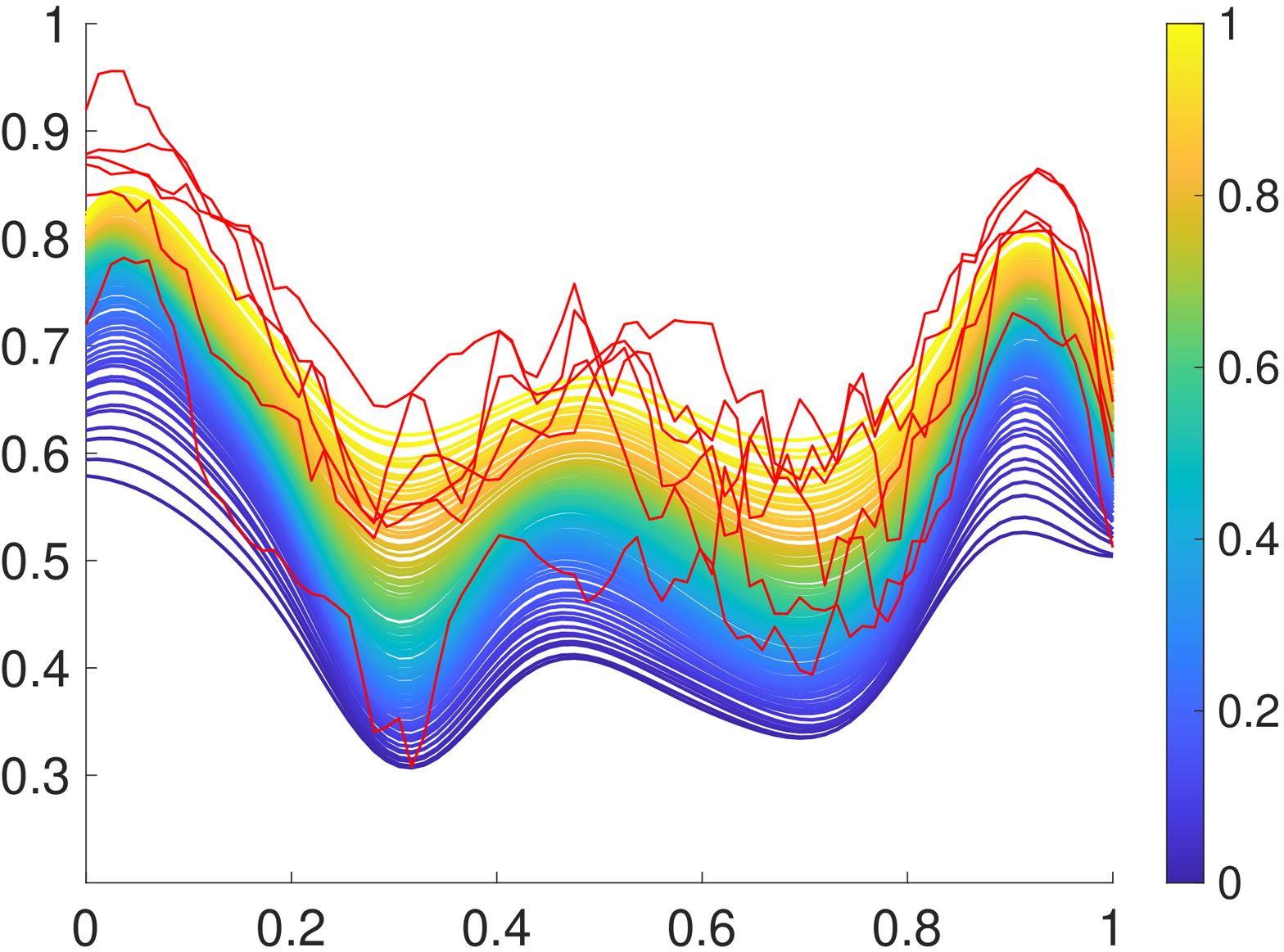} &
\includegraphics[height=1.6in]{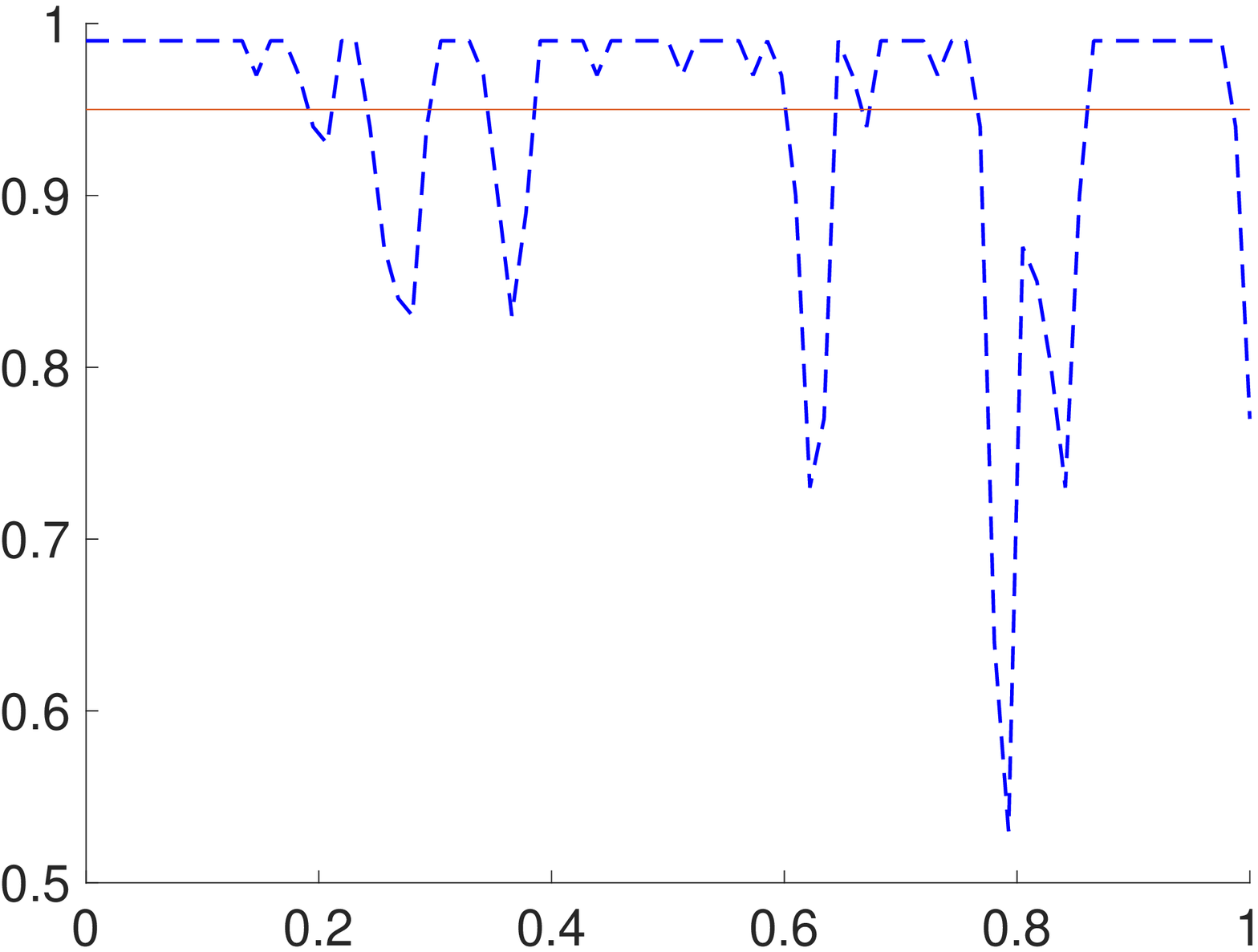}\\
(a) & (b)
\end{tabular}
\caption{Illustration of marginal distribution of $Y|\bf x$ and statistical inference using p-values. (a) Marginal distribution of $Y(s_j)$ given ${\bf x} = (1,\text{ male},\text{ age}=60,\text{ ADAS}=51)$. Red curves show $Y_i$ of subjects having much smaller ADAS scores ($<10$). (b) The p-value curves of one subject in (a) with ADAS$=8$. The solid horizontal line represents $p = 0.95$.  } \label{fig:predtand97}
\end{center}
\end{figure}

\begin{figure}
\begin{center}
\begin{tabular}{cccc}
\includegraphics[height=1.1in]{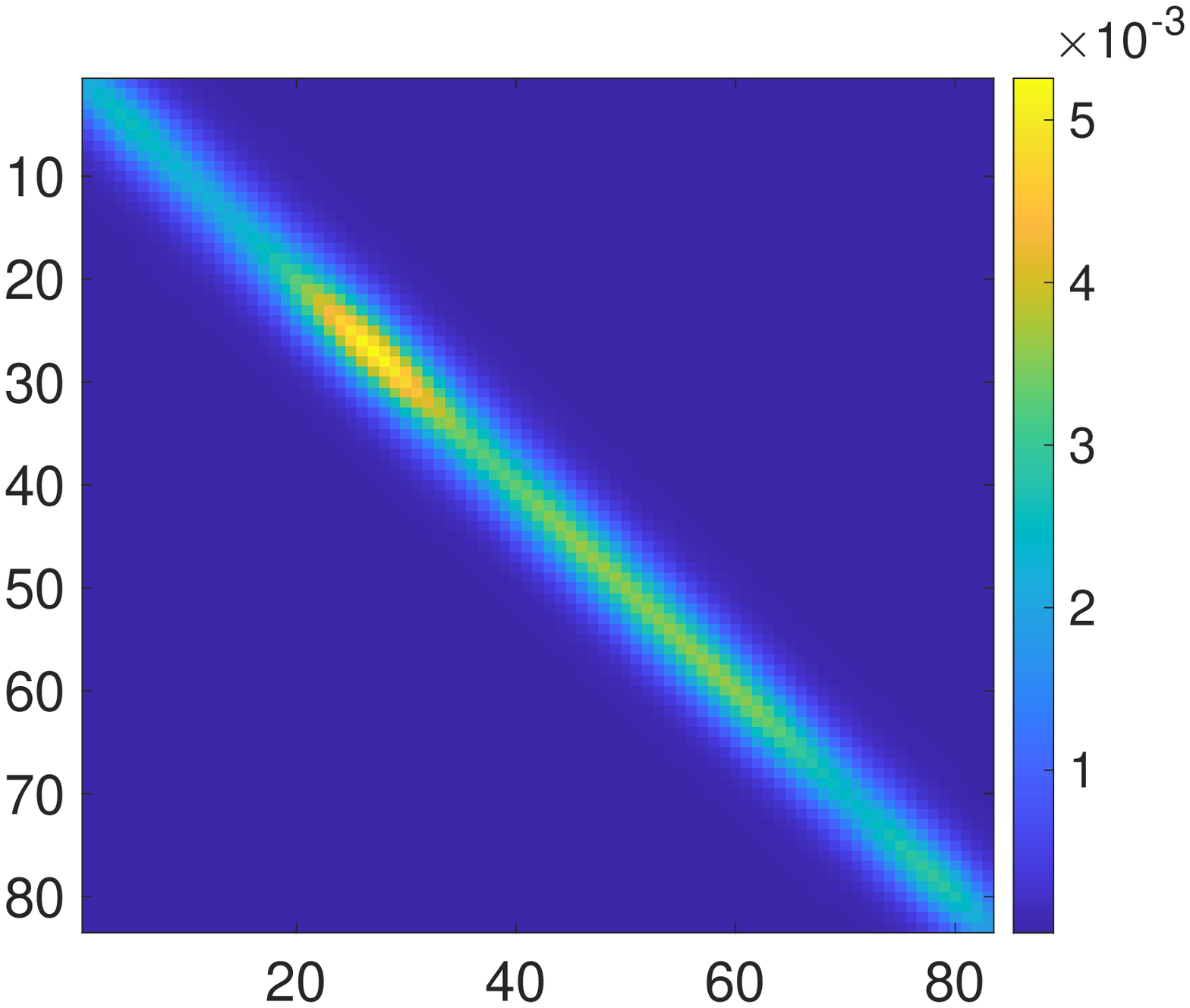} &
\includegraphics[height=1.1in]{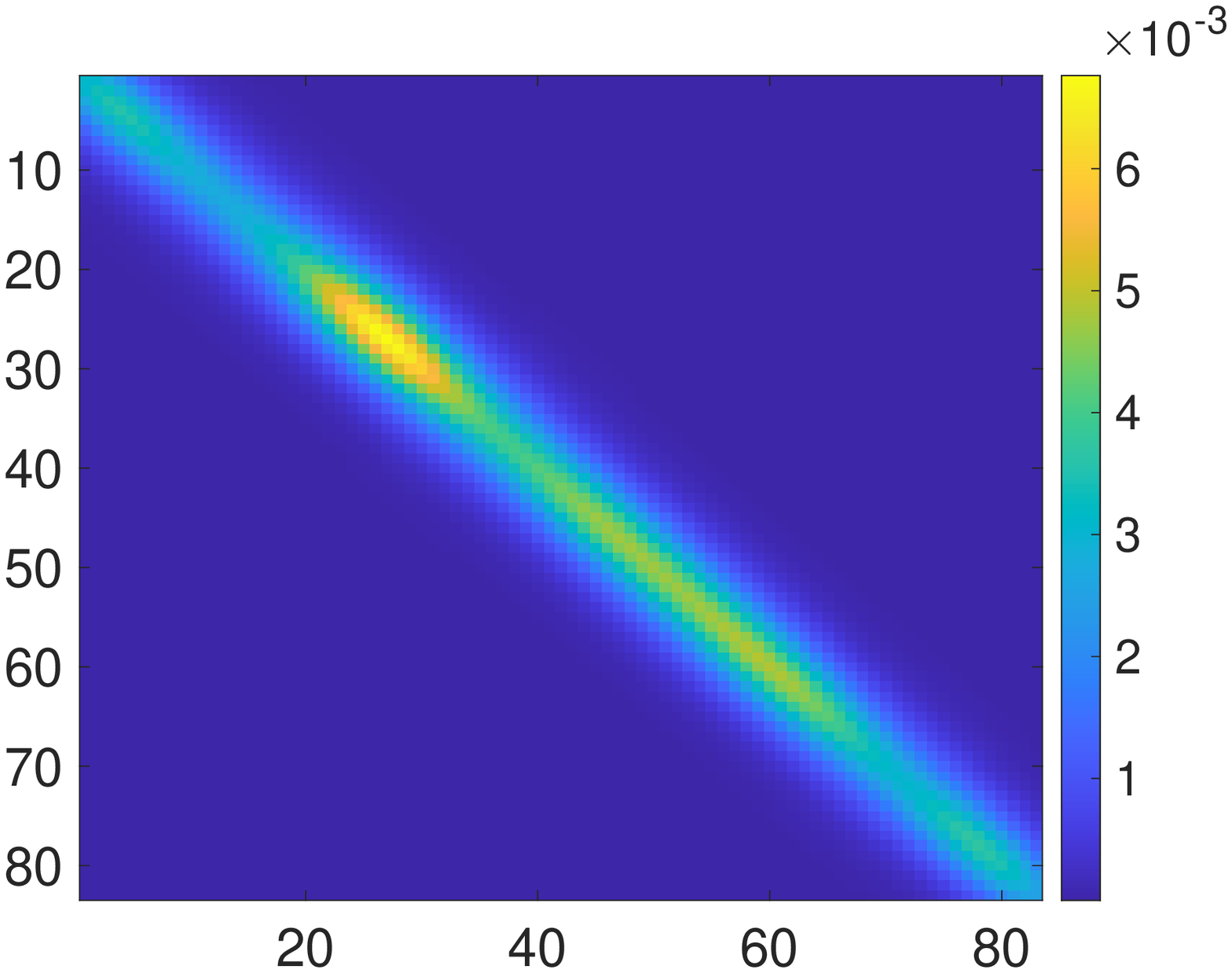}&
\includegraphics[height=1.1in]{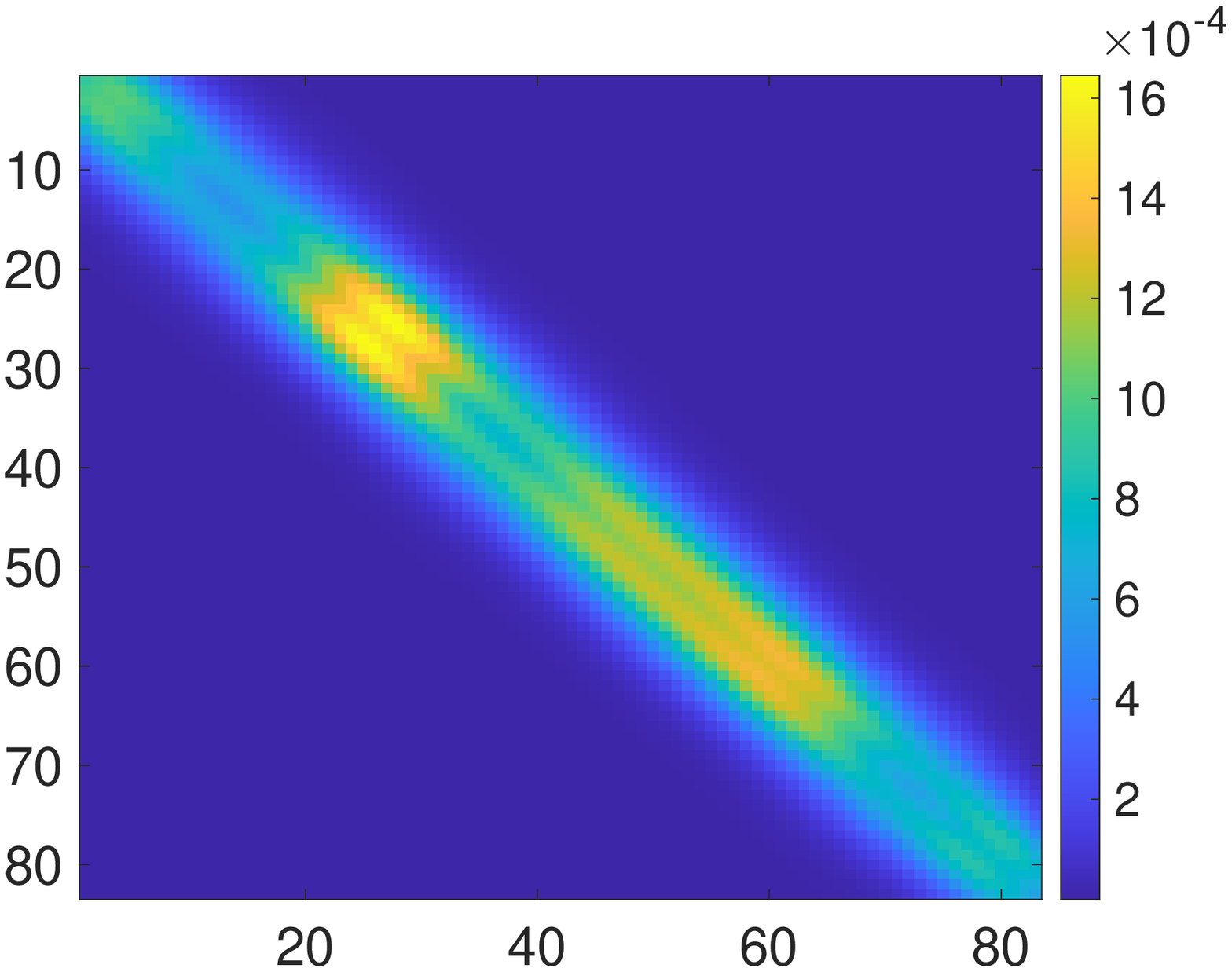} &
\includegraphics[height=1.1in]{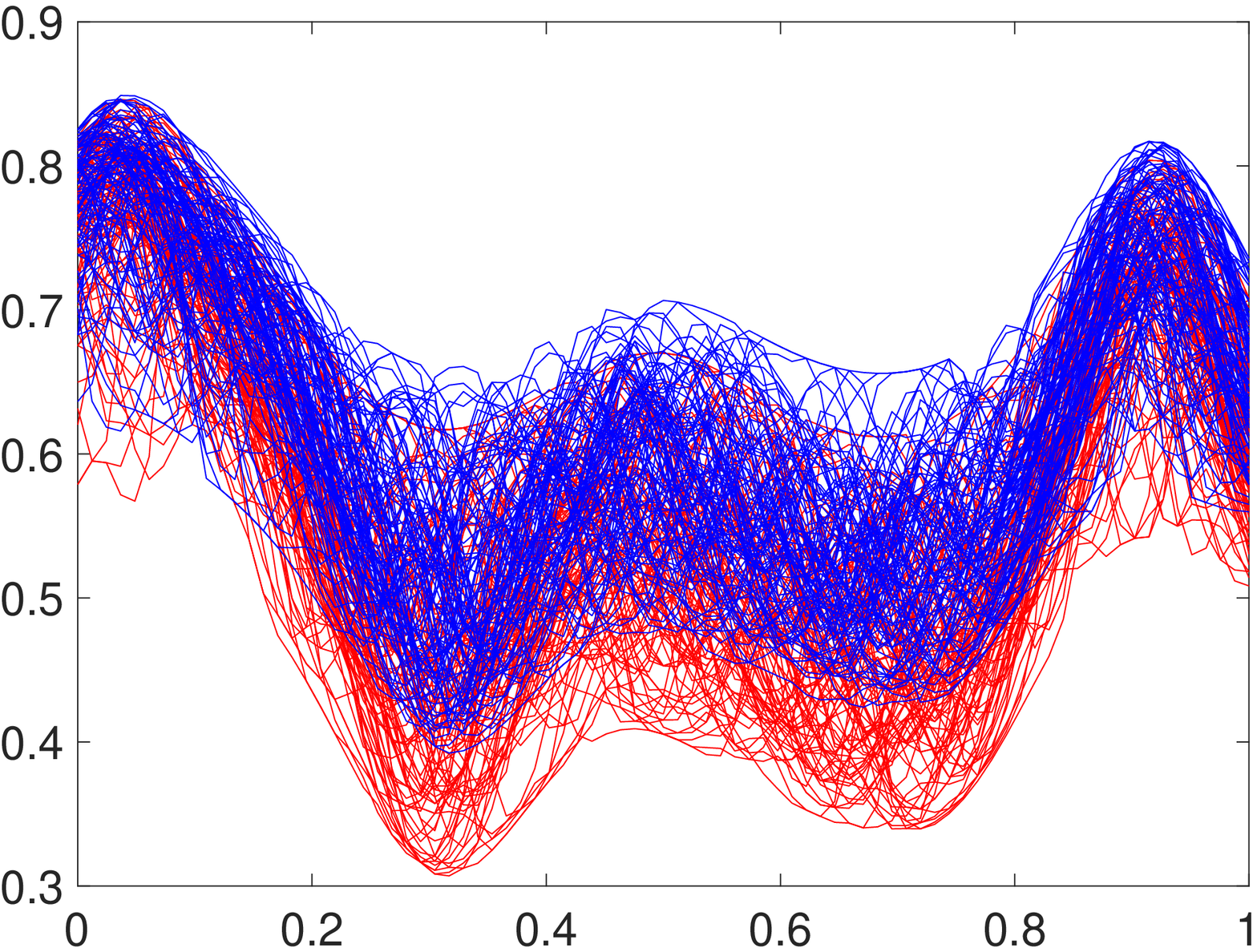}\\
(a)  & (b) & (c) & (d) \\
\end{tabular}
\caption{Results of the copula model in ANDI DTI dataset. (a) $\rho/(\rho-2)\widehat{\bm \Sigma}_{\bm \theta({\bf x}_1)}$ with ${\bf x}_1 =$ (male, age $=61$, ADAS $= 5$). (b) $\rho/(\rho-2)\widehat{\bm \Sigma}_{\bm \theta({\bf x}_2)}$ with ${\bf x}_2 =$ (male, age $=60$, ADAS $= 51$). (c) $\rho/(\rho-2)(\widehat{\bm \Sigma}_{\bm \theta({\bf x}_1)}-\widehat{\bm \Sigma}_{\bm \theta({\bf x}_2)})$.  (d) Random samples from the distributions of $Y|{\bf x}_1$ (blue) and $Y|{\bf x}_2$ (red). } \label{fig:b}
\end{center}
\end{figure}

{Next we show how to use SQR to make  statistical inference.  First, we selected one male patient with age $60$ and ADAS $51$ and calculated FA curves at different quantiles along the midsagittal corpus callosum skeleton with the fitted model. Figure \ref{fig:predtand97} (a) shows the marginal distribution $Y(s_j)$ given ${\bf x} = (1,\text{male},\text{age}=60,\text{ADAS}=51)$. In addition, we plot a subset of subjects having much smaller ADAS scores ($<10$) in red in Figure \ref{fig:predtand97} (a).  From this plot, we see that, in general, people with smaller ADAS scores have higher FA values than the one with ADAS$=51$. FA represents white matter integrity. A smaller FA along the midsagittal corpus callosum indicates deteriorations of the corpus callosum fiber bundle, and thus weakened bilateral communications. Our finding is consistent with the literature \citep{biegon1994human,ardekani2014corpus}.  Second, we carried out statistical inference using $p$-values. Figure \ref{fig:predtand97} (b) shows the p-value curve of the subject with ADAS=8 under the estimated marginal distribution of the subject with ADAS=51. In most locations,   the FA value differs significantly from the two subjects ($1-p<0.05$). Third, Figure \ref{fig:b}  shows more results obtained from the copula model, where panels (a) and (b) show $\varrho/(\varrho-2)\widehat{\bm \Sigma}_{\bm \theta({\bf x}_1)}$ and $\varrho/(\varrho-2)\widehat{\bm \Sigma}_{\bm \theta({\bf x}_2)}$ for ${\bf x}_1 =$ (male, age $=61$, ADAS $= 5$) and ${\bf x}_2 =$ (male, age $=60$, ADAS $= 51$), respectively, panel (c) shows their difference, and panel (d) shows random samples simulated from the distributions of $Y|{\bf x}_1$ (blue) and $Y|{\bf x}_2$ (red). The covariance matrices $\varrho/(\varrho-2)\widehat{\bm \Sigma}_{\bm \theta({\bf x})}$ and their difference in panels (a) (b) and (c) reveal that  subjects with large ADAS values  have longer range correlation between $t_\varrho^{-1}(u_i)$ and $t^{-1}_\varrho(u_j)$ at certain locations, indicating that they have more smoothed FA curves compared with those with  small ADAS values.   Moreover, the results in panel (d) reveals that healthy subjects  in general have higher FA values than AD subjects with the same age and gender.}

\subsubsection{Hippocampus Surface Data}
In another example, we analyze the hippocampal substructures extracted from MRI scans in the ADNI study.  The hippocampus locates in the medial temporal lobe underneath the cortical surface. It belongs to the limbic system and plays important roles in the consolidation of information from short-term memory to long-term memory and spatial navigation \citep{colom2013hippocampal, luders2013global}. The neurodegenerative activity of  AD is evident in the hippocampus region. 

{
In our study, given the MRI scans, the hippocampal substructures were segmented with FSL FIRST \citep{patenaude2011bayesian} and the hippocampal surfaces were automatically reconstructed with the marching cube method \citep{lorensen1987marching}. We used a surface fluid registration based hippocampal subregional analysis package \citep{wang2011surface, shi2013surface, shi2014genetic} that uses isothermal coordinates and fluid registration to generate the correspondence between hippocampal surfaces and the statistics computed on the surface.   This method introduces two cuts on a hippocampal surface to convert it into a genus zero surface with two open boundaries. The locations of the two cuts were at the front and back of the hippocampal surface. By using conformal parameterization, it converts a 3D surface registration problem into a 2D image registration problem. The flow induced in the parameter domain establishes high-order correspondences between 3D surfaces. After the registration, various surface statistics were computed on the registered surface, such as multivariate tensor-based morphometry statistics \citep{wang2011surface} that retain the full tensor information of the deformation Jacobian matrix, and the radial distance \citep{pizer1999segmentation}. This software package has been applied in various studies \citep{wang2011surface,wang2013applying, shi2013surface, shi2014genetic}. The radial distance feature is used in this paper. An example of one subject's left and right hippocampus images are shown in Figure \ref{fig:hippimg}, where the left side of (a) or (b) corresponds to the bottom of the hippocampus in (c), and the right side corresponds to the top. 

\begin{figure}
\begin{center}
\begin{tabular}{ccc}
\includegraphics[height=1.5in]{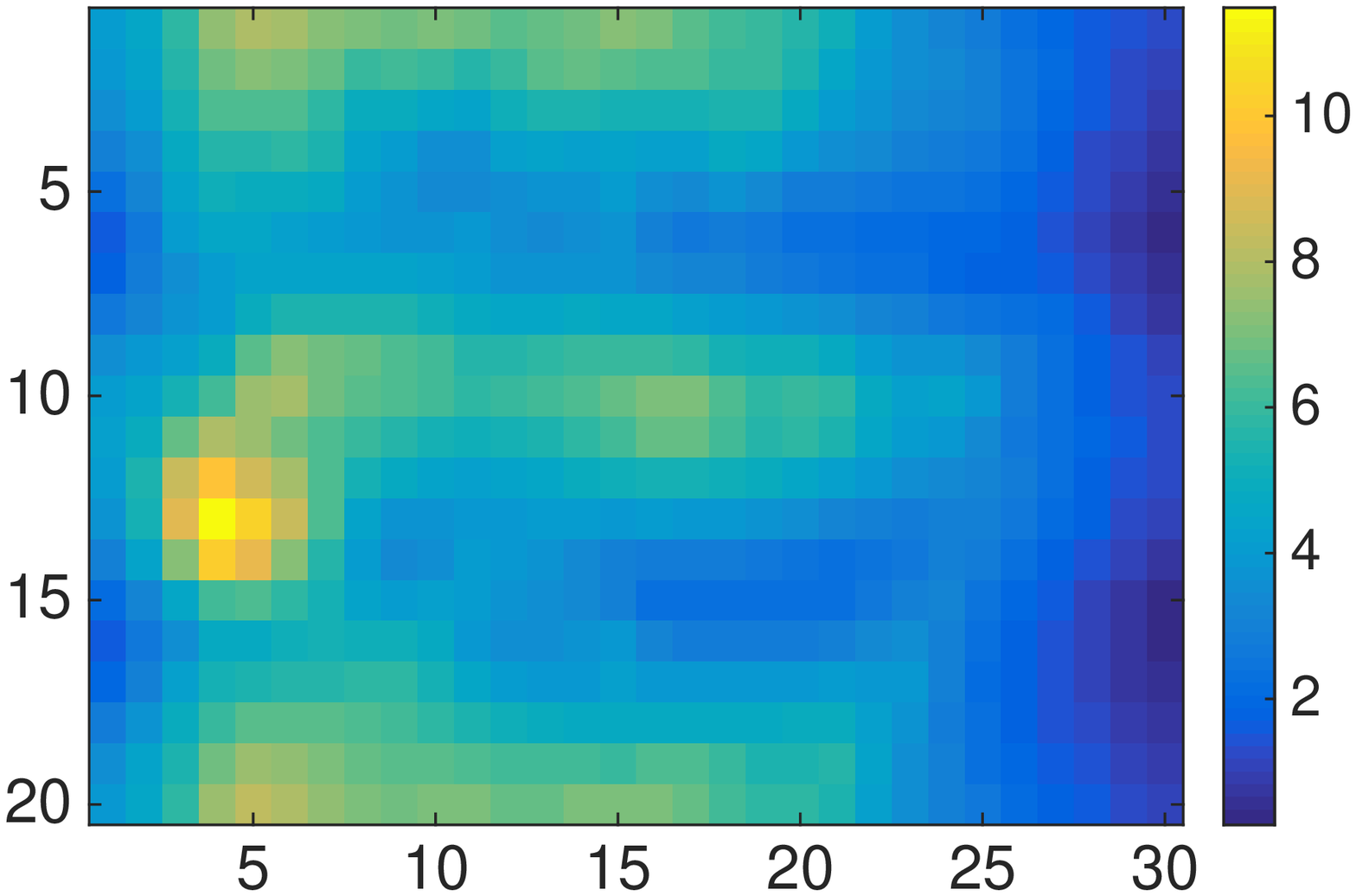} &
\includegraphics[height=1.5in]{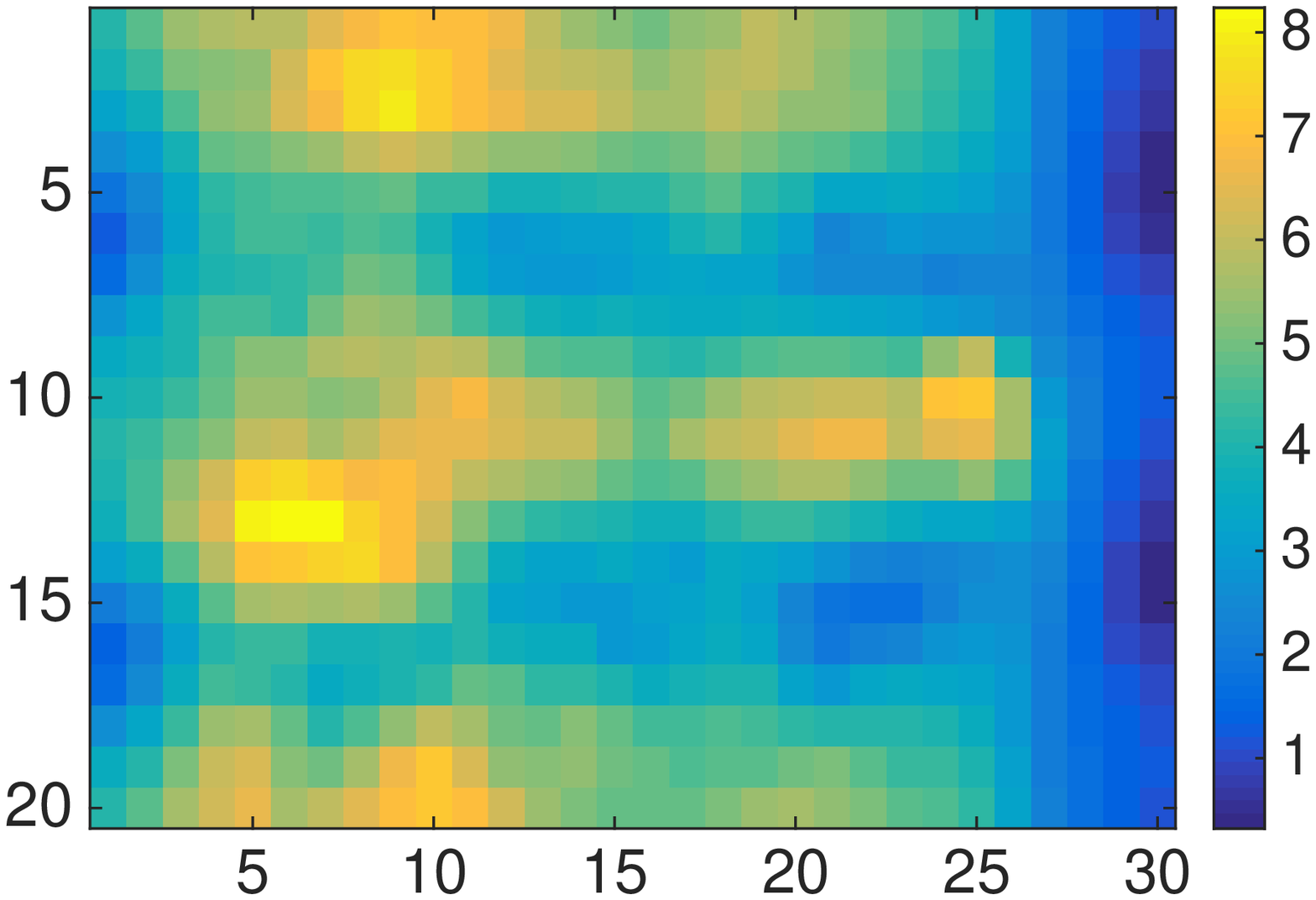}&
\includegraphics[height=1.5in]{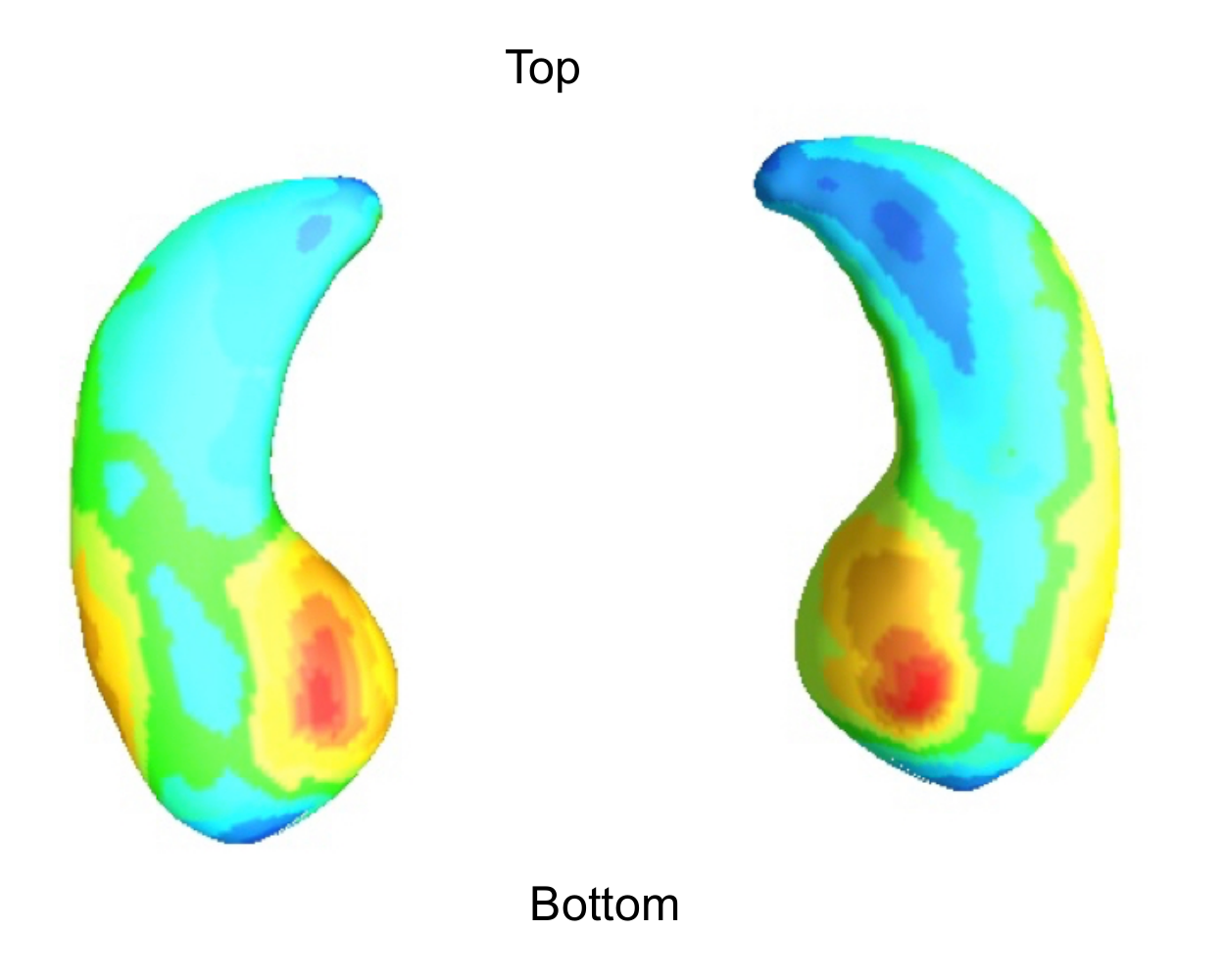}\\
(a) & (b) & (c) \\
\end{tabular}
\caption{Observed conformally mapped left (a) and right (b) hippocampal images, and the original statistics on the surface (c).} \label{fig:hippimg}
\end{center}
\end{figure}

We applied our SQR to the hippocampus data set. We have $403$ subjects, $223$ of which are healthy controls (107 females and 116 males), and 180 of which have AD ( 87 females and 93 males). Scalar covariates include the subject's gender, age, and behavior score ($1$ - $36$, where lower scores correspond to healthy controls and higher scores corresponds to ADs). The response variable is the 2D hippocampus image. Preliminary analysis indicates that gender does not have a significant effect on hippocampus and therefore, in our analysis, we only include two covariates -- age and behavior score, both of which are normalized. In the following, we present results for the right hippocampal surfaces.  Figure \ref{fig:coeffimage1} shows the coefficient images at $\tau = 0.5$, with more results for $\tau = 0.25$ and $0.75$ shown in Figure 2.1 in the Supplement II. Our results indicate that both aging and AD will degenerate the hippocampus, especially the bottom part, and the AD has more adverse effect than the aging \citep{scher2007hippocampal,frisoni2008mapping}.  }

In another experiment, we compared marginal distributions given different covariates. We first got the marginal distribution $Y(s)|{\bf x} = (1.47,2.48)$ (corresponding to age = 85, behavior = 30.3) and then calculated the p-value maps of some randomly selected observations $Y_i$. The results are shown in Figure \ref{fig:coeffimage2}. We can see that the hippocampus in younger adults with good behavior scores have significantly bigger hippocampus (especially the part close to the bottom) than subjects with age 85 and behavior 30.3. %This result is consist with the literature \citep{scher2007hippocampal,frisoni2008mapping}. %With the copula model, given $\bf x$, we obtain the joint distribution of $Y|{\bf x}$. In Supplementary Figure xxx, we show some sampled hippocampal surfaces with different $\bf x$'s.

\begin{figure}
\begin{center}
\begin{tabular}{ccc}
\includegraphics[height=1.2in]{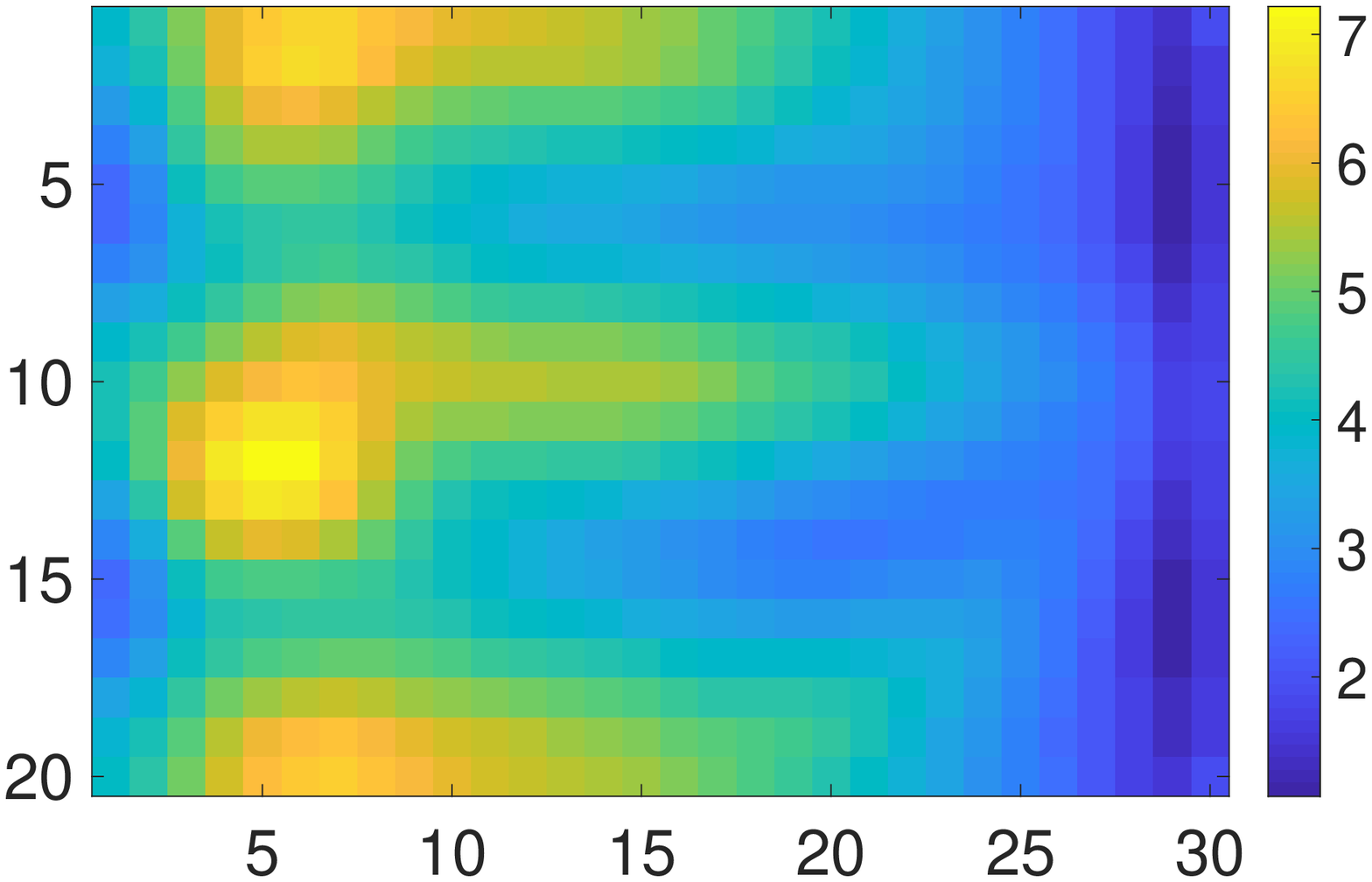} &
\includegraphics[height=1.2in]{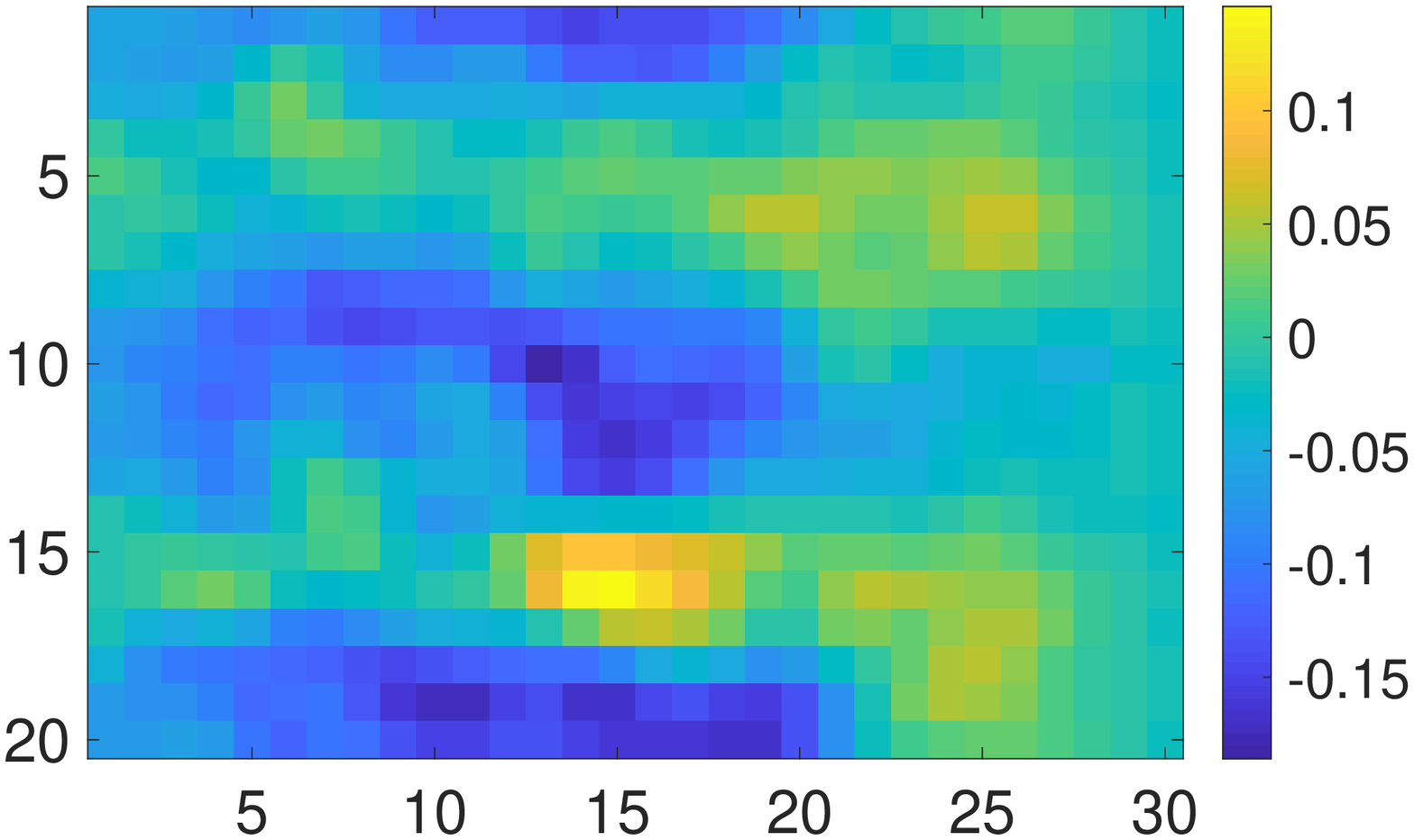}&
\includegraphics[height=1.2in]{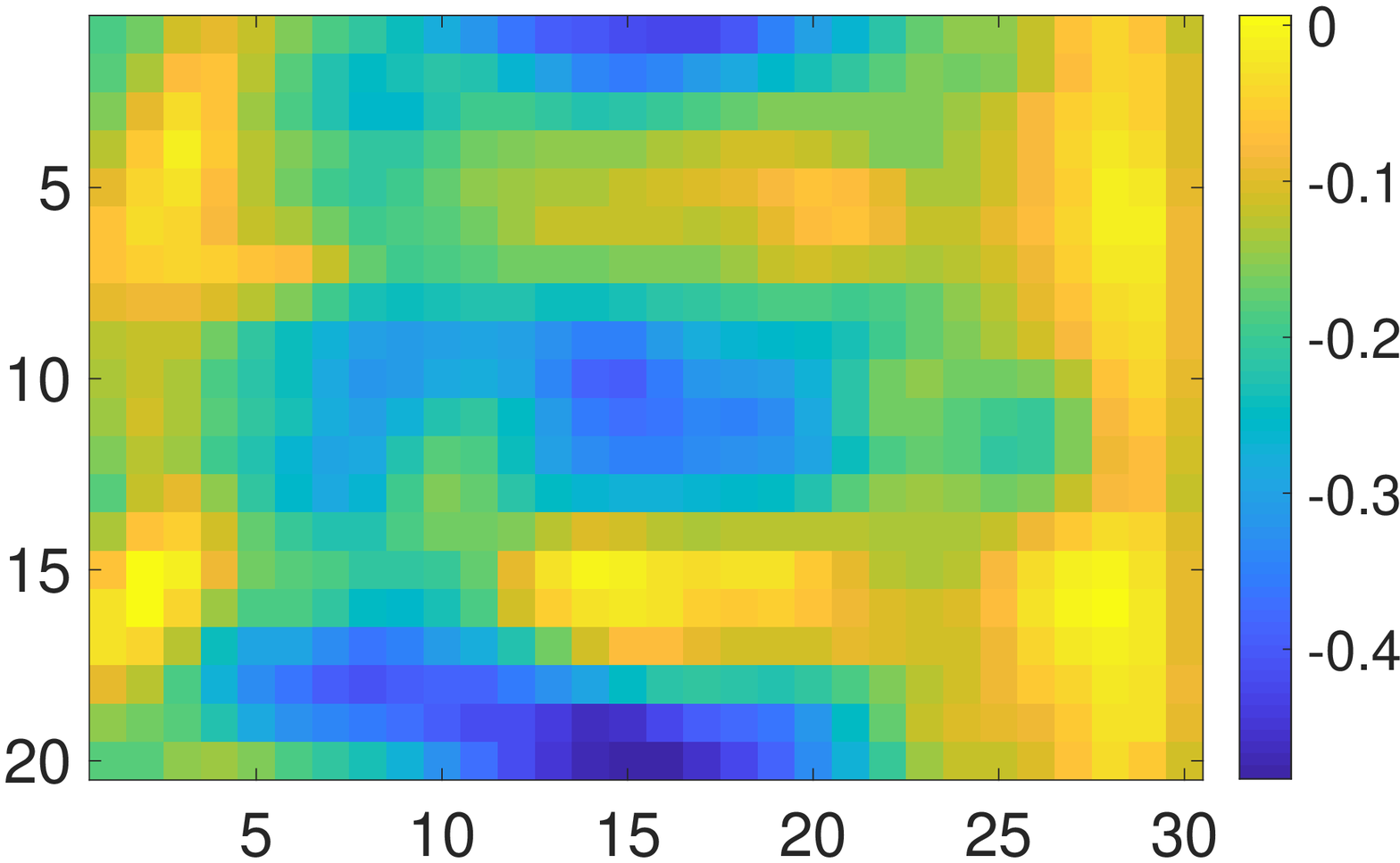}\\
(a) Intercept & (b) $\beta_{age}$ & (c) $\beta_{behavior}$
\end{tabular}
\caption{Coefficient images $\bm \beta$  at $\tau = 0.5$.} \label{fig:coeffimage1}
\end{center}
\end{figure}

\begin{figure}
\begin{center}
\begin{tabular}{ccc}
\includegraphics[height=1.2in]{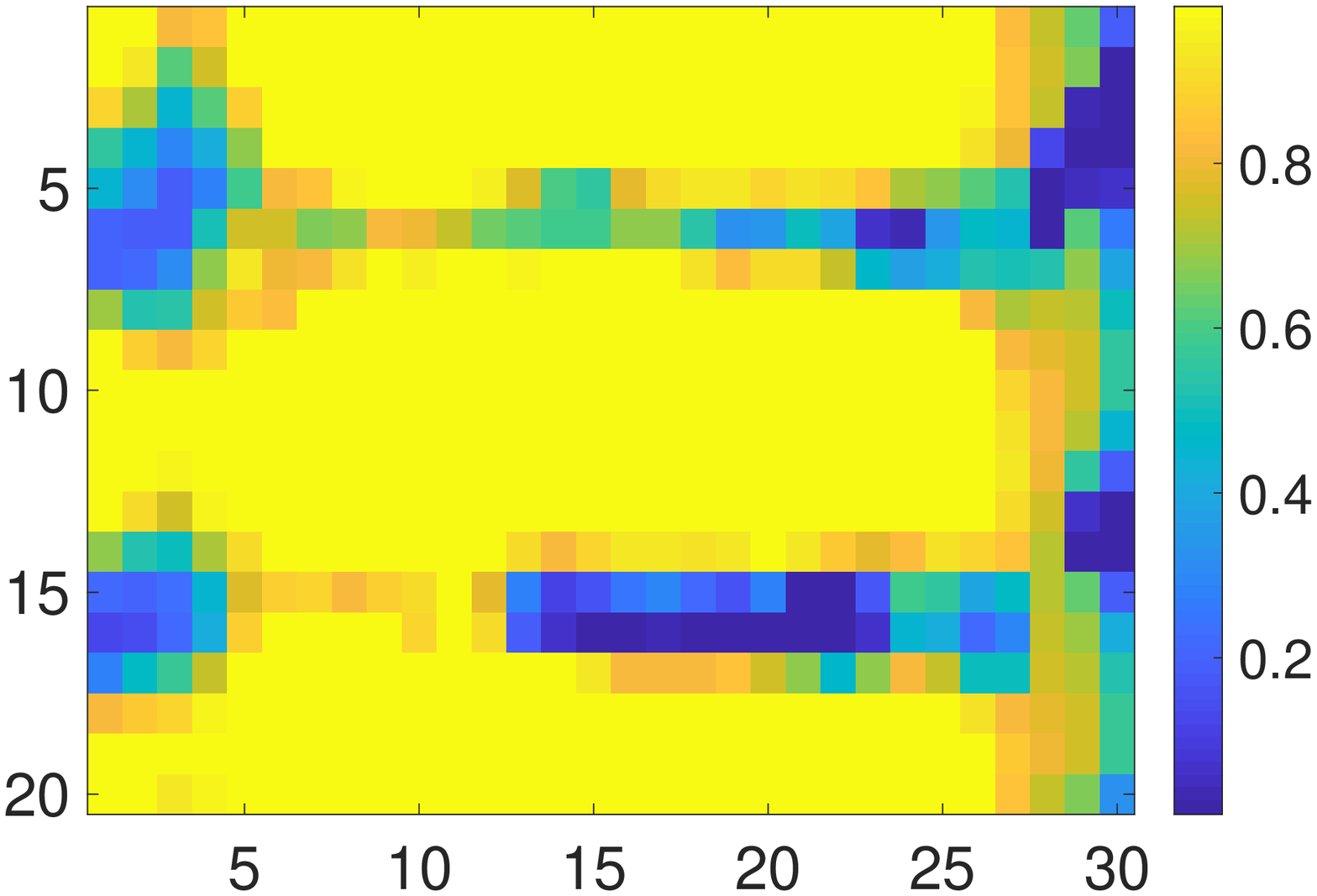} &
\includegraphics[height=1.2in]{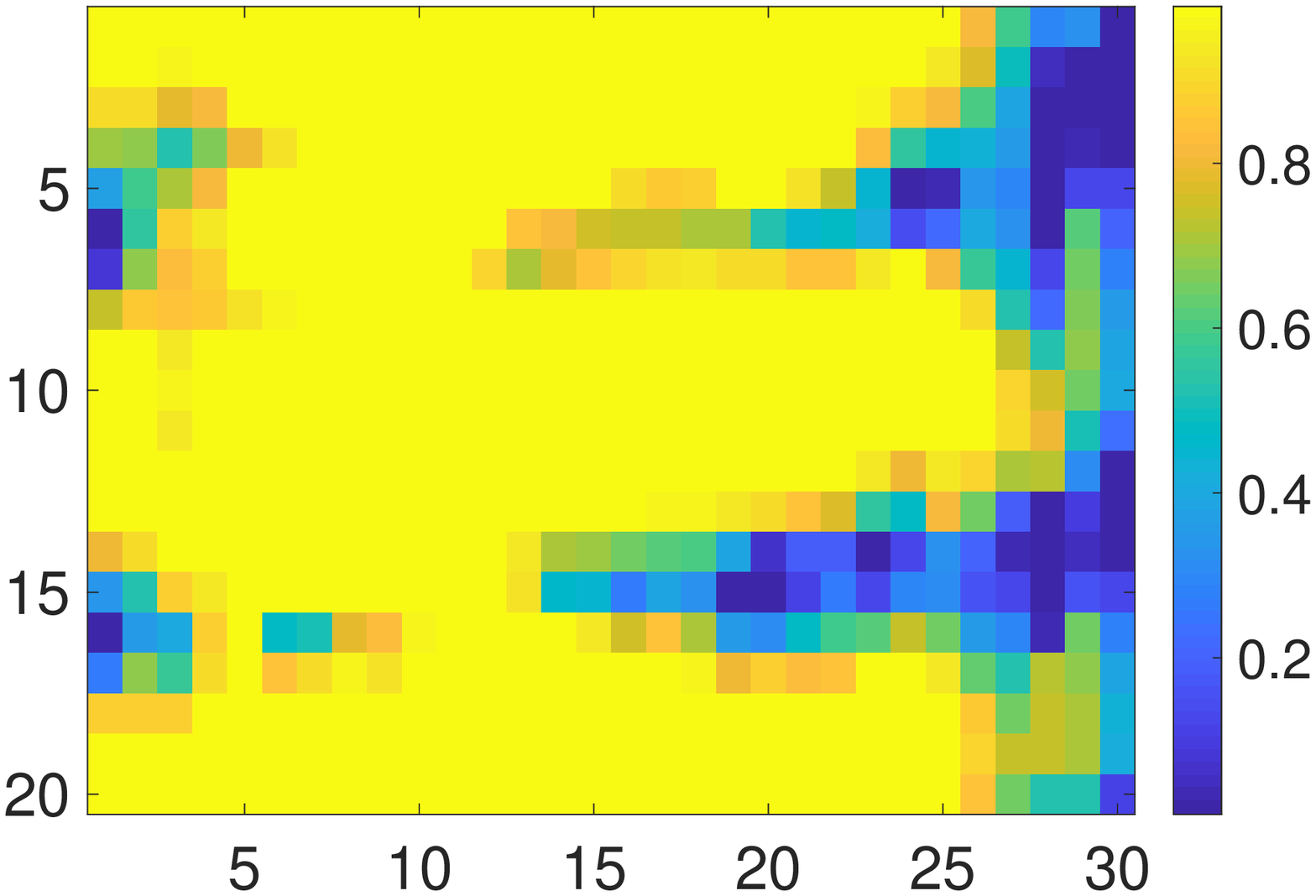}&
\includegraphics[height=1.2in]{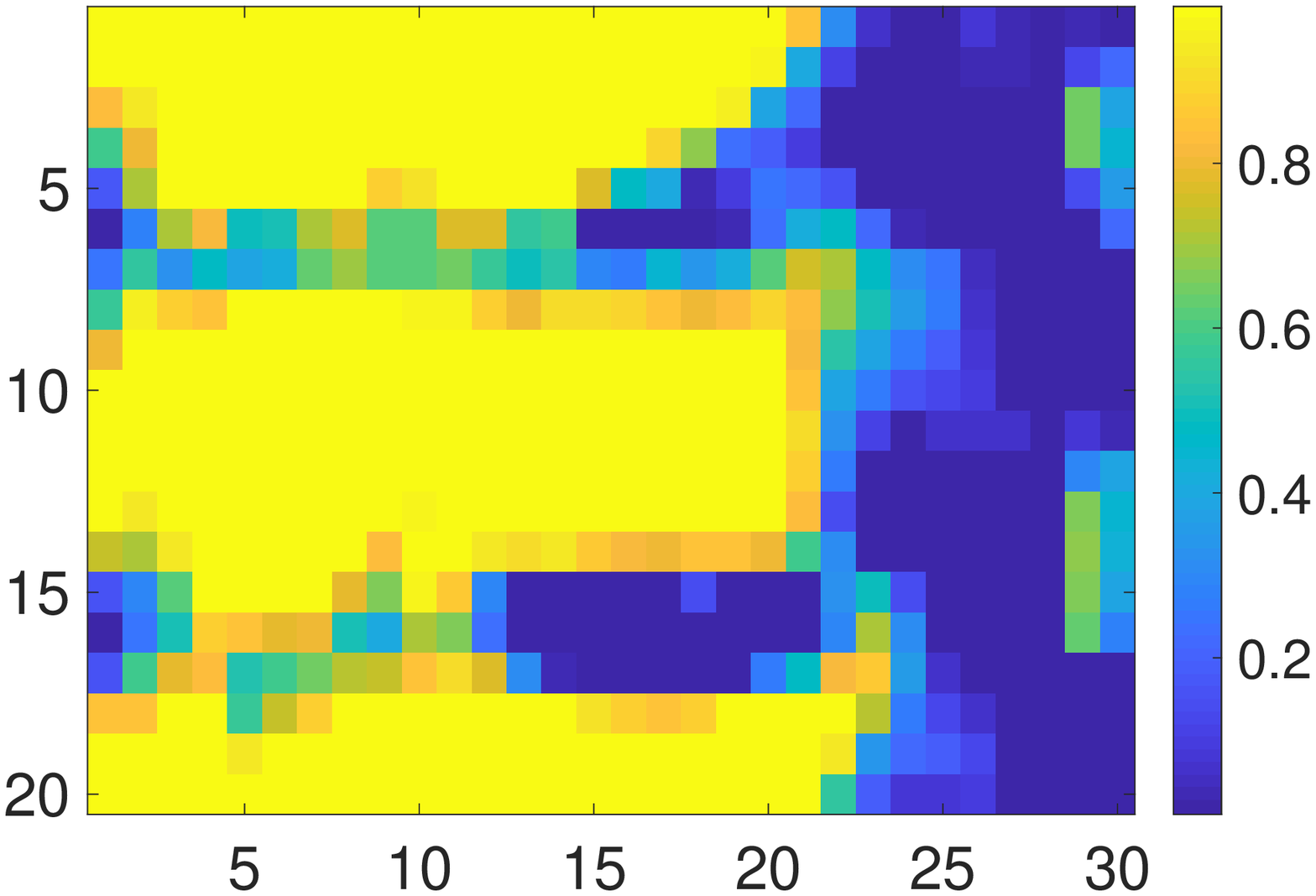}\\
(62,3.3) & (70,3) & (66,2.7) \\
\includegraphics[height=1.2in]{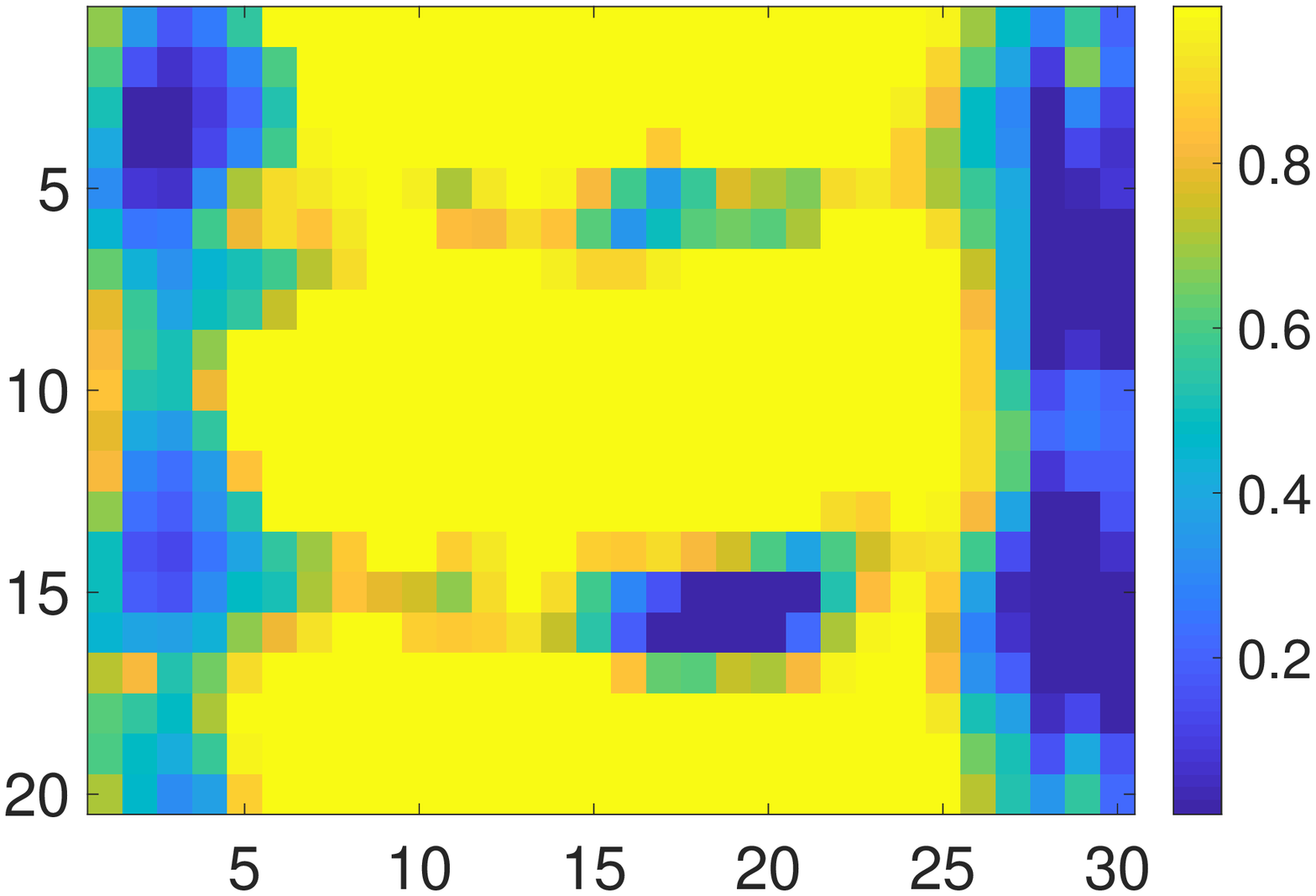} &
\includegraphics[height=1.2in]{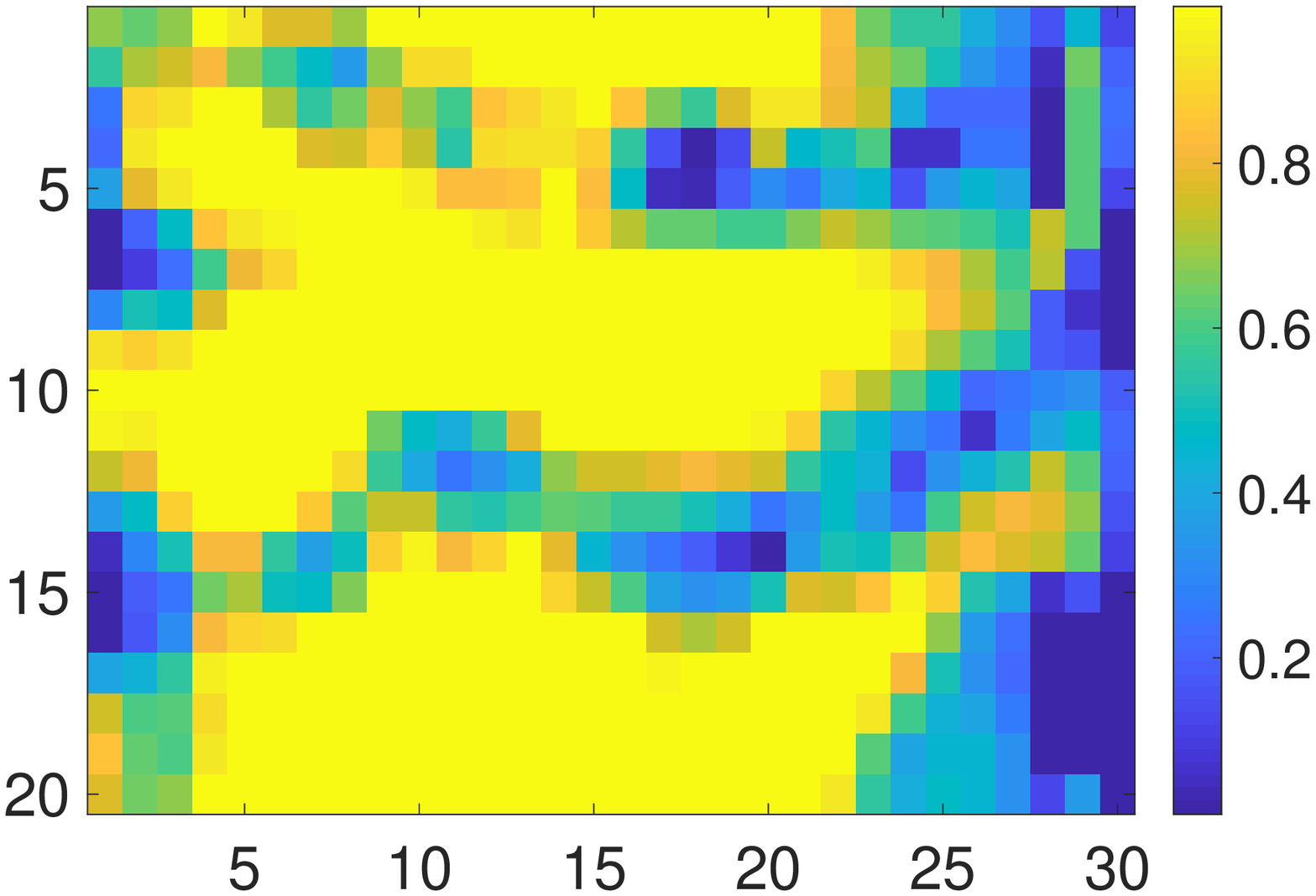}&
\includegraphics[height=1.2in]{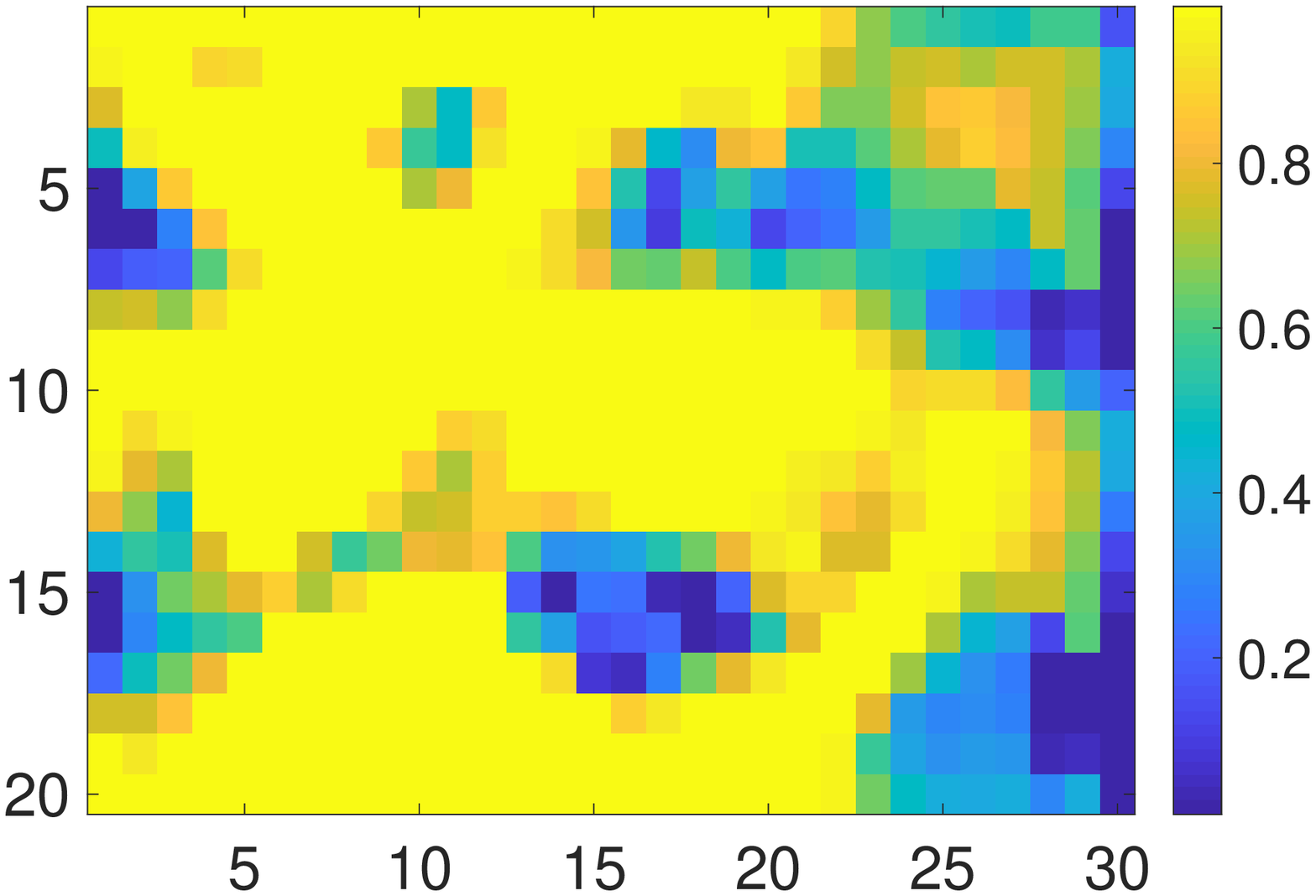}\\
(66,2.7) & (70,5.3) & (53,4) \\
\end{tabular}
\caption{P-value maps for some observed $Y_i$ under the marginal distribution of $Y(s)|{\bf x}$ with ${\bf x} = (1.47,2.48)$, corresponding to an age of 85 and a behavior score of 30.3. Numbers in the parentheses correspond to age and behavior score.} \label{fig:coeffimage2}
\end{center}
\end{figure}

\section{Discussions}

We have developed a class of SQR models for function-on-scalar regression, which explicitly characterizes the conditional spatial distribution of a high-dimensional functional/image response given scalar predictors. We have borrowed the strength from both quantile regression and copula modeling, and have developed an efficient primal-dual algorithm to estimate unknown parameters. Simulations and real data analysis are used to show that SQR is efficient to obtain a comprehensive understanding of the effect of scalar covariates on the functional response at different quantile levels. We also established the optimal rate of convergence on the estimation of the coefficient functions. 

{

Many important issues need to be addressed in future research. First, although we focus on the analysis of one-  and two-dimensional images, it is straightforward to  apply our methods to   $k$-dimensional images with $k>2$. Under this circumstance, the coefficient functions $\beta_\tau(\cdot)$ are $k$-dimensional unknown functions, and the spatial location sampling frequency $m$ will be much larger than that of two-dimensional images. In this case,  the computational efficiency is the key challenge. 
Second, instead of analyzing   image data in a raw space directly,  we  may achieve better   results by working on a transformed space, such as frequency. In this case, some further
development may be necessary.   
Third, it is interesting to consider several alternatives to the quantile used in 
this paper. For instance, we may consider 
  a  new multivariate concept of quantile  based on a directional
version of Koenker and Bassett’s traditional regression quantiles for multivariate location and multiple-output regression
problems \citep{kroneker2017}. 
 In their empirical version, those quantiles can be computed efficiently via linear programming techniques. The contours generated by those quantiles
are shown to coincide with the classical halfspace depth contours associated with the name of Tukey. This depth approach is an interesting alternative and we will leave it for future research.  

% Specifically, let ${\bf Y} = (Y_1, \ldots, Y_m)^T$ be the response and ${\bf x}$ be the covariate. For any $\tau\in (0, 1)$ and any direction ${\bf u}$ in the unit sphere, they can produce a hyperplane $\pi_{\tau{\bf u}}$ which is defined as the classical Koenker and Bassett regression quantile hyperplane of order $\tau$ once $(0,u)$ has been chosen as the
%``vertical direction" in the computation of the relevant $L_1$ deviations. 
%
%We can decompose ${\bf Y}$ into $({\bf u}^T {\bf Y}){\bf Y} + {\bf \Gamma}_{\bf u}({\bf\Gamma}_{\bf u}^T {\bf Y})$, where $({\bf u}, {\bf \Gamma}_{\bf u})$ is an orthogonal matrix.
%
%They showed that $\pi_{\tau{\bf u}}$ can be obtained  as the hyperplanes with the equation ${\bf b}_\tau^T {\bf Y} -{\bf a}_\tau^T {\bf x} = 0$, minimizing, with respect to ${\bf b}$ satisfying ${\bf b}^T{\bf u}=1$ and ${\bf a}$, the $L_1$ criteria $\mathbb E[\rho_\tau({\bf b}^T {\bf Y}-{\bf a}^T{\bf x})]$.

}

%Replacing $u$ by $uL\sqrt{p/ n}$ completes the proof of Part (a).

%\bibliographystyle{harvard}
\bibliographystyle{chicago}
\bibliography{bibfile}

\begin{thebibliography}{}

\bibitem[\protect\citeauthoryear{Ardekani, Bachman, Figarsky, and
  Sidtis}{Ardekani et~al.}{2014}]{ardekani2014corpus}
Ardekani, B.~A., A.~H. Bachman, K.~Figarsky, and J.~J. Sidtis (2014).
\newblock Corpus callosum shape changes in early {Alzheimer’s} disease: an
  {MRI} study using the {OASIS} brain database.
\newblock {\em Brain Structure and Function\/}~{\em 219\/}(1), 343--352.

\bibitem[\protect\citeauthoryear{B{\'a}rdossy}{B{\'a}rdossy}{2006}]{bardossy2006copula}
B{\'a}rdossy, A. (2006).
\newblock Copula-based geostatistical models for groundwater quality
  parameters.
\newblock {\em Water Resources Research\/}~{\em 42\/}(11).

\bibitem[\protect\citeauthoryear{Biegon, Eberling, Richardson, Roos, Wong,
  Reed, and Jagust}{Biegon et~al.}{1994}]{biegon1994human}
Biegon, A., J.~Eberling, B.~Richardson, M.~Roos, S.~Wong, B.~R. Reed, and
  W.~Jagust (1994).
\newblock Human corpus callosum in aging and {Alzheimer's} disease: a magnetic
  resonance imaging study.
\newblock {\em Neurobiology of aging\/}~{\em 15\/}(4), 393--397.

\bibitem[\protect\citeauthoryear{Bouy\'e and Salmon}{Bouy\'e and
  Salmon}{2013}]{bouye2013}
Bouy\'e, E. and M.~Salmon (2013).
\newblock Dynamic copula quantile regressions and tail area dynamic dependence
  in forex markets.
\newblock {\em Copulae and Multivariate Probability Distributions in
  Finance\/}, 125--154.

\bibitem[\protect\citeauthoryear{Bowman}{Bowman}{2010}]{bowman2010functional}
Bowman, A. (2010).
\newblock Functional data analysis with {R} and {MATLAB}.
\newblock {\em Journal of Statistical Software\/}~{\em 34\/}(1), 1--2.

\bibitem[\protect\citeauthoryear{Boyd, Parikh, Chu, Peleato, and Eckstein}{Boyd
  et~al.}{2011}]{boyd2011distributed}
Boyd, S., N.~Parikh, E.~Chu, B.~Peleato, and J.~Eckstein (2011).
\newblock Distributed optimization and statistical learning via the alternating
  direction method of multipliers.
\newblock {\em Foundations and Trends{\textregistered} in Machine
  learning\/}~{\em 3\/}(1), 1--122.

\bibitem[\protect\citeauthoryear{Cai and Hall}{Cai and Hall}{2006}]{tony2006}
Cai, T. and P.~Hall (2006).
\newblock Prediction in functional linear regression.
\newblock {\em The Annals of Statistics,\/}~{\em 34\/}(5), 2159--2179.

\bibitem[\protect\citeauthoryear{Cai and Yuan}{Cai and
  Yuan}{2012}]{cai2012minimax}
Cai, T.~T. and M.~Yuan (2012).
\newblock Minimax and adaptive prediction for functional linear regression.
\newblock {\em Journal of the American Statistical Association\/}~{\em
  107\/}(499), 1201--1216.

\bibitem[\protect\citeauthoryear{Cai and Xu}{Cai and
  Xu}{2008}]{cai2008nonparametric}
Cai, Z. and X.~Xu (2008).
\newblock Nonparametric quantile estimations for dynamic smooth coefficient
  models.
\newblock {\em Journal of the American Statistical Association\/}~{\em
  103\/}(484), 1595--1608.

\bibitem[\protect\citeauthoryear{Chen, Koenker, and Xiao}{Chen
  et~al.}{2009}]{chen2009}
Chen, X., R.~Koenker, and Z.~Xiao (2009).
\newblock Copula based nonlinear quantile autoregression.
\newblock {\em The Econometrics Journal\/}~{\em 12}, S50--S67.

\bibitem[\protect\citeauthoryear{Colom, Stein, Rajagopalan, Mart{\'\i}nez,
  Hermel, Wang, {\'A}lvarez-Linera, Burgaleta, Quiroga, Shih, and
  Thompson}{Colom et~al.}{2013}]{colom2013hippocampal}
Colom, R., J.~L. Stein, P.~Rajagopalan, K.~Mart{\'\i}nez, D.~Hermel, Y.~Wang,
  J.~{\'A}lvarez-Linera, M.~Burgaleta, M.~{\'A}. Quiroga, P.~C. Shih, and P.~M.
  Thompson (2013).
\newblock Hippocampal structure and human cognition: Key role of spatial
  processing and evidence supporting the efficiency hypothesis in females.
\newblock {\em Intelligence\/}~{\em 41\/}(2), 129--140.

\bibitem[\protect\citeauthoryear{Crambes, Kneip, and Sarda}{Crambes
  et~al.}{2009}]{crambes2009smoothing}
Crambes, C., A.~Kneip, and P.~Sarda (2009).
\newblock Smoothing splines estimators for functional linear regression.
\newblock {\em The Annals of Statistics\/}~{\em 37\/}(1), 35--72.

\bibitem[\protect\citeauthoryear{Cressie}{Cressie}{1985}]{cressie1985fitting}
Cressie, N. (1985).
\newblock Fitting variogram models by weighted least squares.
\newblock {\em Mathematical Geology\/}~{\em 17\/}(5), 563--586.

\bibitem[\protect\citeauthoryear{Cressie}{Cressie}{2015}]{Cressie2015}
Cressie, N.~A. (2015).
\newblock {\em Statistics for Spatial Data, Revised Edition}.
\newblock Wiley.

\bibitem[\protect\citeauthoryear{De~Angelis, Pardalos, and Toraldo}{De~Angelis
  et~al.}{1997}]{de1997quadratic}
De~Angelis, P.~L., P.~M. Pardalos, and G.~Toraldo (1997).
\newblock Quadratic programming with box constraints.
\newblock In {\em Developments in Global Optimization}, pp.\  73--93. Springer.

\bibitem[\protect\citeauthoryear{De~Backer, Ghouch, and Van~Keilegom}{De~Backer
  et~al.}{2017}]{backer2017}
De~Backer, M., A.~E. Ghouch, and I.~Van~Keilegom (2017).
\newblock Semiparametric copula quantile regression for complete or censored
  data.
\newblock {\em Electronic Journal of Statistics\/}~{\em 11}, 1660--1698.

\bibitem[\protect\citeauthoryear{Demarta and McNeil}{Demarta and
  McNeil}{2005}]{demarta2005t}
Demarta, S. and A.~J. McNeil (2005).
\newblock The t copula and related copulas.
\newblock {\em International statistical review\/}~{\em 73\/}(1), 111--129.

\bibitem[\protect\citeauthoryear{Du and Wang}{Du and
  Wang}{2014}]{du2014penalized}
Du, P. and X.~Wang (2014).
\newblock Penalized likelihood functional regression.
\newblock {\em Statistica Sinica\/}~{\em 24\/}(2), 1017--1041.

\bibitem[\protect\citeauthoryear{Efron}{Efron}{1986}]{efron1986biased}
Efron, B. (1986).
\newblock How biased is the apparent error rate of a prediction rule?
\newblock {\em Journal of the American Statistical Association\/}~{\em
  81\/}(394), 461--470.

\bibitem[\protect\citeauthoryear{Ferraty and Vieu}{Ferraty and
  Vieu}{2006}]{ferraty2006nonparametric}
Ferraty, F. and P.~Vieu (2006).
\newblock {\em Nonparametric Functional Data Analysis: Methods, Theory,
  Applications and Implementations}.
\newblock Springer.

\bibitem[\protect\citeauthoryear{Frisoni, Ganzola, Canu, R{\"u}b, Pizzini,
  Alessandrini, Zoccatelli, Beltramello, Caltagirone, and Thompson}{Frisoni
  et~al.}{2008}]{frisoni2008mapping}
Frisoni, G.~B., R.~Ganzola, E.~Canu, U.~R{\"u}b, F.~B. Pizzini,
  F.~Alessandrini, G.~Zoccatelli, A.~Beltramello, C.~Caltagirone, and P.~M.
  Thompson (2008).
\newblock Mapping local hippocampal changes in {A}lzheimer's disease and normal
  ageing with {MRI} at 3 {Tesla}.
\newblock {\em Brain\/}~{\em 131\/}(12), 3266--3276.

\bibitem[\protect\citeauthoryear{Genton}{Genton}{1998}]{genton1998variogram}
Genton, M.~G. (1998).
\newblock Variogram fitting by generalized least squares using an explicit
  formula for the covariance structure.
\newblock {\em Mathematical Geology\/}~{\em 30\/}(4), 323--345.

\bibitem[\protect\citeauthoryear{Goldsmith and Kitago}{Goldsmith and
  Kitago}{2016}]{goldsmith2016assessing}
Goldsmith, J. and T.~Kitago (2016).
\newblock Assessing systematic effects of stroke on motor control by using
  hierarchical function-on-scalar regression.
\newblock {\em Journal of the Royal Statistical Society: Series C (Applied
  Statistics)\/}~{\em 65\/}(2), 215--236.

\bibitem[\protect\citeauthoryear{Greven and Scheipl}{Greven and
  Scheipl}{2017}]{greven2017general}
Greven, S. and F.~Scheipl (2017).
\newblock A general framework for functional regression modelling.
\newblock {\em Statistical Modelling\/}~{\em 17\/}(1-2), 1--35.

\bibitem[\protect\citeauthoryear{Gutenbrunner and Jureckov{\'a}}{Gutenbrunner
  and Jureckov{\'a}}{1992}]{gutenbrunner1992regression}
Gutenbrunner, C. and J.~Jureckov{\'a} (1992).
\newblock Regression rank scores and regression quantiles.
\newblock {\em The Annals of Statistics\/}, 305--330.

\bibitem[\protect\citeauthoryear{Gutenbrunner, Jure{\v{c}}kov{\'a}, Koenker,
  and Portnoy}{Gutenbrunner et~al.}{1993}]{gutenbrunner1993tests}
Gutenbrunner, C., J.~Jure{\v{c}}kov{\'a}, R.~Koenker, and S.~Portnoy (1993).
\newblock Tests of linear hypotheses based on regression rank scores.
\newblock {\em Journaltitle of Nonparametric Statistics\/}~{\em 2\/}(4),
  307--331.

\bibitem[\protect\citeauthoryear{Guttorp and Gneiting}{Guttorp and
  Gneiting}{2006}]{guttorp2006studies}
Guttorp, P. and T.~Gneiting (2006).
\newblock Studies in the history of probability and statistics {XLIX} on the
  {M}at{\'e}rn correlation family.
\newblock {\em Biometrika\/}~{\em 93\/}(4), 989--995.

\bibitem[\protect\citeauthoryear{Hall and Horowitz}{Hall and
  Horowitz}{2007}]{hall2017}
Hall, P. and J.~L. Horowitz (2007).
\newblock Methodology and convergence rates for functional linear regression.
\newblock {\em The Annals of Statistics\/}~{\em 35\/}(1), 70--91.

\bibitem[\protect\citeauthoryear{Hallin, Lu, and Yu}{Hallin
  et~al.}{2009}]{hallin2009}
Hallin, M., Z.~L. Lu, and K.~Yu (2009).
\newblock Local linear spatial quantile regression.
\newblock {\em Bernoulli\/}, 659--686.

\bibitem[\protect\citeauthoryear{Inano, Takao, Hayashi, Abe, and Ohtomo}{Inano
  et~al.}{2011}]{inano2011effects}
Inano, S., H.~Takao, N.~Hayashi, O.~Abe, and K.~Ohtomo (2011).
\newblock Effects of age and gender on white matter integrity.
\newblock {\em American Journal of Neuroradiology\/}~{\em 32\/}(11),
  2103--2109.

\bibitem[\protect\citeauthoryear{Ivanescu, Staicu, Scheipl, and
  Greven}{Ivanescu et~al.}{2015}]{ivanescu2015penalized}
Ivanescu, A.~E., A.-M. Staicu, F.~Scheipl, and S.~Greven (2015).
\newblock Penalized function-on-function regression.
\newblock {\em Computational Statistics\/}~{\em 30\/}(2), 539--568.

\bibitem[\protect\citeauthoryear{Kato et~al.}{Kato
  et~al.}{2012}]{kato2012estimation}
Kato, K. et~al. (2012).
\newblock Estimation in functional linear quantile regression.
\newblock {\em The Annals of Statistics\/}~{\em 40\/}(6), 3108--3136.

\bibitem[\protect\citeauthoryear{Kazianka and Pilz}{Kazianka and
  Pilz}{2010}]{Kazianka2010Copula}
Kazianka, H. and J.~Pilz (2010).
\newblock Copula-based geostatistical modeling of continuous and discrete data
  including covariates.
\newblock {\em Stochastic Environmental Research and Risk Assessment\/}~{\em
  24\/}(5), 661--673.

\bibitem[\protect\citeauthoryear{Koay, Chang, Carew, Pierpaoli, and
  Basser}{Koay et~al.}{2006}]{koay2006unifying}
Koay, C.~G., L.-C. Chang, J.~D. Carew, C.~Pierpaoli, and P.~J. Basser (2006).
\newblock A unifying theoretical and algorithmic framework for least squares
  methods of estimation in diffusion tensor imaging.
\newblock {\em Journal of Magnetic Resonance\/}~{\em 182\/}(1), 115--125.

\bibitem[\protect\citeauthoryear{Kochunov, Thompson, Lancaster, Bartzokis,
  Smith, Coyle, Royall, Laird, and Fox}{Kochunov
  et~al.}{2007}]{kochunov2007relationship}
Kochunov, P., P.~M. Thompson, J.~L. Lancaster, G.~Bartzokis, S.~Smith,
  T.~Coyle, D.~R. Royall, A.~Laird, and P.~T. Fox (2007).
\newblock Relationship between white matter fractional anisotropy and other
  indices of cerebral health in normal aging: tract-based spatial statistics
  study of aging.
\newblock {\em Neuroimage\/}~{\em 35\/}(2), 478--487.

\bibitem[\protect\citeauthoryear{Koenker}{Koenker}{2004}]{koenker2004quantile}
Koenker, R. (2004).
\newblock Quantile regression for longitudinal data.
\newblock {\em Journal of Multivariate Analysis\/}~{\em 91\/}(1), 74--89.

\bibitem[\protect\citeauthoryear{Koenker}{Koenker}{2005}]{Koenker2015book}
Koenker, R. (2005).
\newblock {\em Quantile Regression}.
\newblock Cambridge University Press.

\bibitem[\protect\citeauthoryear{Koenker and Bassett~Jr}{Koenker and
  Bassett~Jr}{1978}]{koenker1978regression}
Koenker, R. and G.~Bassett~Jr (1978).
\newblock Regression quantiles.
\newblock {\em Econometrica\/}~{\em 46\/}(1), 33--50.

\bibitem[\protect\citeauthoryear{Koenker, Chernozhukov, He, and Peng}{Koenker
  et~al.}{2017}]{kroneker2017}
Koenker, R., V.~Chernozhukov, X.~He, and L.~Peng (2017).
\newblock {\em Handbook of Quantile Regression}.
\newblock Chapman and Hall/CRC.

\bibitem[\protect\citeauthoryear{Koenker, Ng, and Portnoy}{Koenker
  et~al.}{1994}]{koenker1994quantile}
Koenker, R., P.~Ng, and S.~Portnoy (1994).
\newblock Quantile smoothing splines.
\newblock {\em Biometrika\/}~{\em 81\/}(4), 673--680.

\bibitem[\protect\citeauthoryear{Koenker and Park}{Koenker and
  Park}{1996}]{koenker1996}
Koenker, R. and B.~J. Park (1996).
\newblock An interior point algorithm for nonlinear quantile regression.
\newblock {\em Journal of Econometrics\/}~{\em 71}, 265--283.

\bibitem[\protect\citeauthoryear{Kostov}{Kostov}{2009}]{kostov2009}
Kostov, P. (2009).
\newblock A spatial quantile regression hedonic model of agricultural land
  prices.
\newblock {\em Spatial Economic Analysis\/}~{\em 4\/}(1), 53--72.

\bibitem[\protect\citeauthoryear{Kraus and Czado}{Kraus and
  Czado}{2017}]{kraus2017}
Kraus, D. and C.~Czado (2017).
\newblock D-vine copula based quantile regression.
\newblock {\em Computational Statistics \& Data Analysis\/}~{\em 110}, 1--18.

\bibitem[\protect\citeauthoryear{Lahiri, Lee, and Cressie}{Lahiri
  et~al.}{2002}]{lahiri2002asymptotic}
Lahiri, S.~N., Y.~Lee, and N.~Cressie (2002).
\newblock On asymptotic distribution and asymptotic efficiency of least squares
  estimators of spatial variogram parameters.
\newblock {\em Journal of Statistical Planning and Inference\/}~{\em
  103\/}(1-2), 65--85.

\bibitem[\protect\citeauthoryear{Li, Liu, and Zhu}{Li
  et~al.}{2007}]{li2007quantile}
Li, Y., Y.~Liu, and J.~Zhu (2007).
\newblock Quantile regression in reproducing kernel hilbert spaces.
\newblock {\em Journal of the American Statistical Association\/}~{\em
  102\/}(477), 255--268.

\bibitem[\protect\citeauthoryear{Lorensen and Cline}{Lorensen and
  Cline}{1987}]{lorensen1987marching}
Lorensen, W.~E. and H.~E. Cline (1987).
\newblock Marching cubes: A high resolution {3D} surface construction
  algorithm.
\newblock In {\em ACM Siggraph Computer Graphics}, Volume~21, pp.\  163--169.

\bibitem[\protect\citeauthoryear{Lu, Tang, and Cheng}{Lu et~al.}{2014}]{lu2014}
Lu, Z., Q.~Tang, and L.~Cheng (2014, 02).
\newblock Estimating spatial quantile regression with functional coefficients:
  A robust semiparametric framework.
\newblock {\em Bernoulli\/}~{\em 20\/}(1), 164--189.

\bibitem[\protect\citeauthoryear{Luders, Thompson, Kurth, Hong, Phillips, Wang,
  Gutman, Chou, Narr, and Toga}{Luders et~al.}{2013}]{luders2013global}
Luders, E., P.~M. Thompson, F.~Kurth, J.-Y. Hong, O.~R. Phillips, Y.~Wang,
  B.~A. Gutman, Y.-Y. Chou, K.~L. Narr, and A.~W. Toga (2013).
\newblock Global and regional alterations of hippocampal anatomy in long-term
  meditation practitioners.
\newblock {\em Human Brain Mapping\/}~{\em 34\/}(12), 3369--3375.

\bibitem[\protect\citeauthoryear{Mat{\'e}rn}{Mat{\'e}rn}{2013}]{matern2013spatial}
Mat{\'e}rn, B. (2013).
\newblock {\em Spatial Variation}, Volume~36.
\newblock Springer-Verlag New York.

\bibitem[\protect\citeauthoryear{Matheron}{Matheron}{1963}]{matheron1963principles}
Matheron, G. (1963).
\newblock Principles of geostatistics.
\newblock {\em Economic Geology\/}~{\em 58\/}(8), 1246--1266.

\bibitem[\protect\citeauthoryear{Meyer and Woodroofe}{Meyer and
  Woodroofe}{2000}]{meyer2000degrees}
Meyer, M. and M.~Woodroofe (2000).
\newblock On the degrees of freedom in shape-restricted regression.
\newblock {\em Annals of Statistics\/}, 1083--1104.

\bibitem[\protect\citeauthoryear{Nikoloulopoulos, Joe, and Li}{Nikoloulopoulos
  et~al.}{2012}]{nikoloulopoulos2012vine}
Nikoloulopoulos, A.~K., H.~Joe, and H.~Li (2012).
\newblock Vine copulas with asymmetric tail dependence and applications to
  financial return data.
\newblock {\em Computational Statistics \& Data Analysis\/}~{\em 56\/}(11),
  3659--3673.

\bibitem[\protect\citeauthoryear{Patenaude, Smith, Kennedy, and
  Jenkinson}{Patenaude et~al.}{2011}]{patenaude2011bayesian}
Patenaude, B., S.~M. Smith, D.~N. Kennedy, and M.~Jenkinson (2011).
\newblock A {B}ayesian model of shape and appearance for subcortical brain
  segmentation.
\newblock {\em Neuroimage\/}~{\em 56\/}(3), 907--922.

\bibitem[\protect\citeauthoryear{Pizer, Fritsch, Yushkevich, Johnson, and
  Chaney}{Pizer et~al.}{1999}]{pizer1999segmentation}
Pizer, S.~M., D.~S. Fritsch, P.~A. Yushkevich, V.~E. Johnson, and E.~L. Chaney
  (1999).
\newblock Segmentation, registration, and measurement of shape variation via
  image object shape.
\newblock {\em IEEE Transactions on Medical Imaging\/}~{\em 18\/}(10),
  851--865.

\bibitem[\protect\citeauthoryear{Portnoy}{Portnoy}{1997}]{portnoy1997b}
Portnoy, S. (1997).
\newblock On computation of regression quantiles: Making the {Laplacian}
  tortoise faster.
\newblock {\em Lecture Notes-Monograph Series\/}, 187--200.

\bibitem[\protect\citeauthoryear{Portnoy and Koenker}{Portnoy and
  Koenker}{1997}]{portnoy1997a}
Portnoy, S. and R.~Koenker (1997).
\newblock The {Gaussian} hare and the {Laplacian} tortoise: computability of
  squared-error versus absolute-error estimators.
\newblock {\em Statistical Science\/}~{\em 12}, 279--300.

\bibitem[\protect\citeauthoryear{Ramsay and Silverman}{Ramsay and
  Silverman}{2005}]{ramsay2005functional}
Ramsay, J.~O. and B.~W. Silverman (2005).
\newblock {\em Functional Data Analysis}.
\newblock Springer-Verlag New York.

\bibitem[\protect\citeauthoryear{Ramsay and Silverman}{Ramsay and
  Silverman}{2007}]{ramsay2007applied}
Ramsay, J.~O. and B.~W. Silverman (2007).
\newblock {\em Applied Functional Data Analysis: Methods and Case Studies}.
\newblock Springer-Verlag New York.

\bibitem[\protect\citeauthoryear{Reich}{Reich}{2012}]{reich2012spatiotemporal}
Reich, B.~J. (2012).
\newblock Spatiotemporal quantile regression for detecting distributional
  changes in environmental processes.
\newblock {\em Journal of the Royal Statistical Society: Series C (Applied
  Statistics)\/}~{\em 61\/}(4), 535--553.

\bibitem[\protect\citeauthoryear{Reich, Fuentes, and Dunson}{Reich
  et~al.}{2011}]{reich2011}
Reich, B.~J., M.~Fuentes, and D.~B. Dunson (2011).
\newblock Bayesian spatial quantile regression.
\newblock {\em Journal of the American Statistical Association\/}, 6--20.

\bibitem[\protect\citeauthoryear{Reiss, Huang, and Mennes}{Reiss
  et~al.}{2010}]{reiss2010fast}
Reiss, P.~T., L.~Huang, and M.~Mennes (2010).
\newblock Fast function-on-scalar regression with penalized basis expansions.
\newblock {\em The International Journal of Biostatistics\/}~{\em 6\/}(1).

\bibitem[\protect\citeauthoryear{Rosen, Mohs, and Davis}{Rosen
  et~al.}{1984}]{rosen1984new}
Rosen, W.~G., R.~C. Mohs, and K.~L. Davis (1984).
\newblock A new rating scale for {A}lzheimer's disease.
\newblock {\em The American Journal of Psychiatry\/}~{\em 141\/}(11),
  1356--1364.

\bibitem[\protect\citeauthoryear{Scher, Xu, Korf, White, Scheltens, Toga,
  Thompson, Hartley, Witter, Valentino, and Launer}{Scher
  et~al.}{2007}]{scher2007hippocampal}
Scher, A., Y.~Xu, E.~Korf, L.~White, P.~Scheltens, A.~Toga, P.~Thompson,
  S.~Hartley, M.~Witter, D.~Valentino, and L.~Launer (2007).
\newblock Hippocampal shape analysis in {A}lzheimer’s disease: a
  population-based study.
\newblock {\em Neuroimage\/}~{\em 36\/}(1), 8--18.

\bibitem[\protect\citeauthoryear{Schwarz}{Schwarz}{1978}]{schwarz1978estimating}
Schwarz, G. (1978).
\newblock Estimating the dimension of a model.
\newblock {\em The Annals of Statistics\/}~{\em 6\/}(2), 461--464.

\bibitem[\protect\citeauthoryear{Shi, Lepore, Gutman, Thompson, Baxter,
  Caselli, and Wang}{Shi et~al.}{2014}]{shi2014genetic}
Shi, J., N.~Lepore, B.~A. Gutman, P.~M. Thompson, L.~C. Baxter, R.~J. Caselli,
  and Y.~Wang (2014).
\newblock Genetic influence of apolipoprotein {E4} genotype on hippocampal
  morphometry: An {N=725} surface-based {A}lzheimer's disease neuroimaging
  initiative study.
\newblock {\em Human Brain Mapping\/}~{\em 35\/}(8), 3903--3918.

\bibitem[\protect\citeauthoryear{Shi, Thompson, Gutman, and Wang}{Shi
  et~al.}{2013}]{shi2013surface}
Shi, J., P.~M. Thompson, B.~Gutman, and Y.~Wang (2013).
\newblock Surface fluid registration of conformal representation: Application
  to detect disease burden and genetic influence on hippocampus.
\newblock {\em NeuroImage\/}~{\em 78}, 111--134.

\bibitem[\protect\citeauthoryear{Smith, Jenkinson, Johansen-Berg, Rueckert,
  Nichols, Mackay, Watkins, Ciccarelli, Cader, Matthews, and Behrens}{Smith
  et~al.}{2006}]{smith2006tract}
Smith, S.~M., M.~Jenkinson, H.~Johansen-Berg, D.~Rueckert, T.~E. Nichols, C.~E.
  Mackay, K.~E. Watkins, O.~Ciccarelli, M.~Z. Cader, P.~M. Matthews, and T.~E.
  Behrens (2006).
\newblock Tract-based spatial statistics: Voxelwise analysis of multi-subject
  diffusion data.
\newblock {\em NeuroImage\/}~{\em 31\/}(4), 1487 -- 1505.

\bibitem[\protect\citeauthoryear{Stein}{Stein}{1981}]{stein1981estimation}
Stein, C.~M. (1981).
\newblock Estimation of the mean of a multivariate normal distribution.
\newblock {\em The Annals of Statistics\/}~{\em 9\/}(6), 1135--1151.

\bibitem[\protect\citeauthoryear{Su and Yang}{Su and Yang}{2007}]{su2007}
Su, L. and Z.~Yang (2007).
\newblock Instrumental variable quantile estimation of spatial autoregressive
  models.
\newblock {\em Development Economics Working Papers 22476, East Asian Bureau of
  Economic Research\/}.

\bibitem[\protect\citeauthoryear{Sun, Du, Wang, and Ma}{Sun
  et~al.}{2018}]{sun2017optimal}
Sun, X., P.~Du, X.~Wang, and P.~Ma (2018).
\newblock Optimal penalized function-on-function regression under a reproducing
  kernel {Hilbert} space framework.
\newblock {\em Journal of the American Statistical Association\/}~(DOI:
  10.1080/01621459.2017.1356320).

\bibitem[\protect\citeauthoryear{Wahba}{Wahba}{1990}]{wahba1990spline}
Wahba, G. (1990).
\newblock {\em Spline Models for Observational Data}, Volume~59.
\newblock SIAM.

\bibitem[\protect\citeauthoryear{Wang, Feng, and Dong}{Wang
  et~al.}{2019}]{wang2019}
Wang, H.~J., X.~Feng, and C.~Dong (2019).
\newblock Copula-based quantile regression for longitudinal data.
\newblock {\em Statistica Sinica\/}~{\em 29}, 245--264.

\bibitem[\protect\citeauthoryear{Wang, Zhu, and Zhou}{Wang
  et~al.}{2009}]{wang2009quantile}
Wang, H.~J., Z.~Zhu, and J.~Zhou (2009).
\newblock Quantile regression in partially linear varying coefficient models.
\newblock {\em The Annals of Statistics\/}~{\em 37\/}(6B), 3841--3866.

\bibitem[\protect\citeauthoryear{Wang, Zhu, and Initiative}{Wang
  et~al.}{2017}]{wang2017generalized}
Wang, X., H.~Zhu, and A.~D.~N. Initiative (2017).
\newblock Generalized scalar-on-image regression models via total variation.
\newblock {\em Journal of the American Statistical Association\/}~{\em
  112\/}(519), 1156--1168.

\bibitem[\protect\citeauthoryear{Wang, Song, Rajagopalan, An, Liu, Chou,
  Gutman, Toga, and Thompson}{Wang et~al.}{2011}]{wang2011surface}
Wang, Y., Y.~Song, P.~Rajagopalan, T.~An, K.~Liu, Y.-Y. Chou, B.~Gutman, A.~W.
  Toga, and P.~M. Thompson (2011).
\newblock Surface-based {TBM} boosts power to detect disease effects on the
  brain: an {N=804} {ADNI} study.
\newblock {\em Neuroimage\/}~{\em 56\/}(4), 1993--2010.

\bibitem[\protect\citeauthoryear{Wang, Yuan, Shi, Greve, Ye, Toga, Reiss, and
  Thompson}{Wang et~al.}{2013}]{wang2013applying}
Wang, Y., L.~Yuan, J.~Shi, A.~Greve, J.~Ye, A.~W. Toga, A.~L. Reiss, and P.~M.
  Thompson (2013).
\newblock Applying tensor-based morphometry to parametric surfaces can improve
  {MRI}-based disease diagnosis.
\newblock {\em Neuroimage\/}~{\em 74}, 209--230.

\bibitem[\protect\citeauthoryear{Wyss-Coray}{Wyss-Coray}{2016}]{wyss2016ageing}
Wyss-Coray, T. (2016).
\newblock Ageing, neurodegeneration and brain rejuvenation.
\newblock {\em Nature\/}~{\em 539\/}(7628), 180--186.

\bibitem[\protect\citeauthoryear{Yang and He}{Yang and He}{2015}]{yang2015}
Yang, Y. and X.~He (2015).
\newblock Quantile regression for spatially correlated data: An empirical
  likelihood approach.
\newblock {\em Statistica Sinica\/}, 261--274.

\bibitem[\protect\citeauthoryear{Yao, M{\"u}ller, and Wang}{Yao
  et~al.}{2005}]{yao2005functional}
Yao, F., H.-G. M{\"u}ller, and J.-L. Wang (2005).
\newblock Functional linear regression analysis for longitudinal data.
\newblock {\em The Annals of Statistics\/}, 2873--2903.

\bibitem[\protect\citeauthoryear{Yi and He}{Yi and He}{2009}]{yi2009median}
Yi, G.~Y. and W.~He (2009).
\newblock Median regression models for longitudinal data with dropouts.
\newblock {\em Biometrics\/}~{\em 65\/}(2), 618--625.

\bibitem[\protect\citeauthoryear{Yuan}{Yuan}{2006}]{yuan2006gacv}
Yuan, M. (2006).
\newblock {GACV} for quantile smoothing splines.
\newblock {\em Computational Statistics \& Data Analysis\/}~{\em 50\/}(3),
  813--829.

\bibitem[\protect\citeauthoryear{Yuan, Cai, et~al.}{Yuan
  et~al.}{2010}]{yuan2010reproducing}
Yuan, M., T.~T. Cai, et~al. (2010).
\newblock A reproducing kernel {H}ilbert space approach to functional linear
  regression.
\newblock {\em The Annals of Statistics\/}~{\em 38\/}(6), 3412--3444.

\bibitem[\protect\citeauthoryear{Zhu, Kong, Li, Styner, Gerig, Lin, and
  Gilmore}{Zhu et~al.}{2011}]{zhu2011fadtts}
Zhu, H., L.~Kong, R.~Li, M.~Styner, G.~Gerig, W.~Lin, and J.~H. Gilmore (2011).
\newblock {FADTTS}: functional analysis of diffusion tensor tract statistics.
\newblock {\em NeuroImage\/}~{\em 56\/}(3), 1412--1425.

\bibitem[\protect\citeauthoryear{Zhu, Zhang, Ibrahim, and Peterson}{Zhu
  et~al.}{2007}]{zhu2007statistical}
Zhu, H., H.~Zhang, J.~G. Ibrahim, and B.~S. Peterson (2007).
\newblock Statistical analysis of diffusion tensors in diffusion-weighted
  magnetic resonance imaging data.
\newblock {\em Journal of the American Statistical Association\/}~{\em
  102\/}(480), 1085--1102.

\end{thebibliography}

\end{document}

\maketitle
\section{More Simulation Studies}
\subsection{Robustness to noise}
\begin{figure}
\begin{center}
\begin{tabular}{ccc}
\includegraphics[height=1.2in]{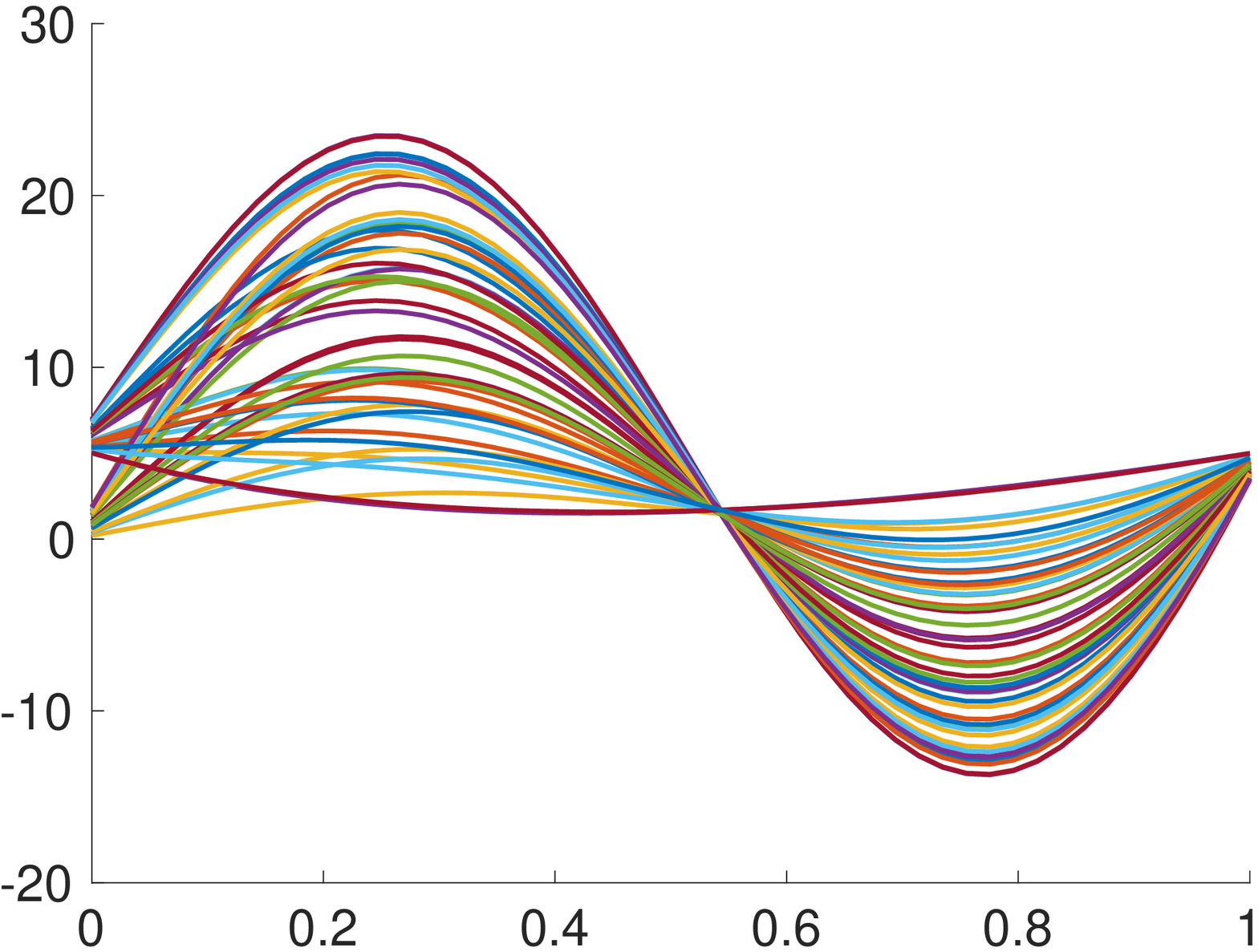} &
\includegraphics[height=1.2in]{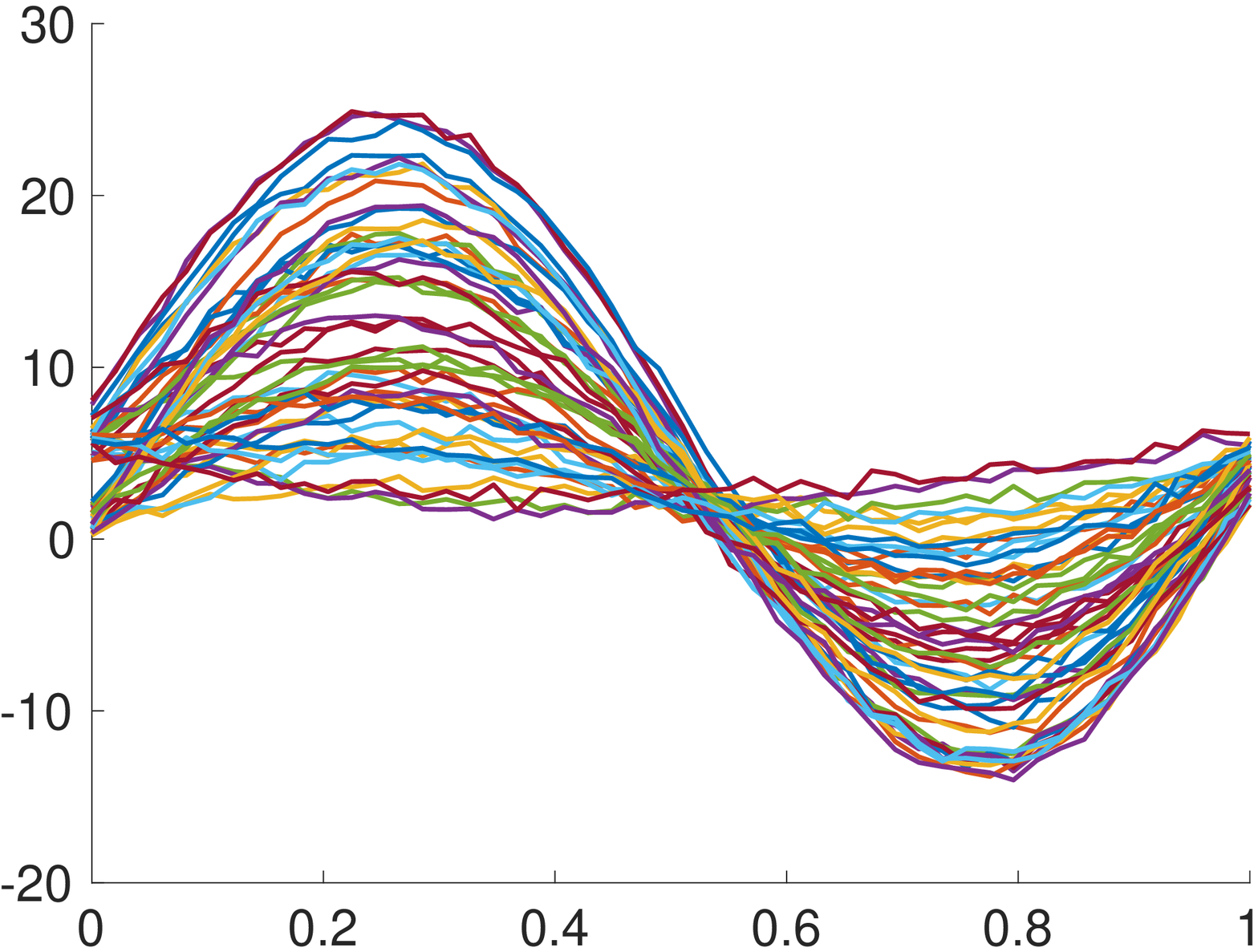}&
\includegraphics[height=1.2in]{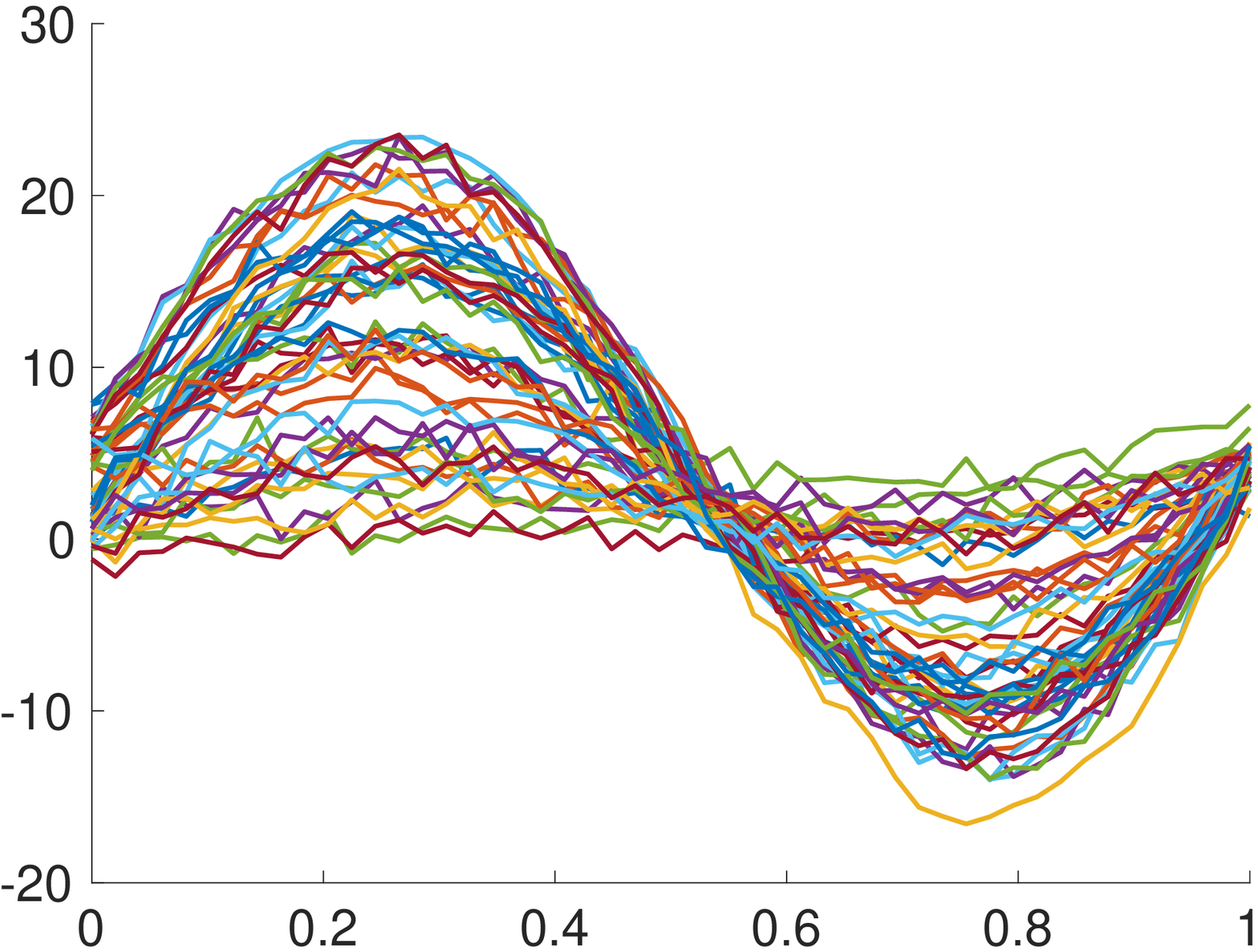}\\
(a) Noise free & (b) Moderate noise & (c) High noise
\end{tabular}
\caption{Simulated functions under different noise levels.} \label{fig:noise}
\end{center}
\end{figure}

We study the robustness of the proposed algorithm by testing its performance at different levels of noise.  The noise $\eta$ in our simulation in Section 5.1 contains two sources of noise, one is i.i.d noise $\epsilon(s_j)$ at each $s_j$ and one is correlated noise $v(\cdot)$.  In the first case, we fix $v(\cdot)$ as a Gaussian process following GP(0,$C_v$), where $C_v(s_j,s_l) = 0.6*\exp(-((s_j - s_l)/0.8)^2)$, and vary $\epsilon(s_j)$ by assuming its distribution as $N(0,0.1)$ (moderate level) and $N(0,0.4)$ (high level).  Figure \ref{fig:noise} shows the simulated $Y$ with no noise, moderate noise and high noise. Table 5.2 in the main paper studies the estimator's accuracy in the moderate noise scenario. Table \ref{tab:rmise_highnoise} here reports the accuracy in the high noise scenario at $\tau=0.5$. 
In another case, we fix  $\epsilon(s_j) \sim N(0,0.4)$ and change $C_v$ to a more complex structure from Mate\'rn family. We let $C_v(s_j,s_l) = \sqrt{C_v(s_j,s_j)}*\sqrt{C_v(s_l,s_l)}*M(\|s_j-s_l\|)$ for $s_j \neq s_l$, where $M(h) = {2^{1-\nu}\over \Gamma(\nu)} (\alpha  \|h\|)^\nu K_\nu(\alpha  \|h\|)$, with $K_\nu$ being the modified Bessel function of the second kind, $\nu=2.5$ and $\alpha=0.5$. We have $C_v(s_j,s_j) = 0.6$. The results for $\tau=0.5$ are displayed in Table \ref{tab:rmise_maternnoise}.

By comparing Table \ref{tab:rmise_highnoise} and \ref{tab:rmise_maternnoise} with the Table 5.2 in the main paper, we find that although  noise can degenerate the performance, we still get reasonable results in the high noise situations. A more complex correlation on noise does not change the performance much. These results  indicate that the developed primal-dual algorithm is robust to noise in general. 

\begin{table*} \centering
\caption{RMISE of the estimators with the primal-dual algorithm under high noise with $\epsilon(s_j) \sim N(0,0.4)$ and $C_v(s_j,s_l) = 0.6*\exp(-((s_j - s_l)/0.8)^2)$}
\label{tab:rmise_highnoise}
\ra{1.0}
\begin{threeparttable}
\begin{tabular}{@ {}lcrrr@{}} \toprule
  (n,m) && $\beta_1$  & $\beta_2$ & $\beta_3$\\
 \midrule
(50,50) &&  $.0788_{(.024)}$   &   $.1024_{(.0321)}$  & $.3224_{(.0507)}$  \\
(50,100) && $.0528_{(.030)}$    &  $.0520_{(.0277)}$   & $.1945_{(.0426)}$ \\
(100,50) &&  $.0491_{(.0168)}$   &  $.0738_{(.0361)}$  & $.2257_{(.0375)}$ \\
(100,100) && $.0437_{(.0192)}$    &  $.0357_{(.0172)}$   & $.1231_{(.0257)}$ \\
(200,50) &&  $.0487_{(.0136)}$   &  $.0265_{(.0160)}$  & $.1468_{(.0215)}$  \\
(200,100) && $.0305_{(.0163)}$    &  $.0218_{(.0123)}$   & $.0818_{(.0165)}$  \\
(500,50) &&  $.0339_{(.0111)}$   &   $.0198_{(.0100)}$  & $.0823_{(.0178)}$  \\
(500,100) &&  $.0278_{(.0102)}$   &   $.0198_{(.0089)}$  & $.0814_{(.0088)}$  \\

\bottomrule
\end{tabular}
%\begin{tablenotes}
%\small
%\item * indicates missing value due to significant computing time. 
%\end{tablenotes}
\end{threeparttable}
\end{table*}

\begin{table*} \centering
\caption{RMISE of the estimators with the primal-dual algorithm under high noise with $\epsilon(s_j) \sim N(0,0.4)$ and $C_v(s_j,s_l) = 0.6*M(\|s_j-s_l\|)$}
\label{tab:rmise_maternnoise}
\ra{1.0}
\begin{threeparttable}
\begin{tabular}{@ {}lcrrr@{}} \toprule
  (n,m) && $\beta_1$  & $\beta_2$ & $\beta_3$\\
 \midrule
(50,50) &&   $0.0713_{(.039)}$ &   $0.1127_{(.042)}$  &  $0.3059_{(.060)}$ \\ 
(50,100) &&  $0.0474_{(.031)}$  &  $0.0527_{(.031)} $ & $ 0.1719_{(.043)}$ \\
(100,50) &&   $0.0590_{(.027)} $ &  $0.0626_{(.029)} $ &  $0.2215_{(.054)}$ \\
(100,100) && $0.0304_{(.020)}$&   $0.0392_{(.020)} $ &  $0.1208_{(.032)}$ \\
(200,50) &&  $0.0427_{(.021)}$   & $0.0417_{(.024)} $ &  $0.1362_{(.035)} $ \\
(200,100) && $0.0305_{(.010)}$  &  $0.0188_{(.010)}$   & $0.0786_{(.020)}$  \\
(500,50) &&   $0.0349_{(.011)}$  &  $0.0188_{(.008)} $ &  $0.0873_{(.018)} $\\
(500,100) &&  $ 0.0306_{(.009)} $&   $0.0154_{(.009)}$    & $0.0806_{(0.015)}$  \\

\bottomrule
\end{tabular}
%\begin{tablenotes}
%\small
%\item * indicates missing value due to significant computing time. 
%\end{tablenotes}
\end{threeparttable}
\end{table*}

\subsection{Irregular grid's effect to the copula model}

\begin{figure}
\begin{center}
\begin{tabular}{cc}
\includegraphics[height=1.2in]{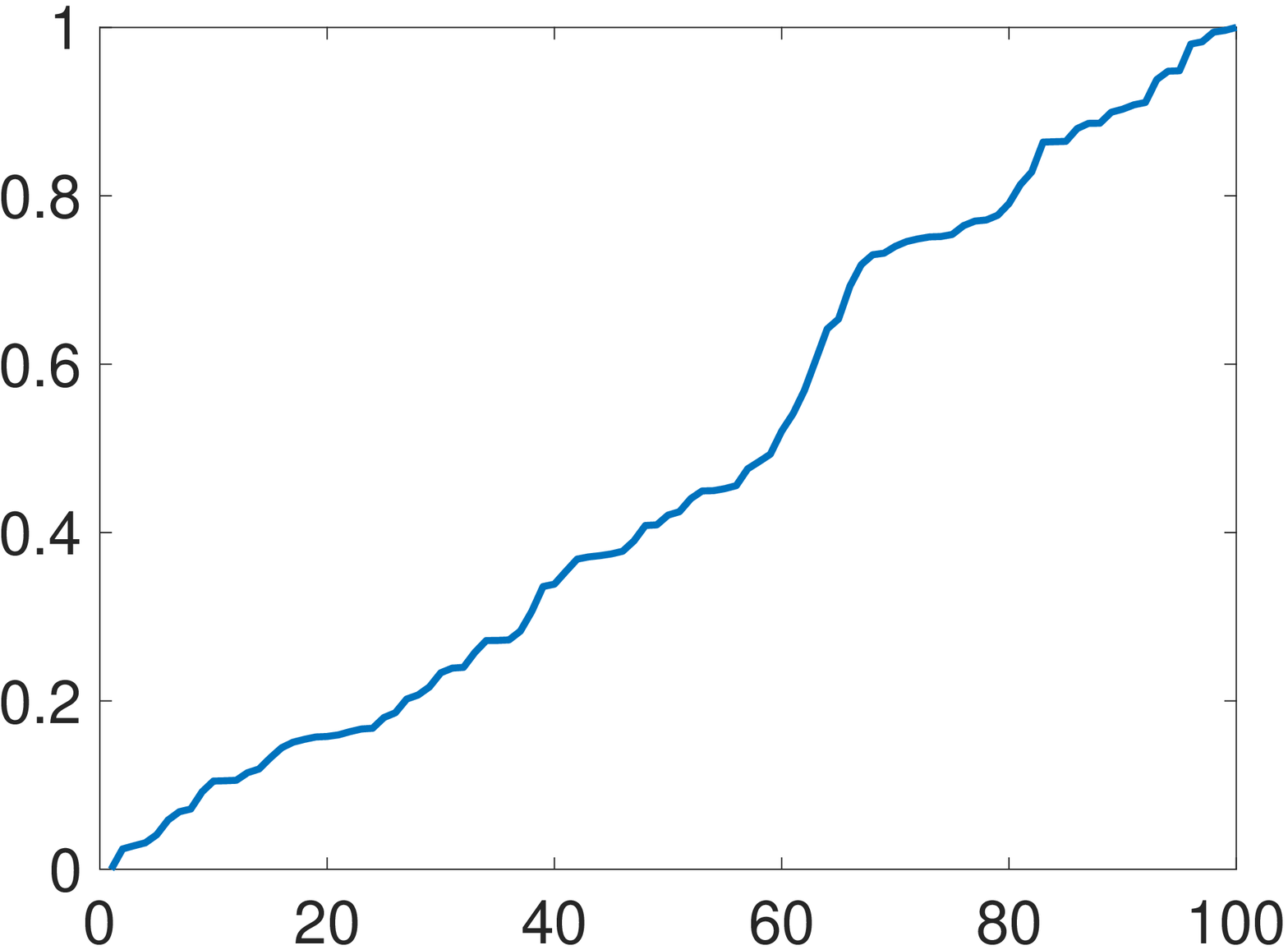} &
\includegraphics[height=1.2in]{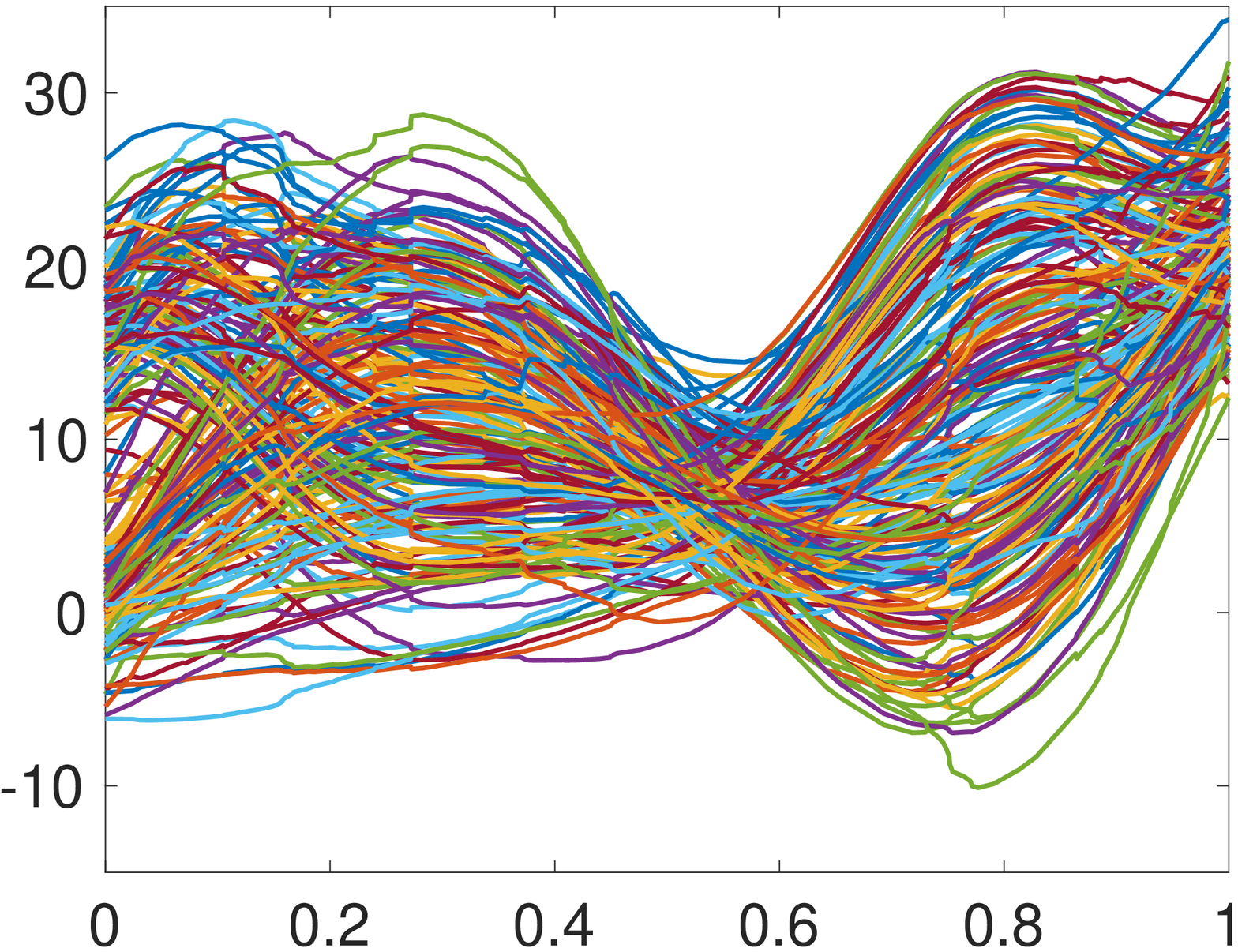}\\
(a) Irregular grid & (b) Simulated Y
\end{tabular}
\caption{Irregular grid and simulated $Y(s)$.} \label{fig:irreg-ys}
\end{center}
\end{figure}

\begin{figure}
\begin{center}
\begin{tabular}{cccc}
\includegraphics[height=1.0in]{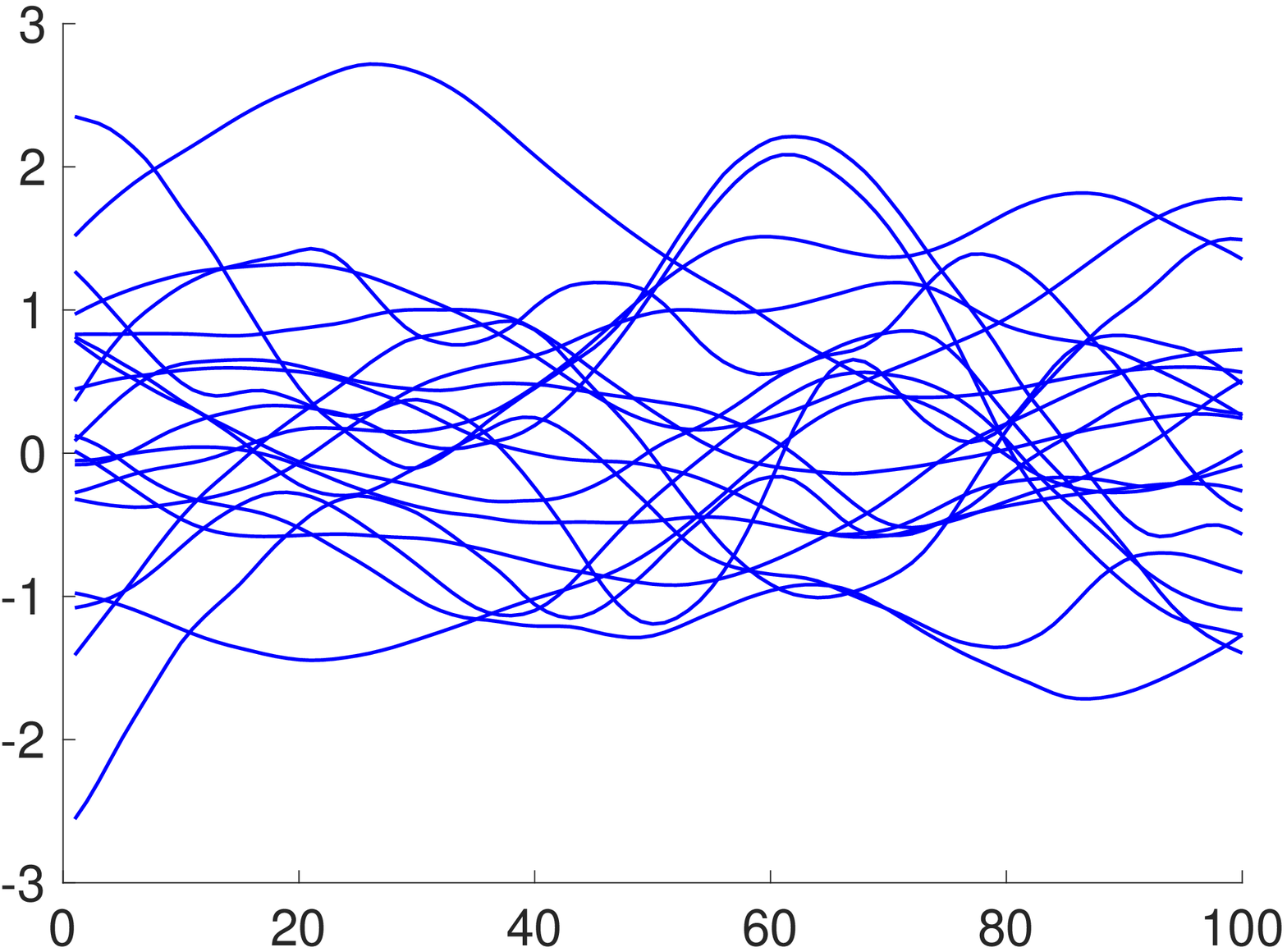} &
\includegraphics[height=1.0in]{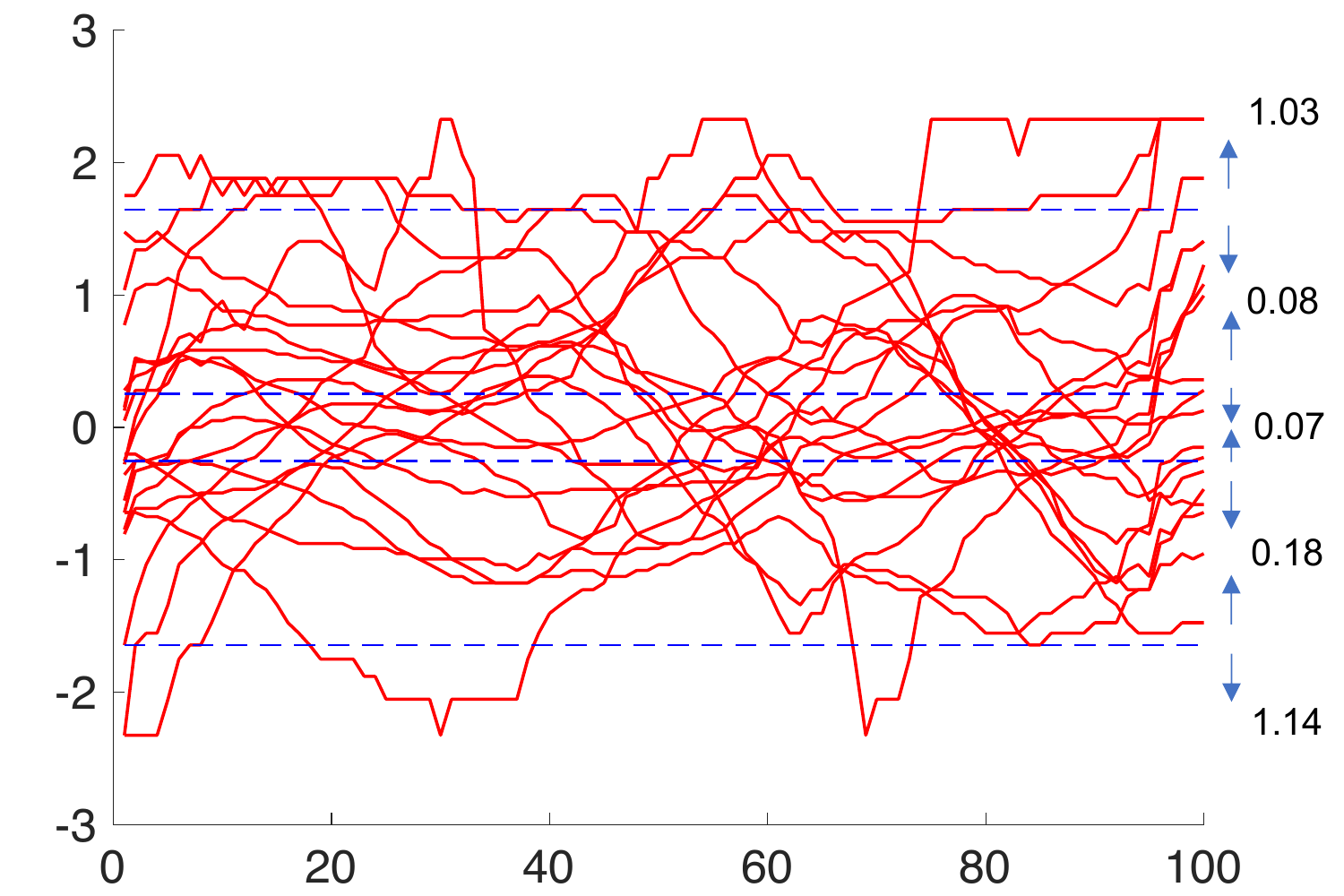}&
\includegraphics[height=1.0in]{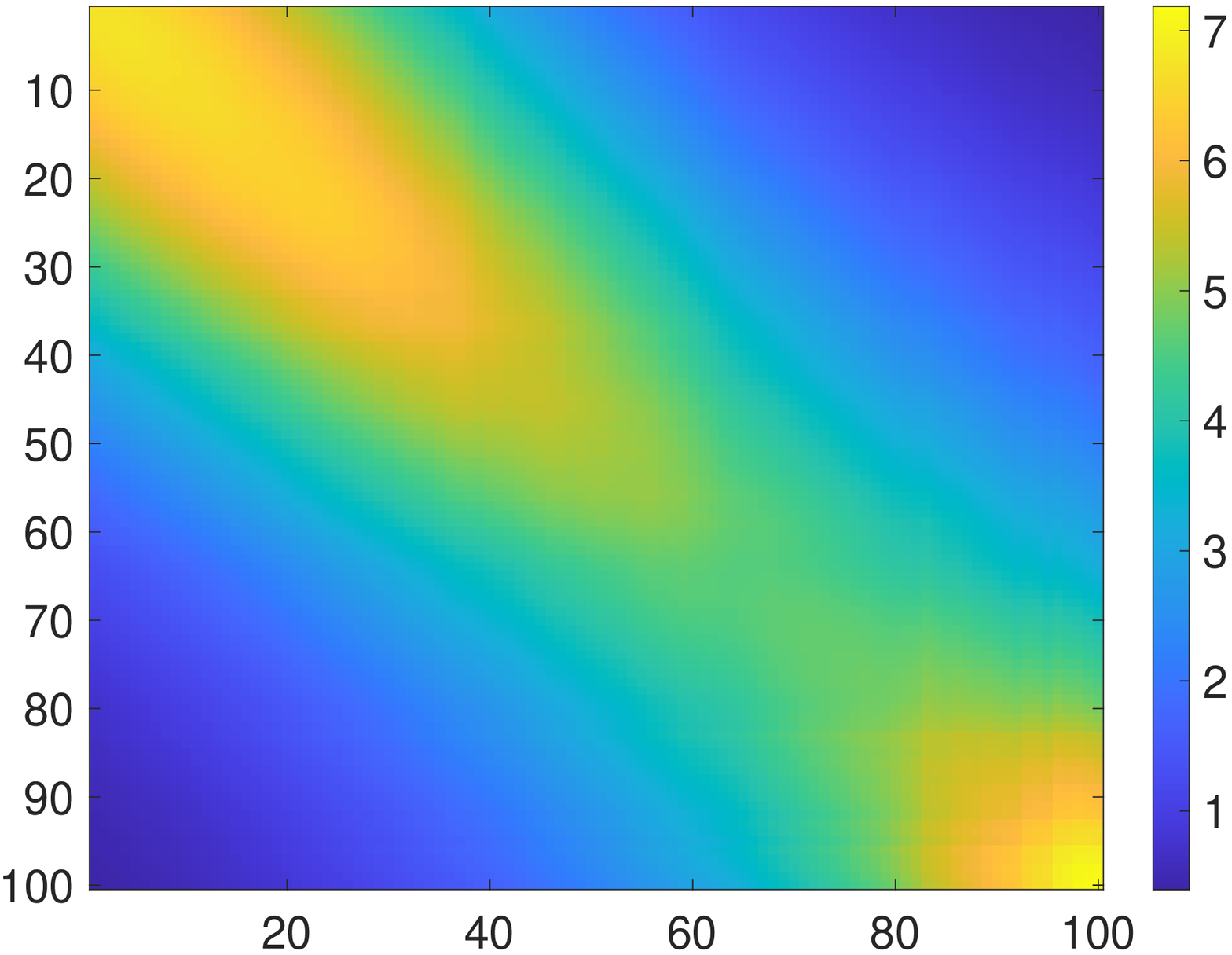} &
\includegraphics[height=1.0in]{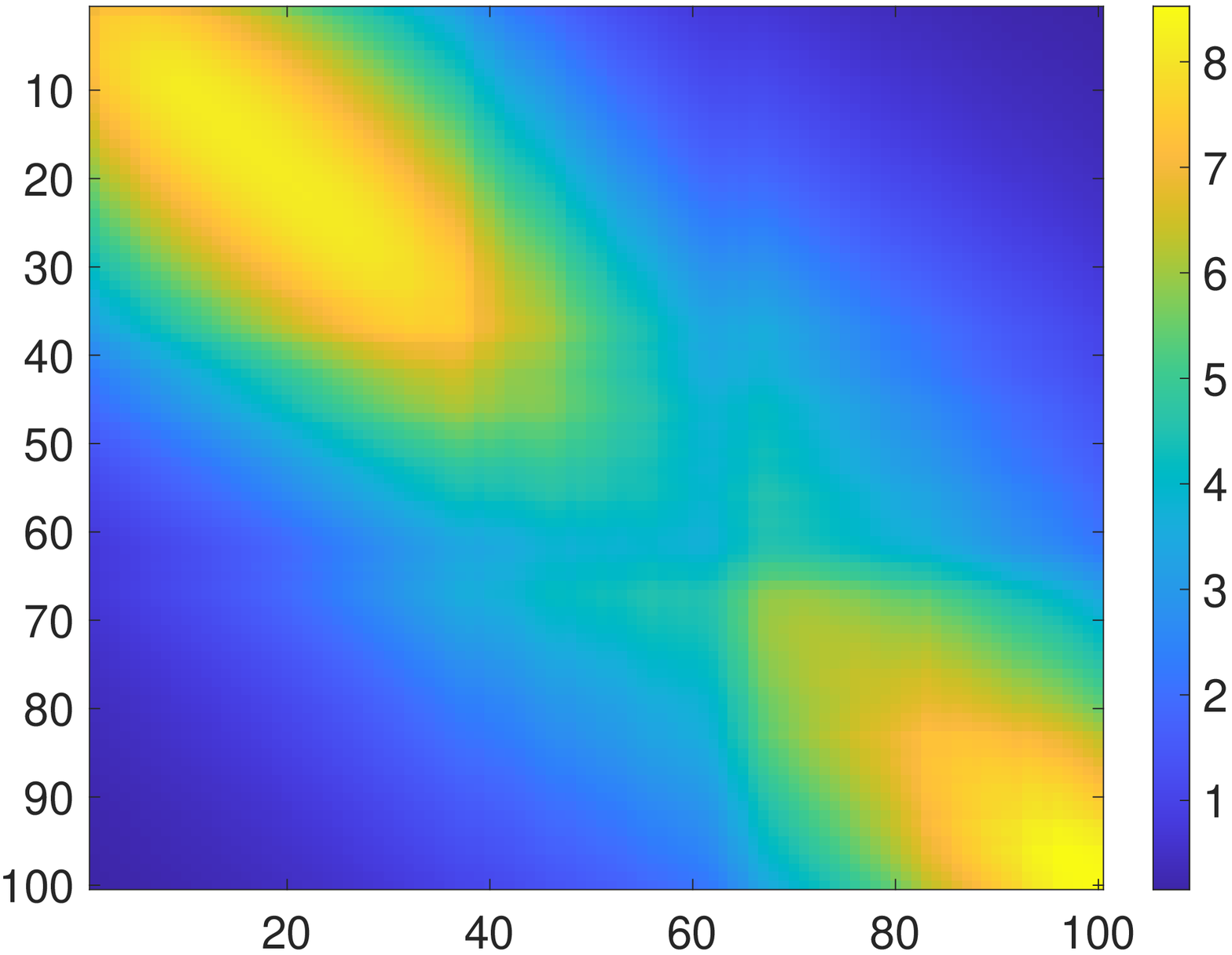}\\
(a) $\Phi^{-1}(U(s))$ & (b) $\hat{\Phi}^{-1}(U(s))$  & (c) true covariance  & (d)  estimated covariance 
\end{tabular}
\caption{Parameters estimation in the copula model with irregular grid.    (a) and (b) show some ground truth quantile functions $\Phi^{-1}(U(s))$ and the corresponding estimated $\hat{\Phi}^{-1}(U(s))$, respectively. The blue lines mark different quantile ranges for calculating difference between $\Phi^{-1}(U(s))$ and $\hat{\Phi}^{-1}(U(s))$.   (c) and (d) show the ground truth covariance function and the estimated covariance function, for ${\bf x} = (1,0,0.3815)^T$, respectively.} \label{fig:irreg-coupla}
\end{center}
\end{figure}

\textcolor{blue}{
It is possible to have irregular grid for observations. To study the effect of the irregular grid on our model, we simulated the grid from random sampling of the interval $[0,1]$ and then took the same parameters as Section 5.2 to study the copula model. A plot of the simulated grid is shown in Figure \ref{fig:irreg-ys} panel (a), and panel (b) shows the simulated $Y(s)$.  In Figure \ref{fig:irreg-coupla} (a) shows the ground truth $20$ functions of $\Phi^{-1}(U(s))$ under the irregular grid, and (b) shows the estimated $\hat{\Phi}^{-1}(U(s))$. Same to the regular grid, here we observe the average difference between $\Phi^{-1}(U(s))$ and $\hat{\Phi}^{-1}(U(s))$ is small in the middle than in the tail. Based on the estimated $\rho=6$, we calculated  $\hat{t}_\rho^{-1}(U(s))$, and estimated the covariance of the t copula $ \rho/(\rho-2)*{\bm \Sigma}_{\bm \theta(\bf x)}$. For ${\bf x} = (1,0,0.0935)^T$, Figure \ref{fig:irreg-coupla} (c) and (d) show the ground truth covariance function and the estimated covariance function, respectively. For other ${\bf x}$, we got similar results. 
}

\section{Comparison of Copula Models Based on Simulated and Real Data}

\begin{figure}
\begin{center}
\begin{tabular}{c}
\includegraphics[width=0.90\textwidth]{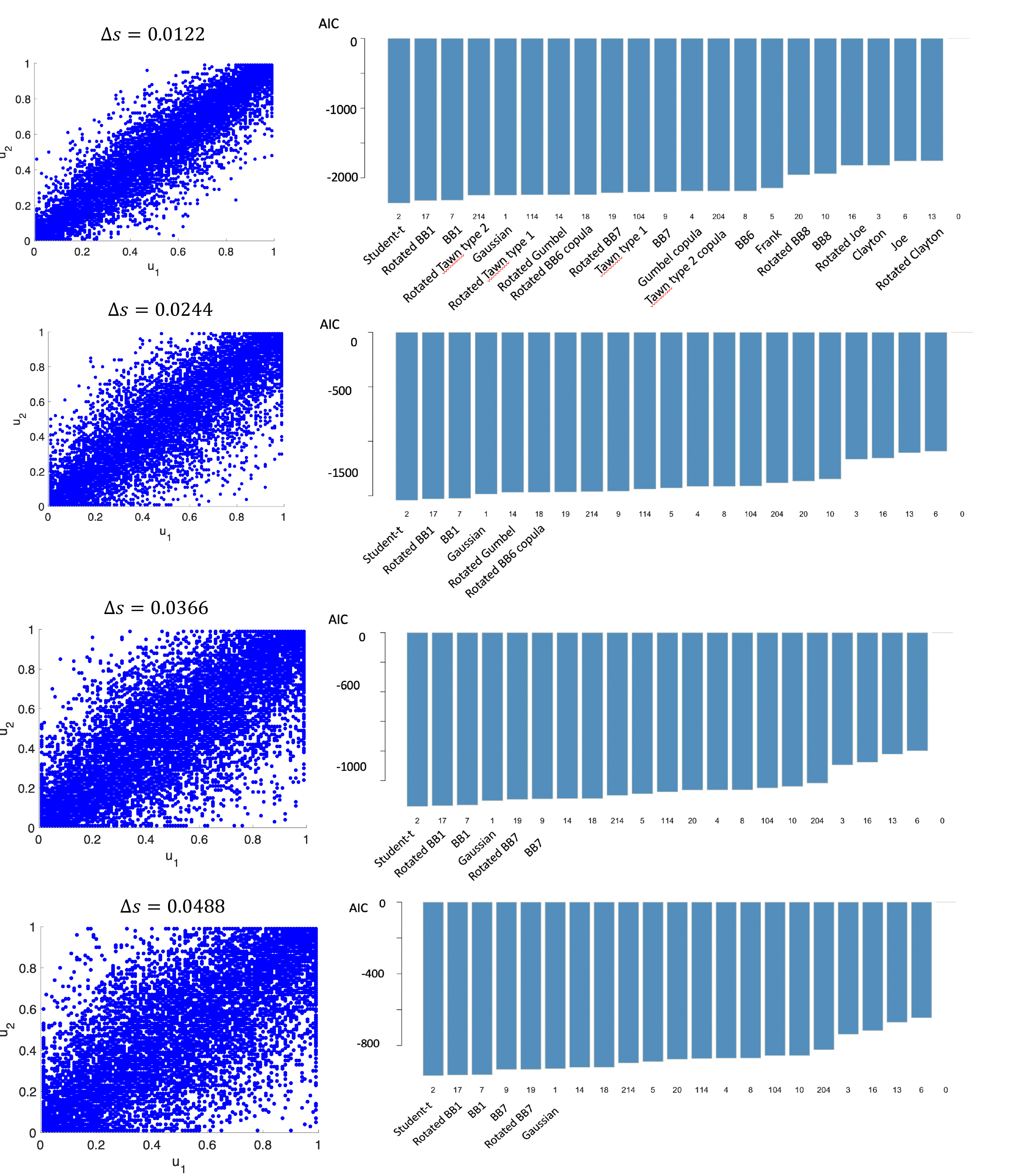} 
\end{tabular}
\caption{ Comparison of different copula models with our DTI FA data. The left panel shows scatter plots of $(\hat{F}_{s_i,\bf{x}}(y_i),\hat{F}_{s_j,\bf{x}}(y_j))$ at different $\Delta s (\Delta s = s_j - s_i)$ and the right panel shows AIC values under different copula models. We compared the following copula models: 1 = Gaussian copula;
2 = Student t copula (t-copula);
3 = Clayton copula;
4 = Gumbel copula;
5 = Frank copula;
6 = Joe copula;
7 = BB1 copula;
8 = BB6 copula;
9 = BB7 copula;
10 = BB8 copula;
13 = rotated Clayton copula ;
14 = rotated Gumbel copula;
16 = rotated Joe copula;
17 = rotated BB1 copula;
18 = rotated BB6 copula;
19 = rotated BB7 copula;
20 = rotated BB8 copula;
104 = Tawn type 1 copula;
114 = rotated Tawn type 1; copula (180 degrees)
204 = Tawn type 2 copula;
214 = rotated Tawn type 2 copula (180 degrees). For more details of these copula models, please refer to \cite{nikoloulopoulos2012vine}. } \label{fig:copula_comp}
\end{center}
\end{figure}

\textcolor{blue}{There are many choices of copula models such as the Student-t copula \citep{demarta2005t}, the non-central $\chi^2$-copula \citep{bardossy2006copula}, and the Gaussian copula \citep{Kazianka2010Copula}. It is a challenging task to determine the optimal choice of the copula. Since our main application is medical image analysis, here we examined 21 different copula models using the real diffusion tensor imaging (DTI) data from the Alzheimer’s Disease Neuroimaging Initiative (ADNI). 
More specifically, we first fitted the spatial quantile regression model (3.1) to the DTI FA data to obtain marginal distribution at each location $s$. For simplicity, we only put intercept and gender into $\bold{x}$. Given a fix $\bf{x}$, e.g., male subjects, we studied the bivariate empirical-observations $(\hat{F}_{s_i,\bf{x}}(y_i),\hat{F}_{s_j,\bf{x}}(y_j))$ with different $\Delta s (\Delta s = s_j - s_i)$ using 21 different copula models, including Gaussian copula, Student t, Clayton copula, Gumbel copula, Frank copula and Joe copula. The model choice is based on AIC. Figure \ref{fig:copula_comp} shows the plots of $(\hat{F}_{s_i,\bf{x}}(y_i),\hat{F}_{s_j,\bf{x}}(y_j))$ and the AICs we got from different models. We can see that, the t-copula outperforms the Gaussian copula and other copula models and therefore the t-copula is used in the paper. }

\begin{figure}
\begin{center}
\begin{tabular}{c}
\includegraphics[width=0.90\textwidth]{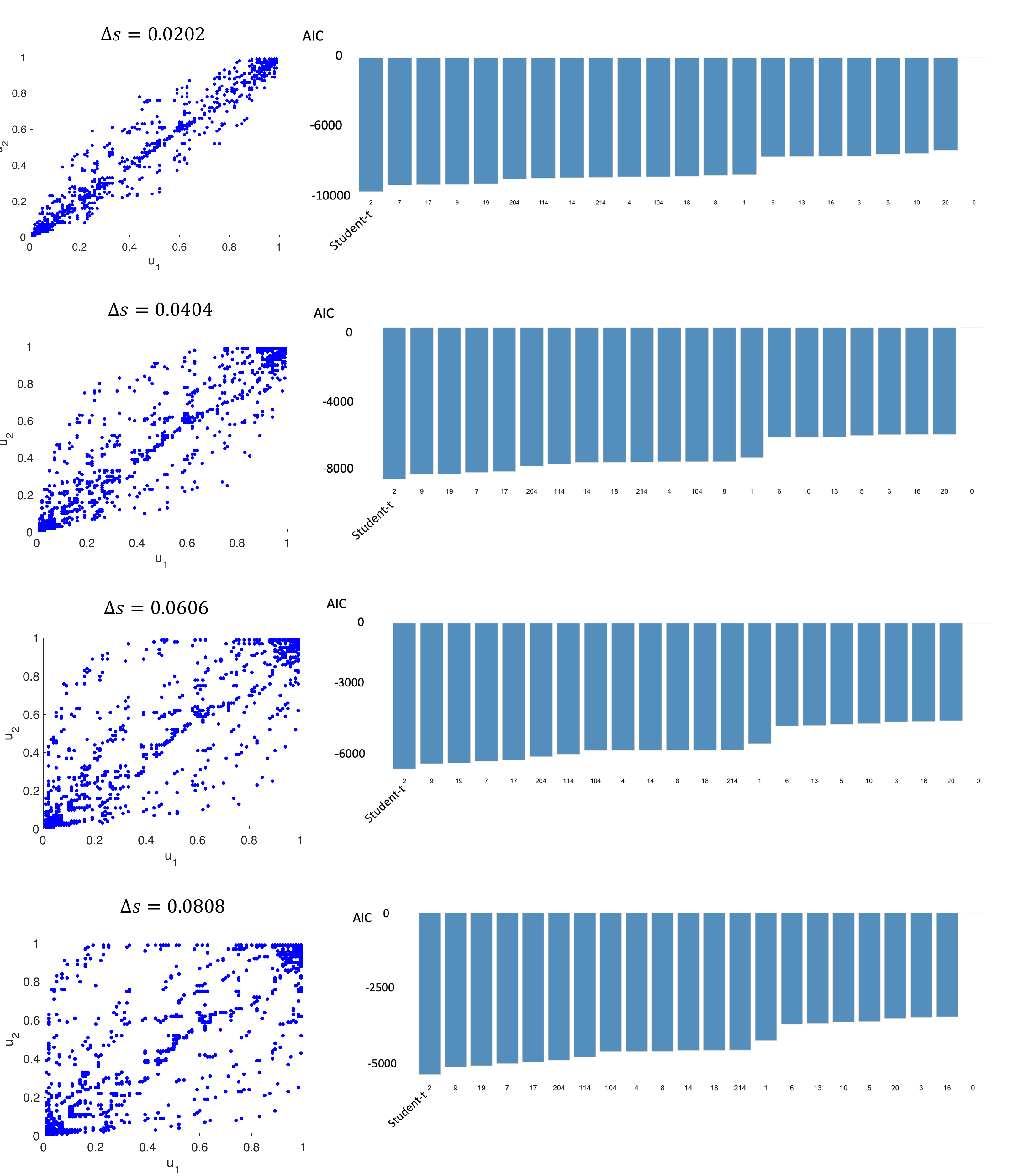} 
\end{tabular}
\caption{ Comparison of different copula models using simulated data from Section 5.2.. The left panel shows scatter plots of $(\hat{F}_{s_i,\bf{x}}(y_i),\hat{F}_{s_j,\bf{x}}(y_j))$ at different $\Delta s (\Delta s = s_j - s_i)$ and the right panel shows AIC values under different copula models. We compared the following copula models: 1 = Gaussian copula;
2 = Student t copula (t-copula);
3 = Clayton copula;
4 = Gumbel copula;
5 = Frank copula;
6 = Joe copula;
7 = BB1 copula;
8 = BB6 copula;
9 = BB7 copula;
10 = BB8 copula;
13 = rotated Clayton copula ;
14 = rotated Gumbel copula;
16 = rotated Joe copula;
17 = rotated BB1 copula;
18 = rotated BB6 copula;
19 = rotated BB7 copula;
20 = rotated BB8 copula;
104 = Tawn type 1 copula;
114 = rotated Tawn type 1; copula (180 degrees)
204 = Tawn type 2 copula;
214 = rotated Tawn type 2 copula (180 degrees). } \label{fig:copula_comp_sim}
\end{center}
\end{figure}

\textcolor{blue}{ A similar study was also performed based on the simulated data in Section 5.2. The  results  are  shown  in  Figure  \ref{fig:copula_comp_sim}.  Although the pair $({F}_{s_i,\bf{x}}(y_i),{F}_{s_j,\bf{x}}(y_j))$ is simulated from Gaussian, t-copula fits the empirical data better in  most  of  the  cases,  justifying  our  choice  of  Student  t  copula. }

\section{Comparison with Function-on-Scalar Mean Regression}
\begin{figure}
\begin{center}
\begin{tabular}{cccc}
\includegraphics[height=1in]{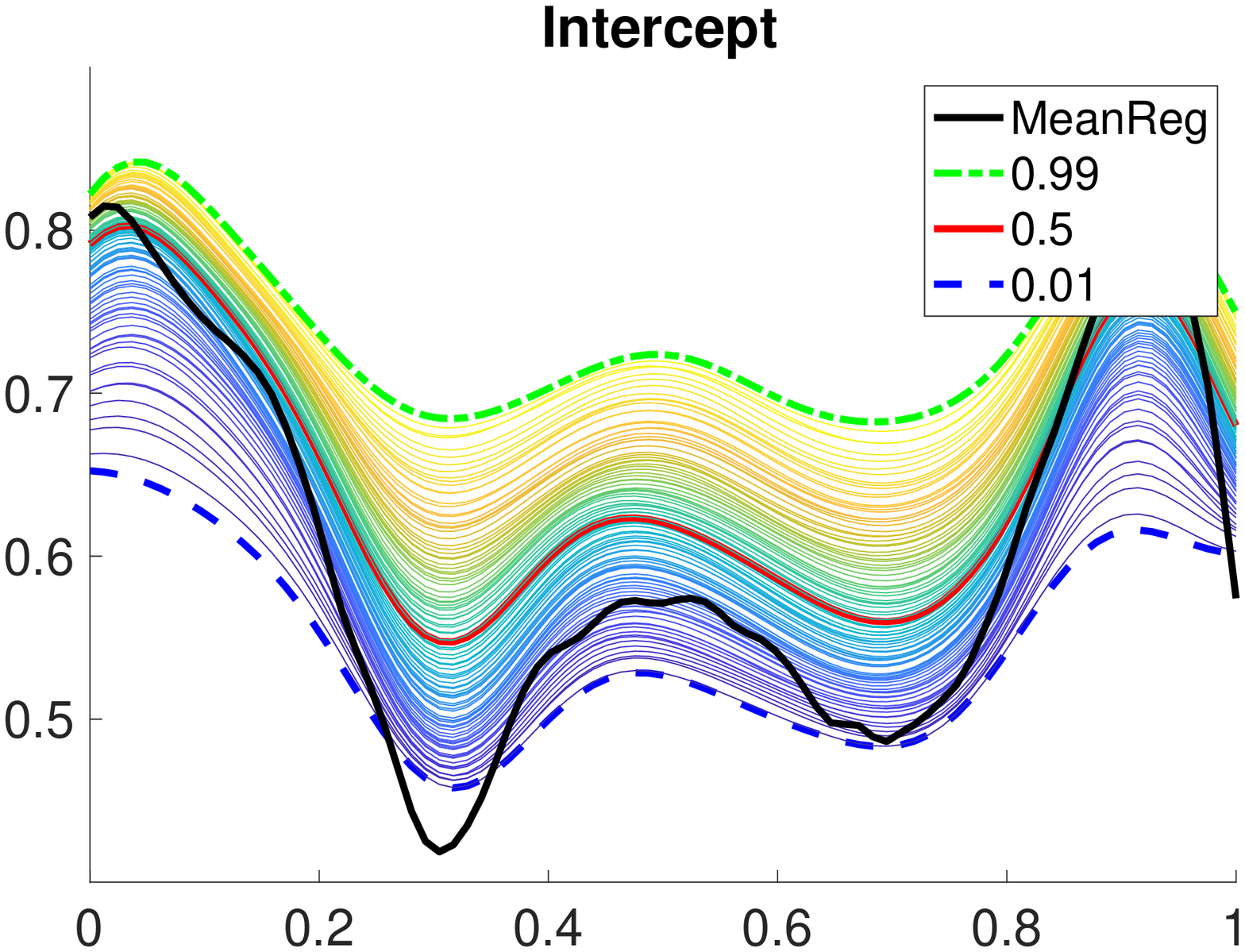} &
\includegraphics[height=1in]{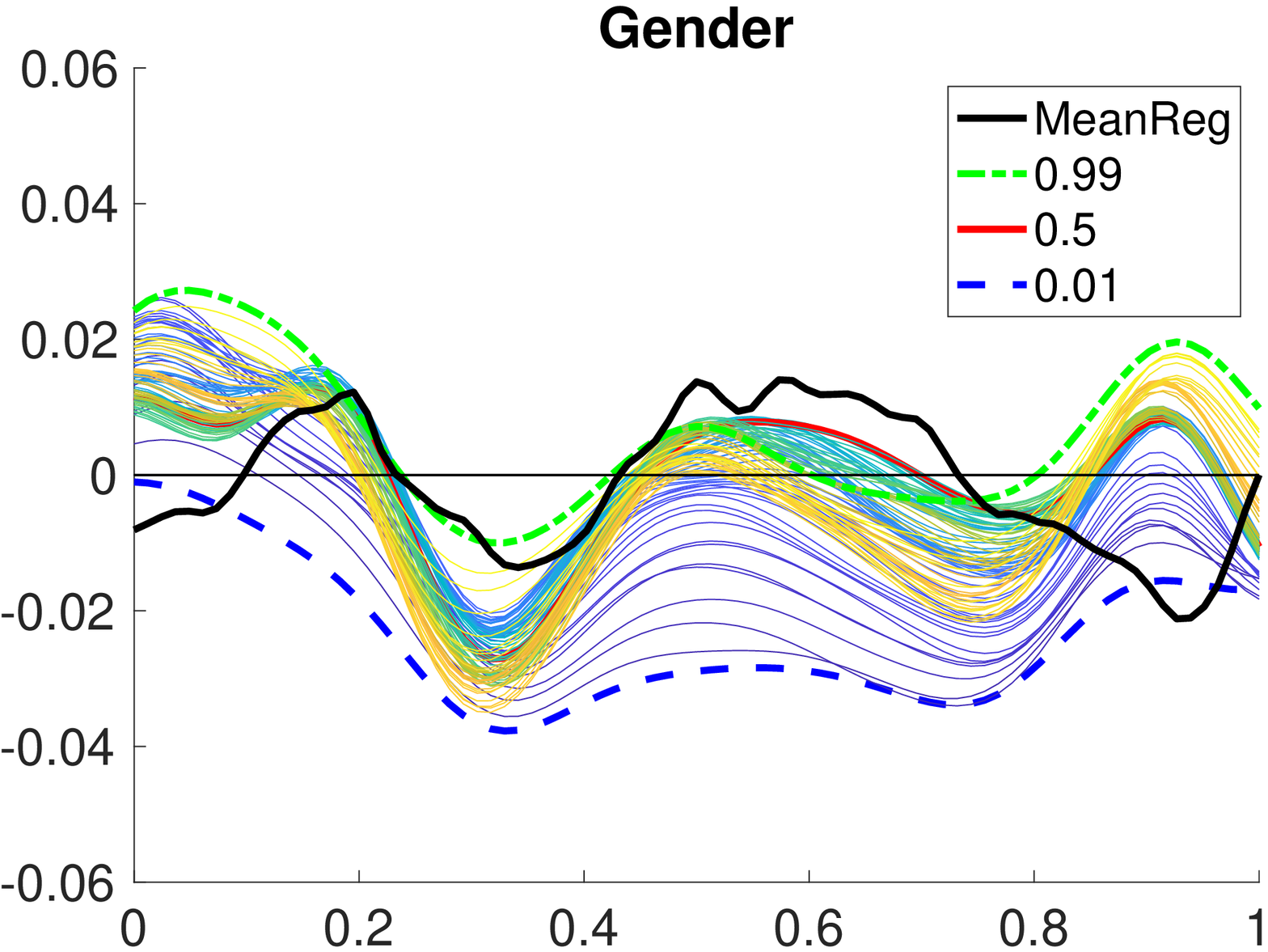}&
\includegraphics[height=1in]{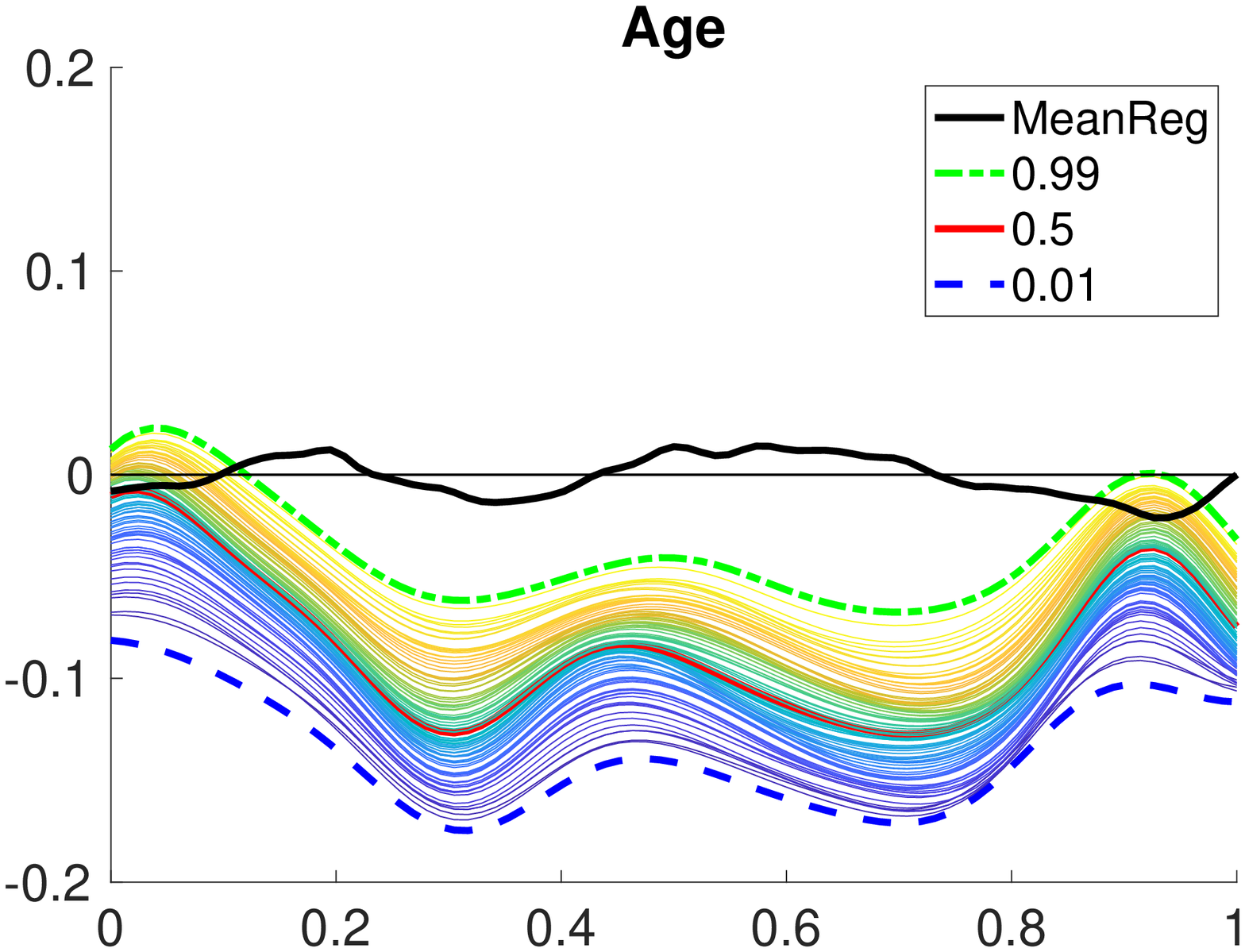}&
\includegraphics[height=1in]{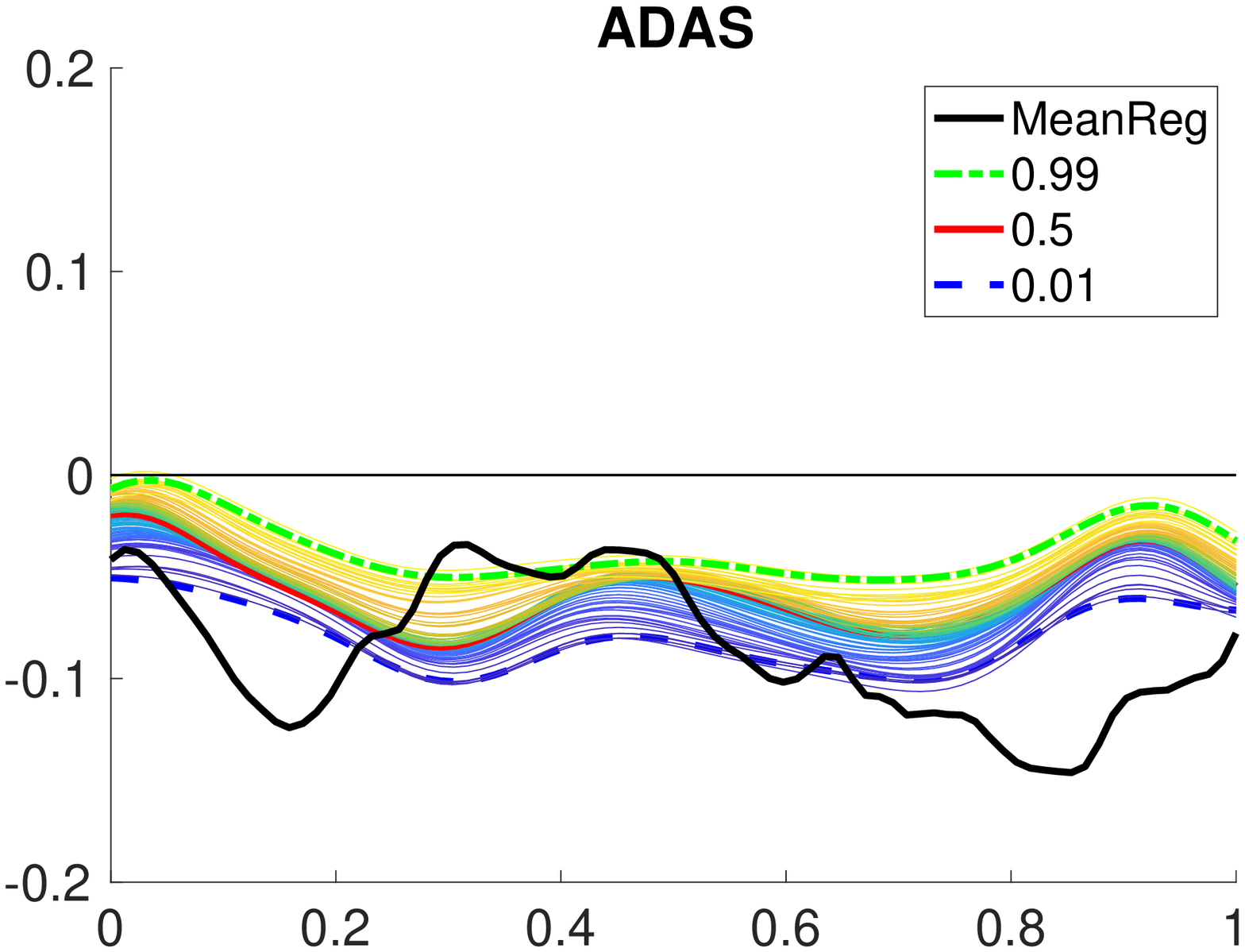} 
\end{tabular}
\caption{Comparison of quantile and mean regression coefficents.} \label{fig:meanreg}
\end{center}
\end{figure}

\textcolor{blue}{
Taking the DTI data in Section 5.3.1 in the main paper, we applied the function-on-scalar regression method developed in \cite{zhu2011fadtts} to study the mean regression. Figure \ref{fig:meanreg} shows the results of both our quantile regression and mean regression. We can see that the interesting structure on different quantiles can not be viewed in the coefficient functions of mean regression. However, the information reflected by different quantiles may be interesting and important. For example, for $\beta_{gender}$,  we observe that there is not much difference between males and females at quantiles ranging from $10$ to $90$. However, if we consider the lower quantiles (e.g., from $1$ to $10$),     FA curves for females along corpus callosum skeleton have smaller values compared with those for males. Biologically, this means the lower FA curves along the midsagittal corpus callcosum in females are smaller than the lower FA curves in males. But their means might not have much difference (confirmed by the mean regression).  Similar phenomenon is observed for the age covariate, that is for people with worse FA values (at lower quantiles), same amount of aging can contribute more on FA value reduction (worse white matter deterioration \citep{kochunov2007relationship}) than those who at high quantiles. For $\beta_{ADAS}$, the deterioration of Alzheimer's Disease (measured by ADAS) is similar across different quantiles.
}

\section{More Results for Hippocampus Surface Data}

\begin{figure}
\begin{center}
\begin{tabular}{ccc}
\includegraphics[height=1.2in]{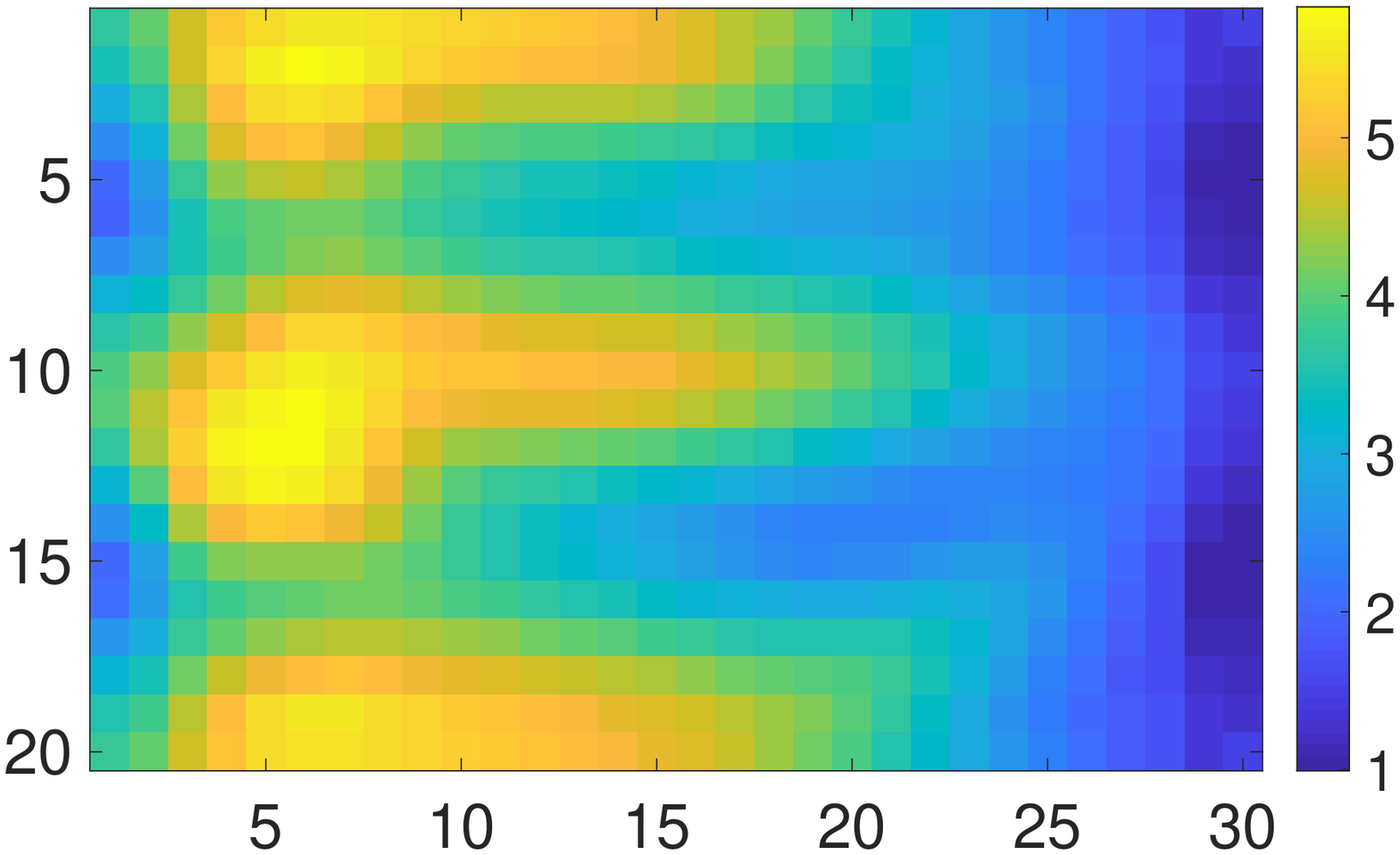} &
\includegraphics[height=1.2in]{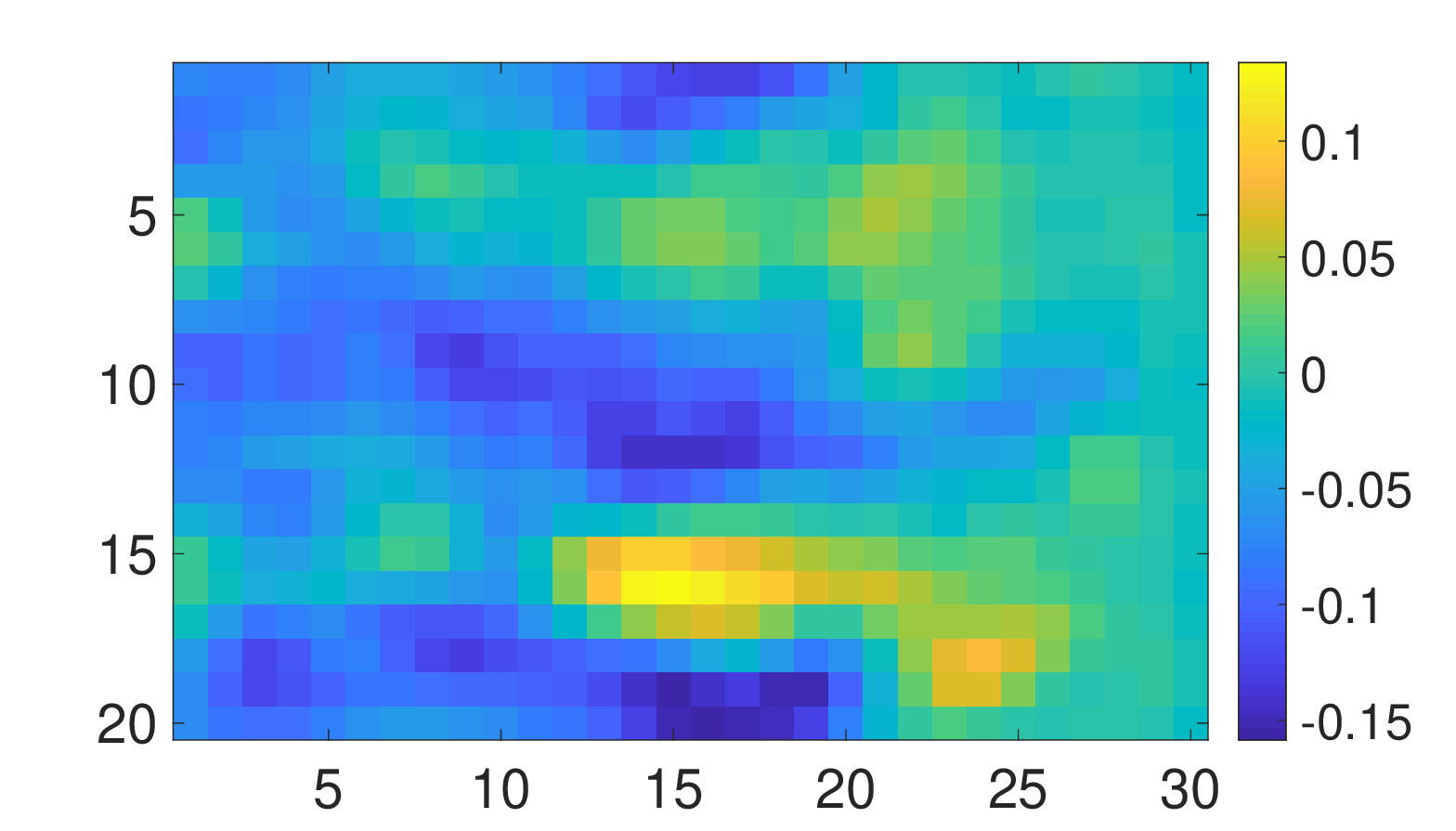}&
\includegraphics[height=1.2in]{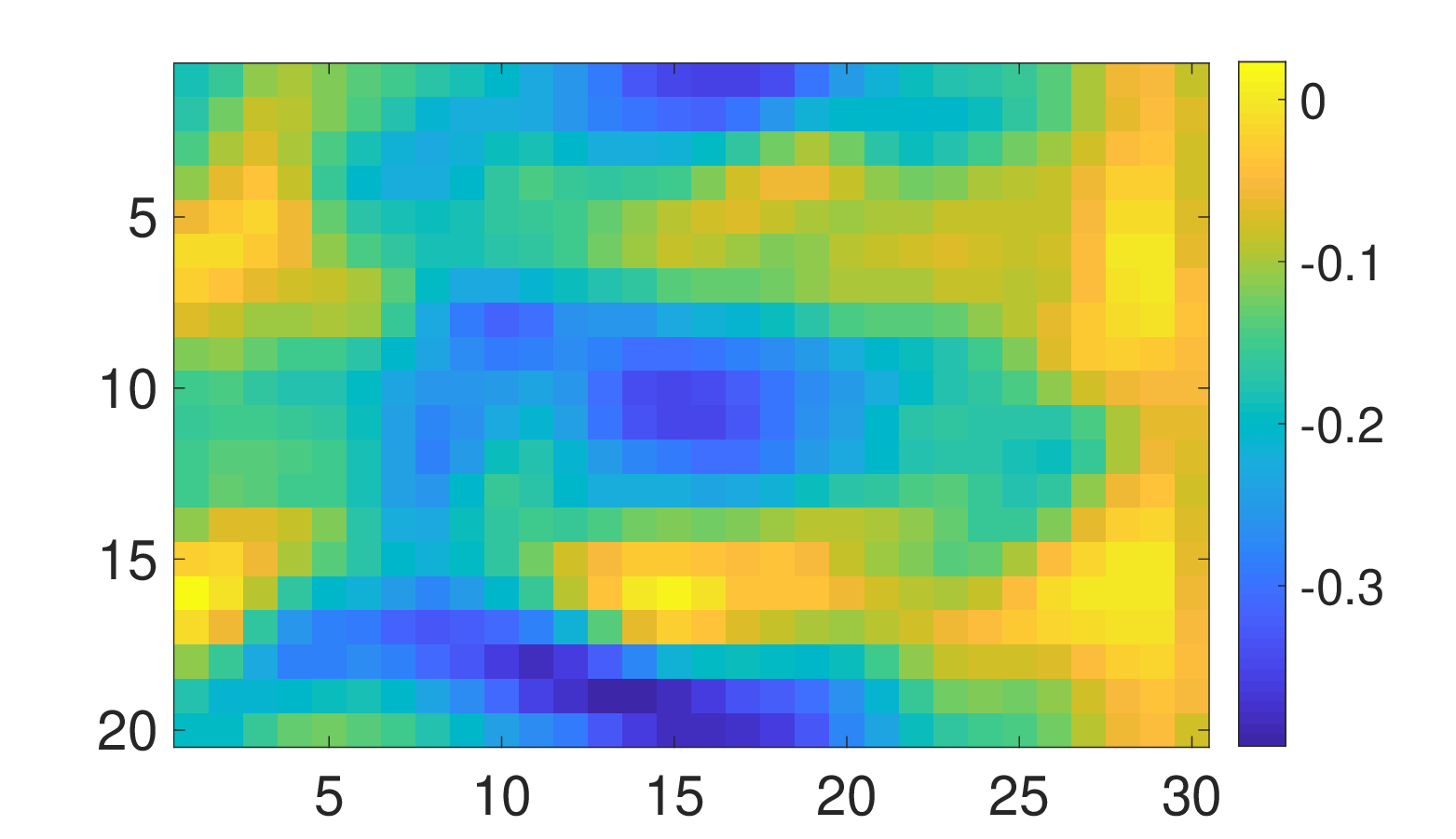}\\
\includegraphics[height=1.2in]{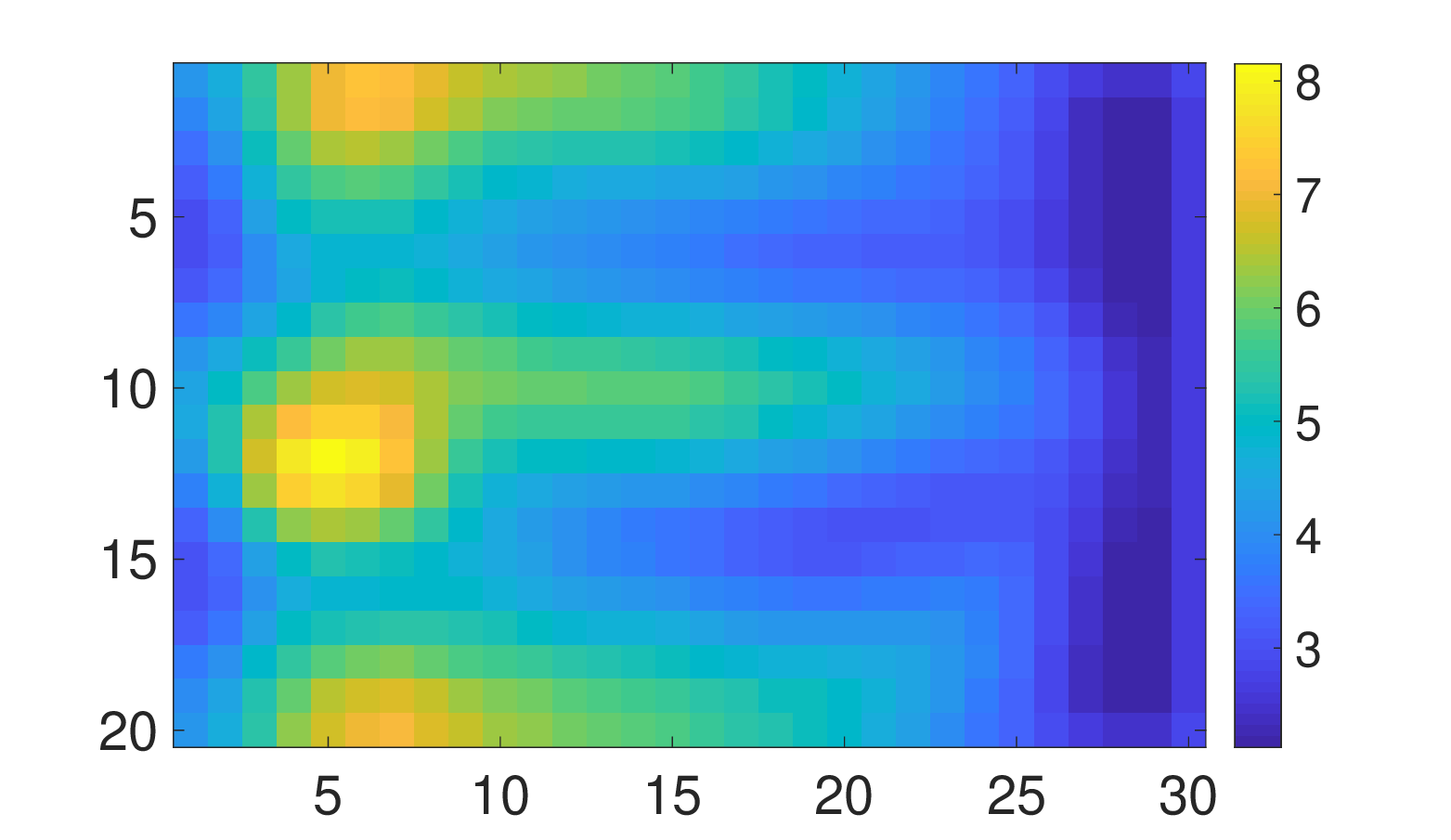} &
\includegraphics[height=1.2in]{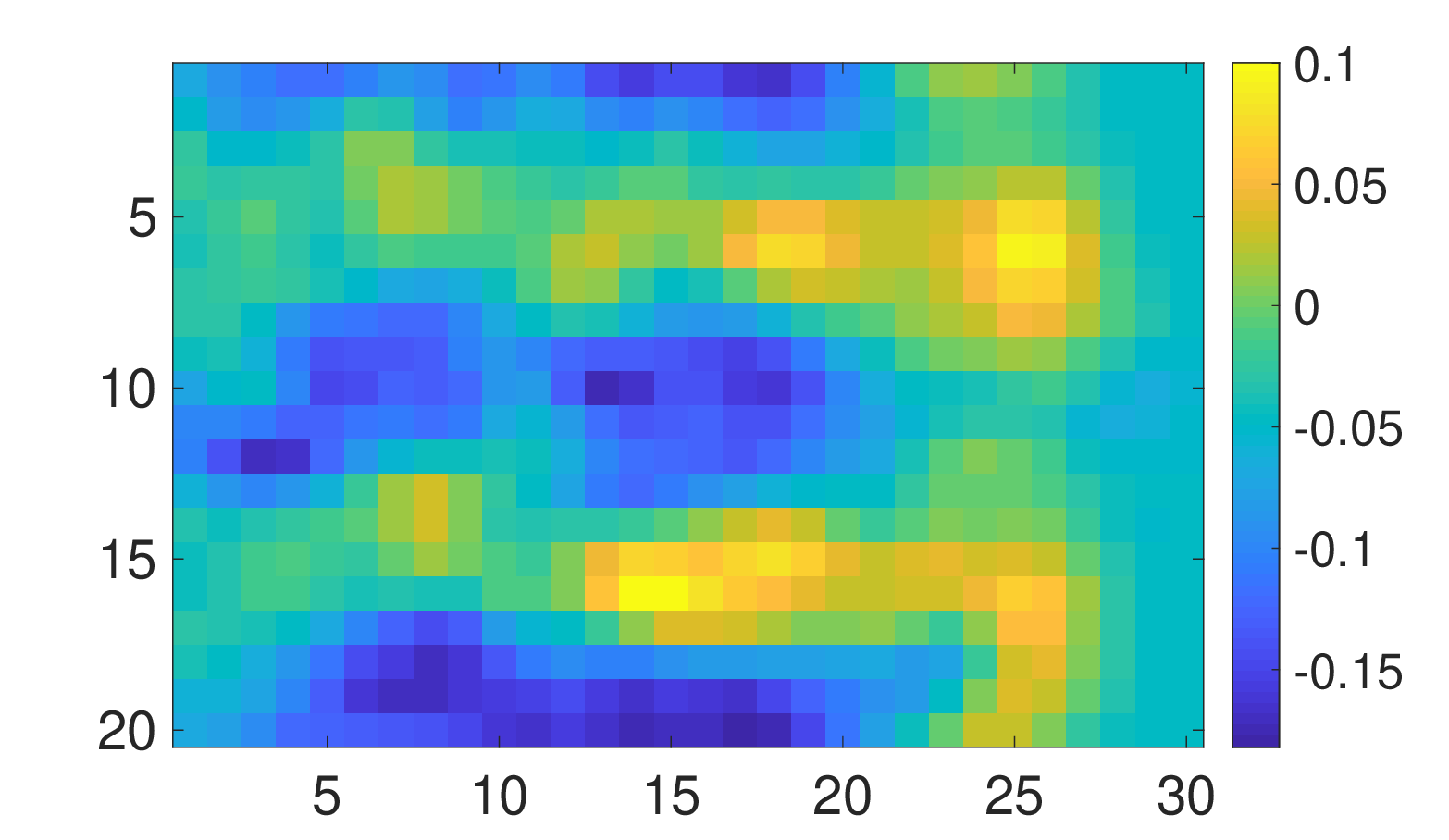}&
\includegraphics[height=1.2in]{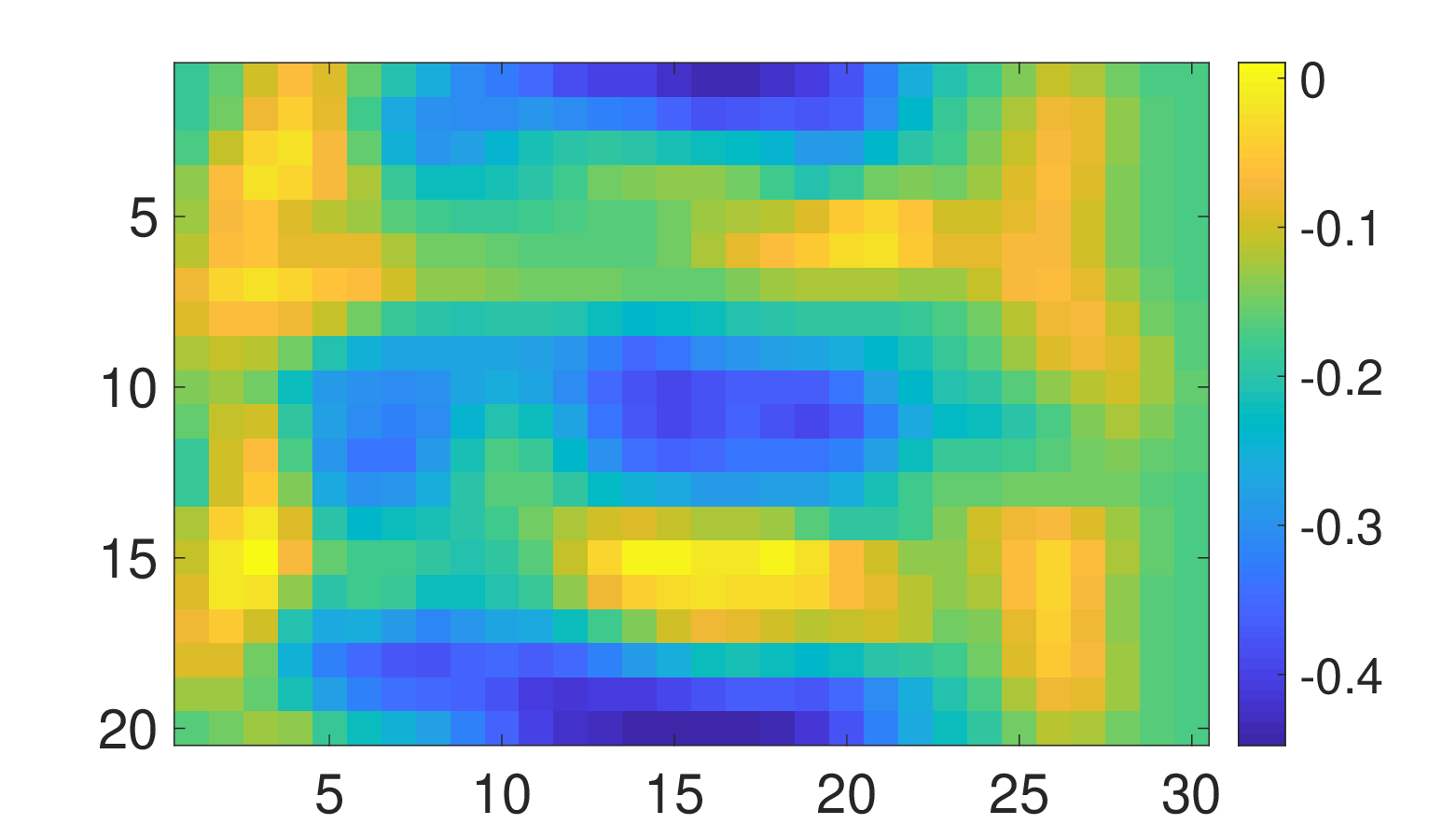}\\
(a) Intercept & (b) $\beta_{age}$ & (c) $\beta_{behavior}$
\end{tabular}
\caption{Coefficient images at $\tau = 0.25$ (first row) and $\tau = 0.75$ (second row).} \label{fig:simulation_beta_app}
\end{center}
\end{figure}

Figure \ref{fig:simulation_beta_app} shows coefficent images at $\tau = 0.25$ and $\tau = 0.75$. At $\tau = .75$ we can see that the behavior score have more adverse effect to the hippocampus volume. 

\bibliographystyle{chicago}
\bibliography{bibfile}